\documentclass[10pt, a4paper]{article}
\pdfoutput=1
\usepackage[utf8]{inputenc}
\usepackage[T1]{fontenc}
\usepackage{lmodern, textcomp}
\usepackage[british]{babel}
\usepackage{csquotes}

\usepackage{eurosym, xspace}
\usepackage[nottoc]{tocbibind}
\usepackage{graphicx, subcaption, float}
\usepackage{authblk}
\usepackage{marginnote}
\usepackage{xparse}

\usepackage{makecell, multirow}
\usepackage{multicol}
\usepackage{enumitem}
\usepackage[multiple]{footmisc}
\usepackage{fancyhdr}
\usepackage{setspace, titlesec}
\usepackage{changepage}
\usepackage{lastpage}

\usepackage{multirow}
\usepackage{booktabs}

\usepackage[usenames, dvipsnames, svgnames, table]{xcolor}

\usepackage[backend=bibtex8, citestyle=numeric-comp, maxbibnames=99,
			sorting=none, sortcase=false, sortcites=true, giveninits=true]{biblatex}

\usepackage{amsmath, amsfonts, amssymb}
\usepackage{mathtools, cancel}
\usepackage{mathrsfs}
\usepackage{siunitx}
\usepackage{listings}

\usepackage{framed, mdframed}
\usepackage[amsmath, hyperref, framed]{ntheorem}

\usepackage[pdftex, breaklinks=true, linktocpage,
	colorlinks=true, urlcolor=blue, linkcolor=blue, citecolor=red
]{hyperref}
\usepackage[all]{hypcap}

\DeclareUnicodeCharacter{00A0}{~}
\DeclareUnicodeCharacter{202F}{~}

\newcommand{\email}[1]{\href{mailto:#1}{\nolinkurl{#1}}}
\newcommand{\emailfoot}[1]{\thanks{\email{#1}}}

\newcounter{draftcommentcnt}
\NewDocumentCommand{\draftcomment}{s O{red} m}{%
	\def\margnote{\IfBooleanTF{#1}{\marginnote}{\marginpar}}%
	\stepcounter{draftcommentcnt}%
	\textcolor{#2}{#3}%
	\margnote{\textcolor{#2}{$\Leftarrow$ \arabic{draftcommentcnt}}}%
}
\newcounter{replycommentcnt}
\NewDocumentCommand{\replycomment}{s O{blue} m}{%
	\def\margnote{\IfBooleanTF{#1}{\marginnote}{\marginpar}}%
	\stepcounter{replycommentcnt}%
	\textcolor{#2}{#3}%
	\margnote{\textcolor{#2}{$\Leftarrow$ \arabic{replycommentcnt}}}%
}

\newcommand{\doi}[1]{\href{http://dx.doi.org/#1}{\nolinkurl{#1}}}

\graphicspath{{images/}}
\numberwithin{equation}{section}

\usepackage[sort&compress, english]{cleveref}

\usepackage[heightrounded,
	top=4cm, bottom=4cm, left=3.5cm, right=3.5cm,
]{geometry}

\lstdefinestyle{boxed}{frame=single, numbers=left}

\input{arxiv.def}

\hypersetup{
	pdftitle={Machine learning for complete intersection Calabi-Yau manifolds: a methodological study},
}

\title{Machine learning for complete intersection Calabi--Yau manifolds: a methodological study}

\author[1]{Harold Erbin\emailfoot{erbin@to.infn.it}}
\author[1]{Riccardo Finotello\emailfoot{riccardo.finotello@to.infn.it}}

\affil[1]{%
	Dipartimento di Fisica, Università di Torino
	\protect\\
	\protect\textsc{Infn} Sezione di Torino and Arnold--Regge Center,
	\protect\\
	Via Pietro Giuria 1, I-10125 Torino, Italy
}

\newcommand{\R}[0]{\ensuremath{\mathbb{R}}}
\newcommand{\C}[0]{\ensuremath{\mathbb{C}}}

\newcommand{\mc}[1]{{\mathcal{#1}}}

\newcommand{\abs}[1]{{|#1|}}
\newcommand{\Abs}[1]{{\left|#1\right|}}
\newcommand{\group}[1]{\mathrm{#1}}

\renewcommand{\P}[0]{\mathbb{P}}

\begin{document}
\maketitle

\begin{abstract}
We revisit the question of predicting both Hodge numbers $h^{1,1}$ and $h^{2,1}$ of complete intersection Calabi--Yau (CICY) $3$-folds using machine learning (ML), considering both the old and new datasets built respectively by Candelas--Dale--Lutken--Schimmrigk / Green--Hübsch--Lutken and by Anderson--Gao--Gray--Lee.
In real-world applications, implementing a ML system rarely reduces to feed the brute data to the algorithm.
Instead, the typical workflow starts with an exploratory data analysis (EDA) which aims at understanding better the input data and finding an optimal representation.
It is followed by the design of a validation procedure and a baseline model.
Finally, several ML models are compared and combined, often involving neural networks with a topology more complicated than the sequential models typically used in physics.
By following this procedure, we improve the accuracy of ML computations for Hodge numbers with respect to the existing literature.
First, we obtain $97\%$ (resp.\ $99\%$) accuracy for $h^{1,1}$ using a neural network inspired by the Inception model for the old dataset, using only $30\%$ (resp.\ $70\%$) of the data for training.
For the new one, a simple linear regression leads to almost $100\%$ accuracy with $30\%$ of the data for training.
The computation of $h^{2,1}$ is less successful as we manage to reach only $50\%$ accuracy for both datasets, but this is still better than the $16\%$ obtained with a simple neural network ( SVM with Gaussian kernel and feature engineering and sequential convolutional network reach at best $36\%$).
This serves as a proof of concept that neural networks can be valuable to study the properties of geometries appearing in string theory.
\end{abstract}

\newpage

\hrule
\tableofcontents
\bigskip
\hrule
\newpage

\section{Introduction}

The last few years have seen a major uprising of machine learning (ML), and more particularly of neural networks~\cite{Goodfellow:2016:DeepLearning, Chollet:2017:DeepLearningPython, Geron:2019:HandsOnMachineLearning}.
This technology is extremely efficient at discovering and predicting patterns and now pervades most fields of applied sciences and of the industry.
In view of its versatility, it is likely that ML will find its way towards high-energy and theoretical physics (see~\cite{Albertsson:2018:MachineLearningHigh, Mehta:2019:HighbiasLowvarianceIntroduction, Ntampaka:2019:RoleMachineLearning, Carleo:2019:MachineLearningPhysical, Buchanan:2019:PowerMachineLearning, Ruehle:2020:DataScienceApplications} for selected reviews).
One of the most critical places where progress can be expected is in understanding the geometries used to describe string compactifications.

String theory is the most developed candidate for a theory of quantum gravity together with the unification of matter and interactions.
However, it predicts ten spacetime dimensions: to recover our four-dimensional Universe, it is necessary to compactify six dimensions.
For string theory to be a fundamental theory of reality, a single compactification should describe the current Universe (obviously, other compactifications may enter at early or later stages since spacetime is dynamical).
Unfortunately, the number of possibilities -- forming the so-called string landscape -- is huge (numbers as high as $\num{e272000}$ have been suggested for some models)~\cite{Lerche:1989:ChiralFourDimensionalHeterotic, Douglas:2003:StatisticsStringMTheory, Ashok:2004:CountingFluxVacua, Douglas:2004:BasicResultsVacuum, Douglas:2007:FluxCompactification, Taylor:2015:FtheoryGeometryMost, Schellekens:2016:BigNumbersString, Halverson:2017:AlgorithmicUniversalityFtheory, Taylor:2018:ScanningSkeleton4D, Constantin:2019:CountingStringTheory}, the mathematical objects entering the compactifications are complex and typical problems are often NP-complete, NP-hard, or even undecidable~\cite{Denef:2007:ComputationalComplexityLandscape, Halverson:2019:ComputationalComplexityVacua, Ruehle:2020:DataScienceApplications}, making an exhaustive classification impossible.
Additionally, there is no single framework to describe all the possible (flux) compactifications.
As a consequence, each class of models must be studied with different methods.
This has prevented any precise connection to the existing and tested theories (in particular, the Standard Model of particle physics) or the proposal of a sharply defined and doable experiment.

Until recently, the string landscape has been studied using different methods: 1) analytic computations for simple examples, 2) general statistics, 3) random scans, 4) algorithmic enumerations of possibilities.
This has been a large endeavor of the string community, and we refer to the reviews~\cite{Grana:2006:FluxCompactificationsString, Lust:2009:SeeingStringLandscape, Ibanez:2012:StringTheoryParticle, Brennan:2018:StringLandscapeSwampland, Halverson:2018:TASILecturesRemnants, Ruehle:2020:DataScienceApplications} and to references therein for more details.
The main objective of such studies is to understand what are the generic predictions of string theory: even if “the” correct compactification has not been found, this helps to narrow down what to look for experimentally.
The first conclusion of these studies is that compactifications giving an effective theory close to the Standard Model are scarce.\footnotemark{}
\footnotetext{%
	This means that the gauge group is not much bigger than $\group{SU}(3) \times \group{SU}(2) \times \group{U}(1)$ and that there are not too many additional particles.
	The current bounds on BSM (Beyond Standard Model) physics put even stronger restrictions.
}%
Each of the four approaches display different limitations: 1) lacks of genericity, 2) is too much general, 3) ignores the structure of the landscape and has few chances to discover rare compactifications, 4) requires too much computational power to move beyond “simple” examples.
As a result, no major phenomenological progress has been seen in the last decade and finding a physical compactification looks still as remote.
In reaction to these difficulties and starting with the seminal paper~\cite{Abel:2014:GeneticAlgorithmsSearch}, new investigations based on ML appeared in the recent years, focusing on different aspects of the string landscape and of the geometries used in compactifications~\cite{Krefl:2017:MachineLearningCalabiYau, Ruehle:2017:EvolvingNeuralNetworks, He:2017:MachinelearningStringLandscape, Carifio:2017:MachineLearningString, Altman:2019:EstimatingCalabiYauHypersurface, Bull:2018:MachineLearningCICY, Cole:2019:TopologicalDataAnalysis, Klaewer:2019:MachineLearningLine, Mutter:2019:DeepLearningHeterotic, Wang:2018:LearningNonHiggsableGauge, Ashmore:2019:MachineLearningCalabiYau, Brodie:2020:MachineLearningLine, Bull:2019:GettingCICYHigh, Cole:2019:SearchingLandscapeFlux, Faraggi:2019:MachineLearningClassification, Halverson:2019:BranesBrainsExploring, He:2019:DistinguishingEllipticFibrations, Bies:2020:MachineLearningAlgebraic, Bizet:2020:TestingSwamplandConjectures, Halverson:2020:StatisticalPredictionsString, Krippendorf:2020:DetectingSymmetriesNeural, Otsuka:2020:DeepLearningKmeans, Parr:2020:ContrastDataMining, Parr:2020:PredictingOrbifoldOrigin} (see also~\cite{Erbin:2018:GANsGeneratingEFT, Betzler:2020:ConnectingDualitiesMachine, Chen:2020:MachineLearningEtudes, Gan:2017:HolographyDeepLearning, Hashimoto:2018:DeepLearningAdSCFT, Hashimoto:2018:DeepLearningHolographic, Hashimoto:2019:AdSCFTDeepBoltzmann, Tan:2019:DeepLearningHolographic, Akutagawa:2020:DeepLearningAdSQCD, Yan:2020:DeepLearningBlack, Comsa:2019:SO8SupergravityMagic, Bobev:2020:CornucopiaAdS5Vacua, Bobev:2020:PropertiesNewN, Krishnan:2020:MachineLearningcal} for related works).
For more context and a summary of the state of the art, the reader is referred to the excellent review~\cite{Ruehle:2020:DataScienceApplications}.
ML is extremely adequate when it comes to pattern search, which motivates two main applications to string theory: 1) explore systematically a space of possibilities (if they are not random, ML should be able to find a pattern, even if it is too complicated to be formulated explicitly), 2) obtain approximate results on distributions from which mathematical formulas can be deduced.

We want to address the question of computing the Hodge numbers $h^{1,1}$ and $h^{2,1}$ (positive integers) for \emph{complete intersection Calabi--Yau} (CICY) $3$-folds~\cite{Green:1987:CalabiYauManifoldsComplete} using different machine learning algorithms.
A CICY is completely specified by its \emph{configuration matrix} (with entries being positive integers), which is the basic input of the algorithms.
The CICY $3$-folds are the simplest Calabi--Yau and they have been well studied.
In particular, they have been completely classified and their topological properties computed~\cite{Candelas:1988:CompleteIntersectionCalabiYau, Green:1989:AllHodgeNumbers, Anderson:2017:FibrationsCICYThreefolds} (see~\cite{Lutken:1988:RecentProgressCalabiYauology, Hubsch:1992:CalabiYauManifoldsBestiary, Anderson:2018:TASILecturesGeometric, He:2020:CalabiYauSpacesString} for reviews).
For these reasons, they provide an excellent sandbox to test ML algorithms in a controlled environment.
More particularly, simple tests show that the task is difficult for simple ML algorithms -- even neural networks -- such that this is an interesting challenge to solve before moving to more difficult problems.

The goal is to predict two positive integers from a matrix of positive integers.
This task is complicated by various redundancies in the description (such as an independence in the permutations of lines and columns).
A simple sequential network taking only the matrix as input performs badly, especially for $h^{2,1}$.
As a consequence, more advanced methods are needed.
While usual physics application of ML reduces to feeding a (big) sequential neural network with raw data, real-world applications are built following a more general workflow~\cite{Coursera:HowWinData, Geron:2019:HandsOnMachineLearning, Skiena:2017:DataScience}: 1) understanding of the problem, 2) exploratory data analysis (EDA), 3) design of a baseline, 4) definition of a validation strategy, 5) feature engineering and selection, 6) design of ML models, 7) ensembling.

While the first step is straightforward, it is still interesting to notice that computations involved in string geometries (using algebraic topology) are far from standard applications of ML algorithms, which makes the problem even more interesting.
EDA aims at understanding better the dataset, in particular, by finding how the variables are distributed, correlated, determining if there are outliers, etc.
This analysis naturally leads to designing new features from the existing ones, which is called \emph{feature engineering}.
Indeed, putting derived features by hand may make the data more easily understandable by the ML algorithms, for example by emphasizing important properties.\footnotemark{}
\footnotetext{%
	While one could expect ML algorithms to generate these features by themselves, this may complicate the learning process.
	So in cases where it is straightforward to compute meaningful derived features, it is often worth considering them.
}%
This phase is followed by \emph{feature selection}, where different set of features are chosen according to the need of each algorithm from step~6).
In between, one needs to set up a validation strategy to ensure that the predictions appropriately reflect the real values, together with a baseline model, which gives a lower bound on the accuracy together with a working pipeline.\footnotemark{}
\footnotetext{%
	For example, the original work on this topic~\cite{He:2017:MachinelearningStringLandscape} did not set up a validation strategy and reported the accuracy over both the training and test data.
	Correcting this problem leads to an accuracy of $37\%$~\cite{Bull:2018:MachineLearningCICY}, which is lower than the linear regression baseline.
}%
For instance, we find that a simple linear regression using the configuration matrix as input gives \SIrange{43.6}{48.8}{\percent} for $h^{1,1}$ and \SIrange{9.6}{10.4}{\percent} for $h^{2,1}$ using from $20\%$ to $80\%$ of data for training.
Hence, any algorithm \emph{must} do better than this to be worth considering.
Finally, we can build different models in step~6), in particular, by considering different topologies of neural networks beyond the simplest sequential models.
The last optional step consists in combining different models together in order to improve the results.
With respect to the whole process, the purpose of this paper is also pedagogical and aims at exemplifying how these steps are performed in an applied ML project.

There is a finite number of $7890$ CICY $3$-folds.
Due to the freedom in representing the configuration matrix, two datasets have been constructed: the “original dataset”~\cite{Candelas:1988:CompleteIntersectionCalabiYau, Green:1989:AllHodgeNumbers} and the “favourable dataset”~\cite{Anderson:2017:FibrationsCICYThreefolds}.
A configuration matrix is said to be favorable if its second cohomology descends completely from the second cohomology of the ambient space: this implies that $h^{1,1}$ equals the number of projective spaces in the ambient space~\cite{Anderson:2017:FibrationsCICYThreefolds, Gray:2014:TopologicalInvariantsFibration}.
In the “favourable dataset”, all configuration matrices are favorable whenever possible (\SI{99.1}{\percent}), whereas in the “original dataset” only \SI{61.8}{\percent} of the matrices are favorable.
Both datasets will be described in more details in \Cref{sec:data:datasets}.

Our analysis continues and generalizes~\cite{He:2017:MachinelearningStringLandscape, Bull:2018:MachineLearningCICY} at different levels.
We compute $h^{2,1}$, which has been ignored in~\cite{He:2017:MachinelearningStringLandscape, Bull:2018:MachineLearningCICY}, where the authors argue that it can be computed from $h^{1,1}$ and from the Euler characteristics (a simple formula exists for the latter).
In our case, we want to push the idea of using ML to learn about the physics (or the mathematics) of CY to its very end: we assume that we do not know anything about the mathematics of the CICY, except that the configuration matrix is sufficient to derive all quantities.
Moreover, we have already mentioned that ML algorithms have rarely been used to derive data in algebraic topology, which can be a difficult task.
For this reason, obtaining also $h^{2,1}$ from ML techniques is an important first step towards using ML for more general problems in string geometries.
In particular, this helps to prepare the study of CICY $4$-folds (classified in~\cite{Gray:2013:AllCompleteIntersection}) for which there are four Hodge numbers which are expected to be even more difficult to compute.
Finally, regression is also more useful for extrapolating results: a classification approach assumes that we already know all the possible values of the Hodge numbers and has difficulties to predict labels which do not appear in the training set.
This is necessary when we move to a dataset for which not all topological quantities have been computed, for instance CY constructed from the Kreuzer--Skarke list of polytopes~\cite{Kreuzer:2002:CompleteClassificationReflexive}.

\begin{figure}[tbp]
    \centering
    \begin{subfigure}[c]{\textwidth}
        \centering
        \includegraphics[width=0.8\textwidth]{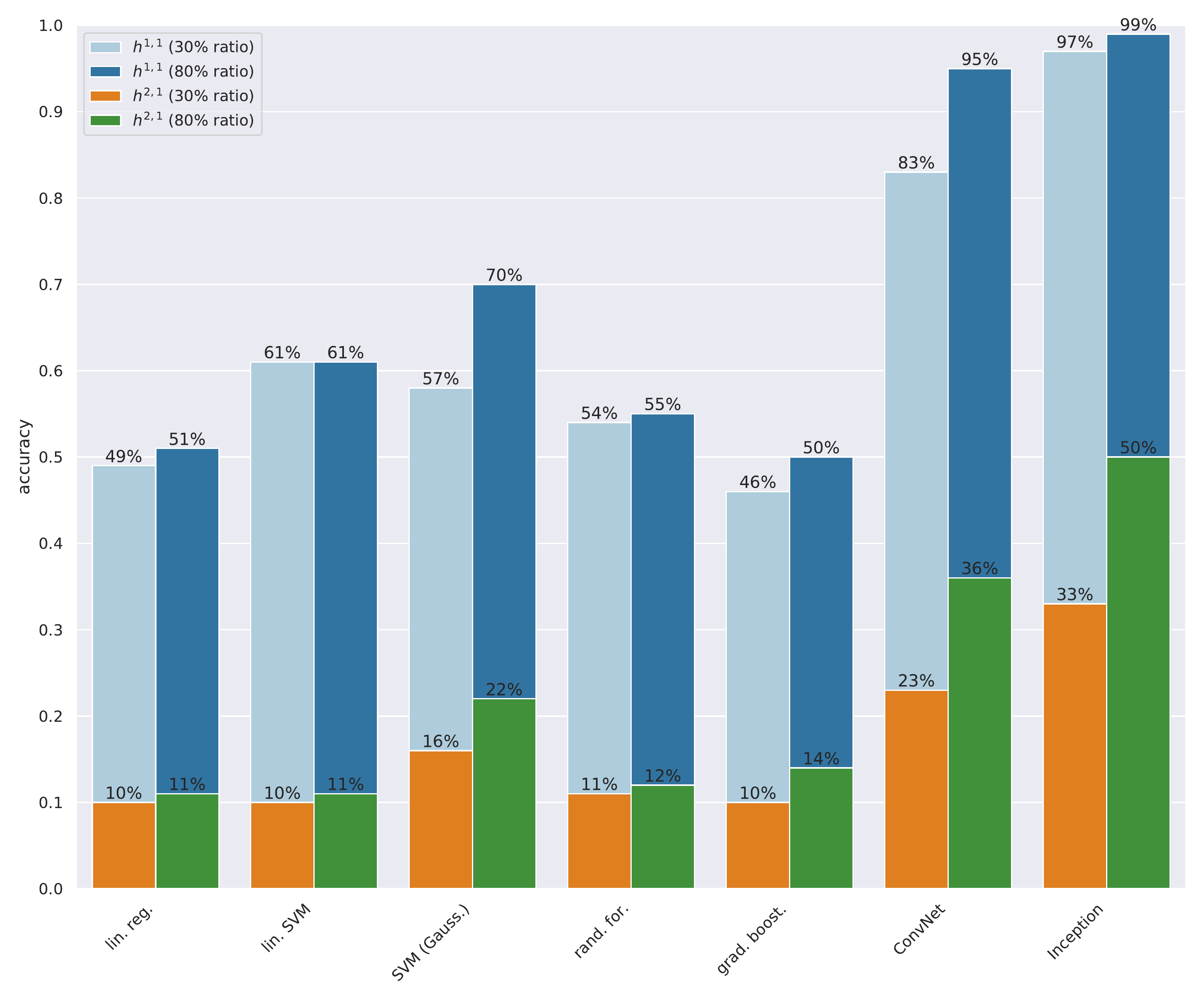}
        \caption{Accuracy reached when trained using only the configuration matrix.}
    \end{subfigure}
    \begin{subfigure}[c]{\textwidth}
        \centering
        \includegraphics[width=0.8\textwidth]{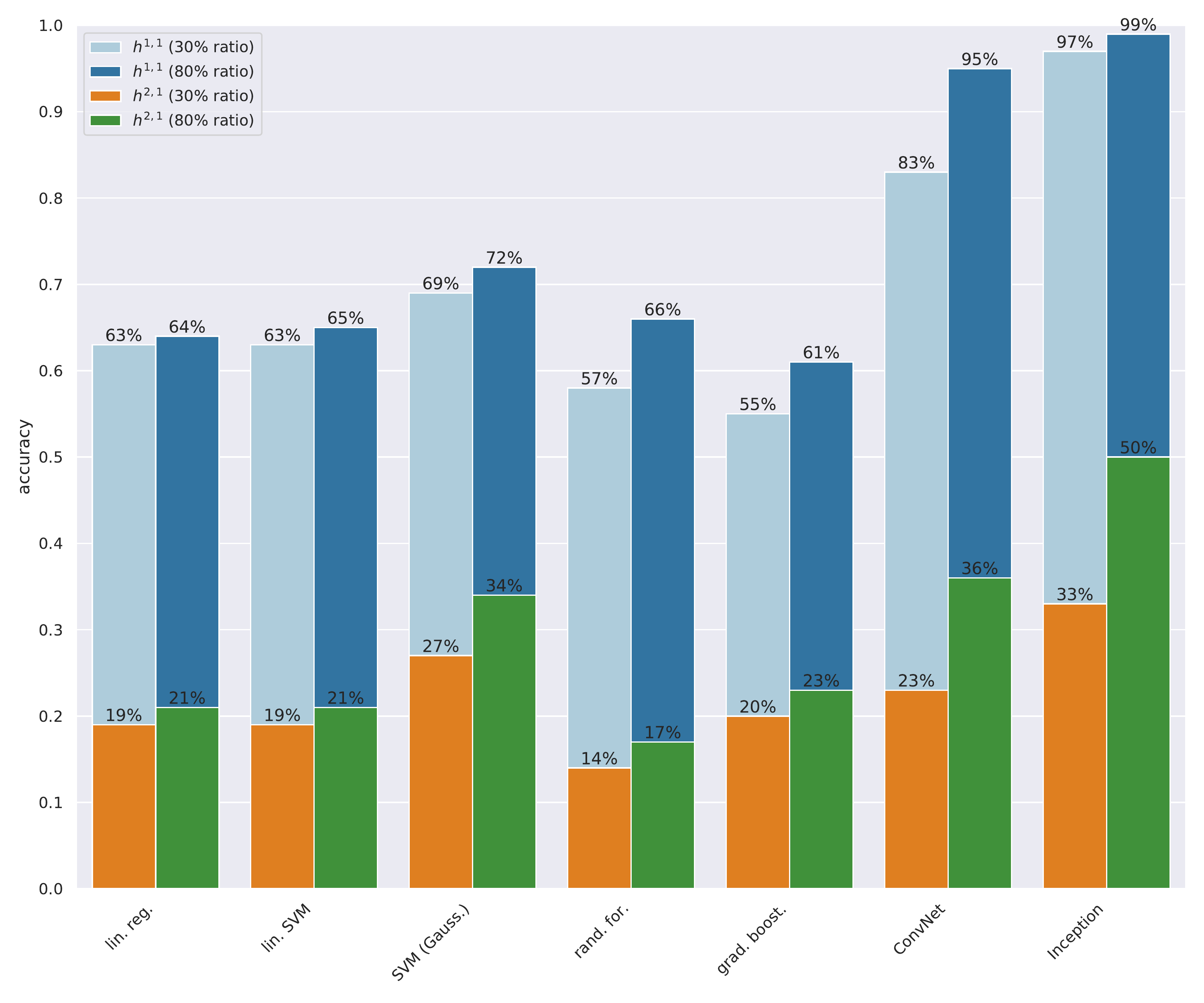}
        \caption{Accuracy reached using the best training set for each algorithm.}
    \end{subfigure}
    \caption{The plots show the best accuracy reached by the models considered in this paper for the old dataset.
    The models are trained to predict separately $h^{1,1}$ and $h^{2,1}$ and using $30\%$ and $80\%$ of the data for training.}
    \label{fig:intro:comparison}
\end{figure}

In this paper, we compare the performances of the following algorithms: linear regression, support vector machines (SVM) with linear and Gaussian kernels, decision trees and ensemble thereof -- random forests and gradient boosting --, and deep neural networks.
The best results obtained with and withou feature engineering are displayed in \Cref{fig:intro:comparison} for the old dataset.
We find that, in all cases except neural networks, using engineered features greatly enhance the performances.
The EDA reveals that the number of projective spaces forming the ambient space (equal to the number of rows) is a particularly distinguished feature.
In fact, all algorithms yield an accuracy of \SIrange{99}{100}{\percent} for $h^{1,1}$ in the favorable dataset.
For the linear regression, this directly gives the well-known results~\cite{Anderson:2017:FibrationsCICYThreefolds} that $h^{1,1}$ equals the number of projective spaces for favorable configuration matrix.
In the case of the original dataset, the best model is a neural network inspired by Google's Inception model~\cite{Szegedy:2014:GoingDeeperConvolutions, Szegedy:2015:RethinkingInceptionArchitecture, Szegedy:2016:Inceptionv4InceptionResNetImpact}, which allows to reach nearly \SI{100}{\percent} accuracy.
This neural network is further studied in~\cite{Erbin:2020:InceptionCICY}.
The algorithms are not as successful for $h^{2,1}$, with the Inception model giving again the best result, close to $\SI{50}{\percent}$ accuracy -- which is still much better that what the baseline or simple models do.
We leave improving the computation of $h^{2,1}$ and interpreting what the different algorithms learn for a future work.

The data analysis and ML are programmed in Python using standard open-source packages: \texttt{pandas}~\cite{McKinney:2011:Pandas}, \texttt{matplotlib}~\cite{Hunter:2007:Matplotlib}, \texttt{seaborn}~\cite{Waskom:2017:Seaborn}, \texttt{scikit-learn}~\cite{Pedregosa:2011:ScikitLearn}, \texttt{scikit-optimize}~\cite{Head:2018:ScikitOptimize}, \texttt{tensorflow}~\cite{Package:Tensorflow} (and its high level API \texttt{Keras}~\cite{Package:Keras}).
The code and its description are available on \href{https://thesfinox.github.io/ml-cicy/}{Github}.

This paper is organized as follows.
In \Cref{sec:data}, we first recall the definition of Calabi--Yau manifolds (\Cref{sec:data:cy}) and describe the two existing CICY datasets (\Cref{sec:data:datasets}).
We then engineer new features before performing an EDA for both datasets (\Cref{sec:data:eda}), reproducing some well-known figures from the literature.
Then, in \Cref{sec:ml}, we implement the different ML algorithms.
Our paper culminates in the description of the Inception-like neural network in \Cref{sec:ml:nn:inception} where we reach the highest accuracy.
Finally, we discuss our results in \Cref{sec:conclusion}.
\Cref{app:ml-algo} contains details on the different algorithms used in this paper.

\section{Data Analysis}
\label{sec:data}

In this section, we introduce \emph{Calabi--Yau} (CY) manifolds before describing the two datasets of CICY manifolds (\Cref{sec:data:datasets}).
Since the CICY have been completely classified, they provide a good opportunity for testing ideas from ML in a controlled setting.
In order to select the most appropriate learning algorithm, we perform a preliminary \emph{exploratory data analysis} (EDA) in \Cref{sec:data:eda}.

\subsection{Calabi--Yau Manifolds}
\label{sec:data:cy}

A CY $n$-fold is a $n$-dimensional complex manifold $X$ with $\group{SU}(n)$ holonomy (they have $2n$ real dimensions).
An equivalent definition is the vanishing of its first Chern class.
A standard reference for the physicist is~\cite{Hubsch:1992:CalabiYauManifoldsBestiary} (see also~\cite{Anderson:2018:TASILecturesGeometric, He:2020:CalabiYauSpacesString} for useful references).

The most relevant case for superstring compactifications are CY $3$-folds.
Indeed, superstrings are well-defined only in $10$ dimensions: in order to recover a $4$-dimensional theory, it is necessary to compactify $6$ dimensions~\cite{Hubsch:1992:CalabiYauManifoldsBestiary}.
Importantly, the compactification on a CY leads to the breaking of a large part of the supersymmetry, which is phenomenologically more realistic.

Calabi--Yau manifolds are characterized by a certain number of topological properties, the most salient being the Hodge numbers $h^{1,1}$ and $h^{2,1}$, counting respectively the Kähler and complex structure deformations, and the Euler characteristics\footnotemark{}
\footnotetext{%
	In full generality, the Hodge numbers $h^{p,q}$ count the numbers of harmonic $(p, q)$-forms.
}%
\begin{equation}
	\chi = 2 (h^{1,1} - h^{2,1}).
	\label{eq:cy:euler}
\end{equation}
Interestingly, topological properties of the manifold directly translates into features of the $4$-dimensional effective action (in particular, the number of fields, the representations and the gauge symmetry)~\cite{Hubsch:1992:CalabiYauManifoldsBestiary, Becker:2006:StringTheoryMTheory}.\footnotemark{}
\footnotetext{%
	Another reason for sticking to topological properties is that there is no CY for which the metric is known.
	Hence, it is not possible to perform explicitly the Kaluza--Klein reduction in order to derive the $4$-dimensional theory.
}%
In particular, the Hodge numbers count the number of chiral multiplets (in heterotic compactifications) and the number of hyper- and vector multiplets (in type II compactifications): these are related to the number of fermion generations ($3$ in the Standard Model) and is thus an important measure of the distance to the Standard Model.

The simplest CYs are constructed by considering the complete intersection of hypersurfaces in a product $\mc A$ of projective spaces $\P^{n_i}$ (called the ambient space)~\cite{Green:1987:CalabiYauManifoldsComplete, Green:1987:PolynomialDeformationsCohomology, Candelas:1988:CompleteIntersectionCalabiYau, Green:1989:AllHodgeNumbers, Anderson:2017:FibrationsCICYThreefolds, Anderson:2018:TASILecturesGeometric}:
\begin{equation}
	\mc A = \P^{n_1} \times \cdots \times \P^{n_m}.
\end{equation}
Such hypersurfaces are defined by homogeneous polynomial equations: a Calabi--Yau $X$ is described by the solution to the system of equations, i.e.\ by the intersection of all these surfaces (that the intersection is “complete” means that the hypersurface is non-degenerate).

To gain some intuition, consider the case of a single projective space $\P^n$ with (homogeneous) coordinates $Z^I$, $I = 0, \ldots, n$.
In this case, a codimension $1$ subspace is obtained by imposing a single homogeneous polynomial equation of degree $a$ on the coordinates
\begin{equation}
	\begin{gathered}
	p_a(Z^0, \ldots, Z^n)
		= P_{I_1 \cdots I_a} Z^{I_1} \cdots Z^{I_a}
		= 0,
	\\
	p_a(\lambda Z^0, \ldots, \lambda Z^n) = \lambda^a \, p_a(Z^0, \ldots, Z^n).
	\end{gathered}
\end{equation}
Each choice of the polynomial coefficients $P_{I_1 \cdots I_a}$ leads to a different manifold.
However, it can be shown that the manifolds are (generically) topologically equivalent.
Since we are interested only in classifying the CY as topological manifolds and not as complex manifolds, the information about $P_{I_1 \cdots I_a}$ can be forgotten and it is sufficient to keep track only on the dimension $n$ of the projective space and of the degree $a$ of the equation.
The resulting hypersurface is denoted equivalently as $[\P^n \mid a] = [n \mid a]$.
Finally, $[\P^n \mid a]$ is $3$-dimensional if $n = 4$ (the equation reduces the dimension by one), and it is a CY (the “quintic”) if $a = n + 1 = 5$ (this is required for the vanishing of its first Chern class).
The simplest representative of this class if Fermat's quintic defined by the equation
\begin{equation}
	\sum_{I=0}^{4} (Z^I)^5 = 0.
\end{equation}

This construction can be generalized to include $m$ projective spaces and $k$ equations, which can mix the coordinates of the different spaces.
A CICY $3$-fold $X$ as a topological manifold is completely specified by a \emph{configuration matrix} denoted by the same symbol as the manifold:
\begin{equation}
	X =
		\left[
		\begin{array}{c|ccc}
			\mathbb P^{n_1} & a_1^1 & \cdots & a_k^1
			\\
			\vdots & \vdots & \ddots & \vdots
			\\
			\mathbb P^{n_m} & a_1^m & \cdots & a_k^m
		\end{array}
		\right]
\end{equation}
where the coefficients $a^r_\alpha$ are positive integers and satisfy the following constraints
\begin{equation}
	\label{eq:cicy-constraints}
	\dim_\C X = \sum_{r=1}^{m} n_r - k = 3,
	\qquad
	\forall r: \quad
		n_r + 1 = \sum_{\alpha=1}^k a_\alpha^r.
\end{equation}
The first relation states that the dimension of the ambient space minus the number of equations equals the dimension of the CY $3$-fold.
The second set of constraints arise from the vanishing of its first Chern class; they imply that the $n_i$ can be recovered from the matrix elements.

In this case also, two manifolds described by the same configuration matrix but different polynomials are equivalent as real manifold (they are diffeomorphic) -- and thus as topological manifolds --, but they are different as complex manifolds.
Hence, it makes sense to write only the configuration matrix.

A given topological manifold is not described by a unique configuration matrix.
First, any permutation of the lines and columns leave the intersection unchanged (it amounts to relabelling the projective spaces and equations).
Secondly, two intersections can define the same manifold.
The ambiguity in the line and column permutations is often fixed by imposing some ordering of the coefficients.
Moreover, in most cases, there is an optimal representation of the manifold $X$, called favourable~\cite{Anderson:2017:FibrationsCICYThreefolds}: in such a form, topological properties of $X$ can be more easily derived from the ambient space $\mc A$.

\subsection{Datasets}
\label{sec:data:datasets}

Simple arguments~\cite{Green:1987:CalabiYauManifoldsComplete, Candelas:1988:CompleteIntersectionCalabiYau, Lutken:1988:RecentProgressCalabiYauology} show that the number of CICY is necessarily finite due to the constraints \eqref{eq:cicy-constraints} together with identities between complete intersection manifolds.
The classification of the CICY $3$-folds has been tackled in~\cite{Candelas:1988:CompleteIntersectionCalabiYau}, which established a dataset of $7890$ CICY.\footnotemark{}
\footnotetext{%
	However, there are redundancies in this set~\cite{Candelas:1988:CompleteIntersectionCalabiYau, Anderson:2008:MonadBundlesHeterotic, Anderson:2017:FibrationsCICYThreefolds}; this fact will be ignored in this paper.
}%
The topological properties of each of these manifolds have been computed in~\cite{Green:1989:AllHodgeNumbers}.
More recently, a new classification has been performed~\cite{Anderson:2017:FibrationsCICYThreefolds} in order to find the favourable representation of each manifold whenever it is possible.

\begin{figure}[htp]
	\centering

	\begin{subfigure}[c]{.45\linewidth}
		\centering
		\includegraphics[width=\textwidth, trim={0 0.45in 6in 0}, clip]{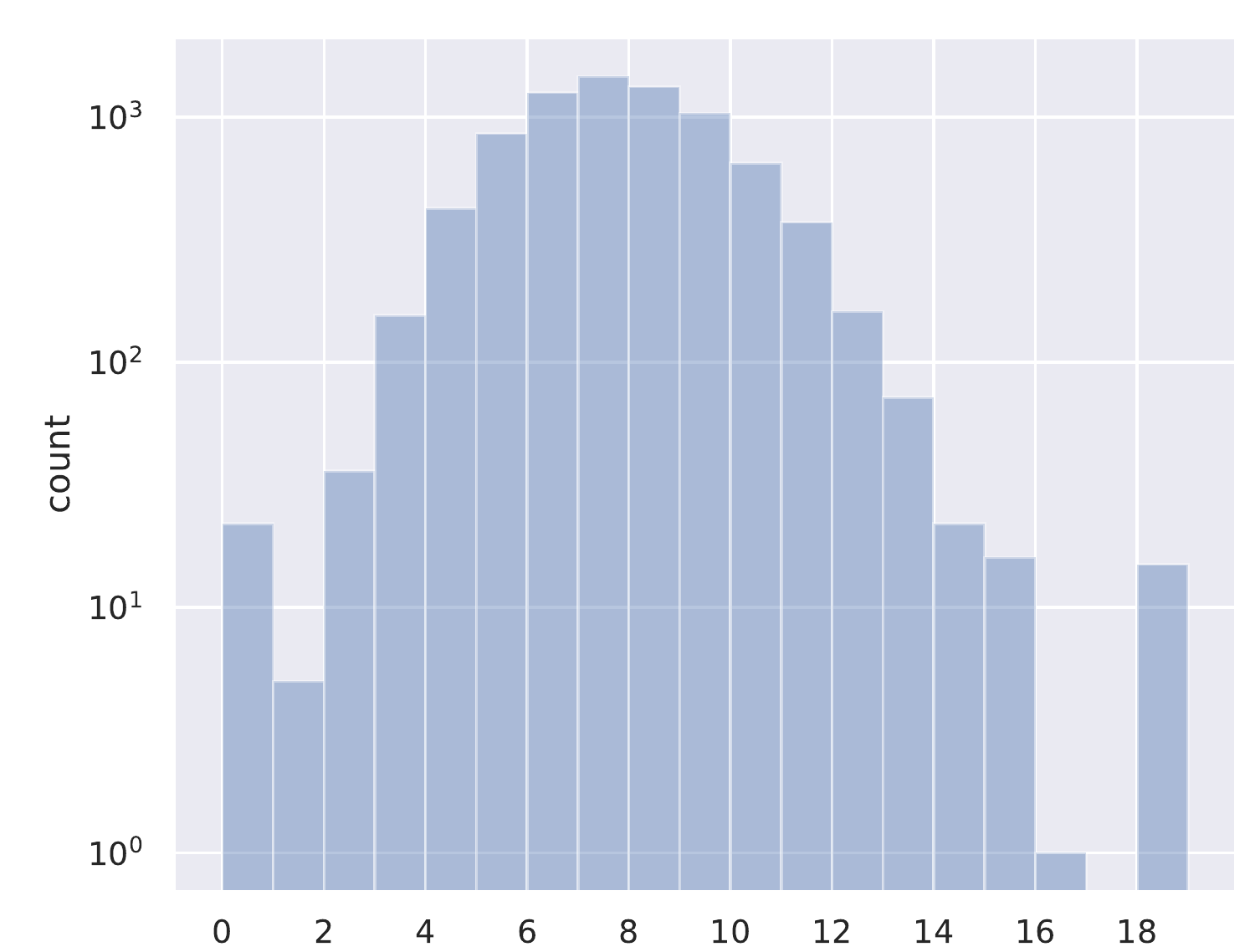}
		\caption{$h^{1,1}$}
		\label{fig:data:hist-h11}
	\end{subfigure}
	\quad
	\begin{subfigure}[c]{.45\linewidth}
		\centering
		\includegraphics[width=\textwidth, trim={6in 0.45in 0 0}, clip]{images/label-distribution_orig}
		\caption{$h^{2,1}$}
		\label{fig:data:hist-h21}
	\end{subfigure}

	\caption{Distribution of the Hodge numbers (log scale).}
	\label{fig:data:hist-hodge}
\end{figure}

Below we show a list of the CICY properties and of their configuration matrices:
\begin{itemize}
	\item general properties
	\begin{itemize}
		\item number of configurations: $7890$

		\item number of product spaces (block diagonal matrix): $22$

		\item $h^{11} \in [0, 19]$, $18$ distinct values (\Cref{fig:data:hist-h11})

		\item $h^{21} \in [0, 101]$, $65$ distinct values (\Cref{fig:data:hist-h21})

		\item unique Hodge number combinations: $266$
	\end{itemize}

	\item “original dataset”~\cite{Candelas:1988:CompleteIntersectionCalabiYau, Green:1989:AllHodgeNumbers}

	\begin{itemize}
		\item maximal size of the configuration matrices: $12 \times 15$

		\item number of favourable matrices (excluding product spaces): $4874$ ($\num{61.8}\%$)

		\item number of non-favourable matrices (excluding product spaces): $2994$

		\item number of different ambient spaces: $235$
	\end{itemize}

	\item “favourable dataset”~\cite{Anderson:2017:FibrationsCICYThreefolds}

	\begin{itemize}
		\item maximal size of the configuration matrices: $15 \times 18$

		\item number of favourable matrices (excluding product spaces): $7820$ ($\num{99.1}\%$)

		\item number of non-favourable matrices (excluding product spaces): $48$

		\item number of different ambient spaces: $126$
	\end{itemize}
\end{itemize}

The configuration matrix completely encodes the information of the CICY and all topological quantities can be derived from it.
However, the computations are involved and there is often no closed-form expression.
This situation is typical in algebraic geometry, and it can be even worse for some problems, in the sense that it is not even known how to compute the desired quantity (think to the metric of CYs).
For these reasons, it is interesting to study how we can retrieve these properties using ML algorithms.
In the current paper, following~\cite{He:2017:MachinelearningStringLandscape, Bull:2018:MachineLearningCICY}, we focus on the computation of the Hodge numbers with the initial scheme:
\begin{equation}
	\text{Input: configuration matrix}
	\quad \longrightarrow \quad
	\text{Output: Hodge numbers}
\end{equation}
To provide a good test case for the use of ML in context where the mathematical theory is not completely understood, we will make no use of known formulas.

\subsection{Exploratory Data Analysis}
\label{sec:data:eda}

A typical ML project does not consist of feeding the raw data -- here, the configuration matrix -- to the algorithm.
It is instead preceded by a phase of exploration in order to better understand the data, which in turn can help to design the learning algorithms.
We call \emph{features} properties given as inputs, and \emph{labels} the targets of the predictions.
There are several phases in the exploratory data analysis (EDA):
\begin{enumerate}
	\item \emph{feature engineering}: new features are derived from the inputs;

	\item \emph{feature selection}: the most relevant features are chosen to explain the targets;

	\item \emph{data augmentation}: new training data is generated from the existing ones;

	\item \emph{data diminution}: part of the training data is not used.
\end{enumerate}
For pragmatical introductions, the reader is refereed to~\cite{Coursera:HowWinData, Skiena:2017:DataScience}.

Engineered features are redundant, by definition, but can help the algorithm learn more efficiently by providing an alternative formulation and by drawing attention on salient characteristics.
A simple example is the following: given a series of numbers, one can compute different statistics -- median, mean, variance, etc. -- and add them to the inputs.
It may happen that the initial series becomes then irrelevant once this new information is introduced.

Another approach to improve the learning process is to augment or decrease the number of training samples artificially.
For example, one can use invariances of the inputs to generate more training data.
This does not help in our case because the entries of the configuration matrices are partially ordered.
Another possibility is to remove outliers which can damage the learning process by driving the algorithm far from the best solution.
If there are few of them, it is better to ignore them altogether during training since an algorithm which is not robust to outliers will in any case make bad predictions (a standard illustration is given by the Pearson and Spearman correlation coefficients, with the first not being robust to outliers~\cite{Skiena:2017:DataScience}).

Finding good features and selecting those to keep requires trials and errors.
In general, it is not necessary to keep track of all steps, but we feel that it is useful to do so in this paper for a pedagogical purpose.

Before starting the EDA, the first step should be to split the data into training and validation sets to avoid biasing the choices of the algorithm and the strategy: the EDA should be performed only on the training set.
However, the dataset we consider is complete and quite uniform: a subset of it would display the same characteristics as the entire set.
To give a general overview of the properties -- which can be useful for the reader interested in understanding the statistics of the CICY and for applications to string compactifications -- we work with the full dataset.

\subsubsection{Engineering}

\begin{figure}[htp]
	\centering

	\begin{subfigure}[c]{.45\linewidth}
		\centering
		\includegraphics[width=\textwidth, trim={6in 0 0 0}, clip]{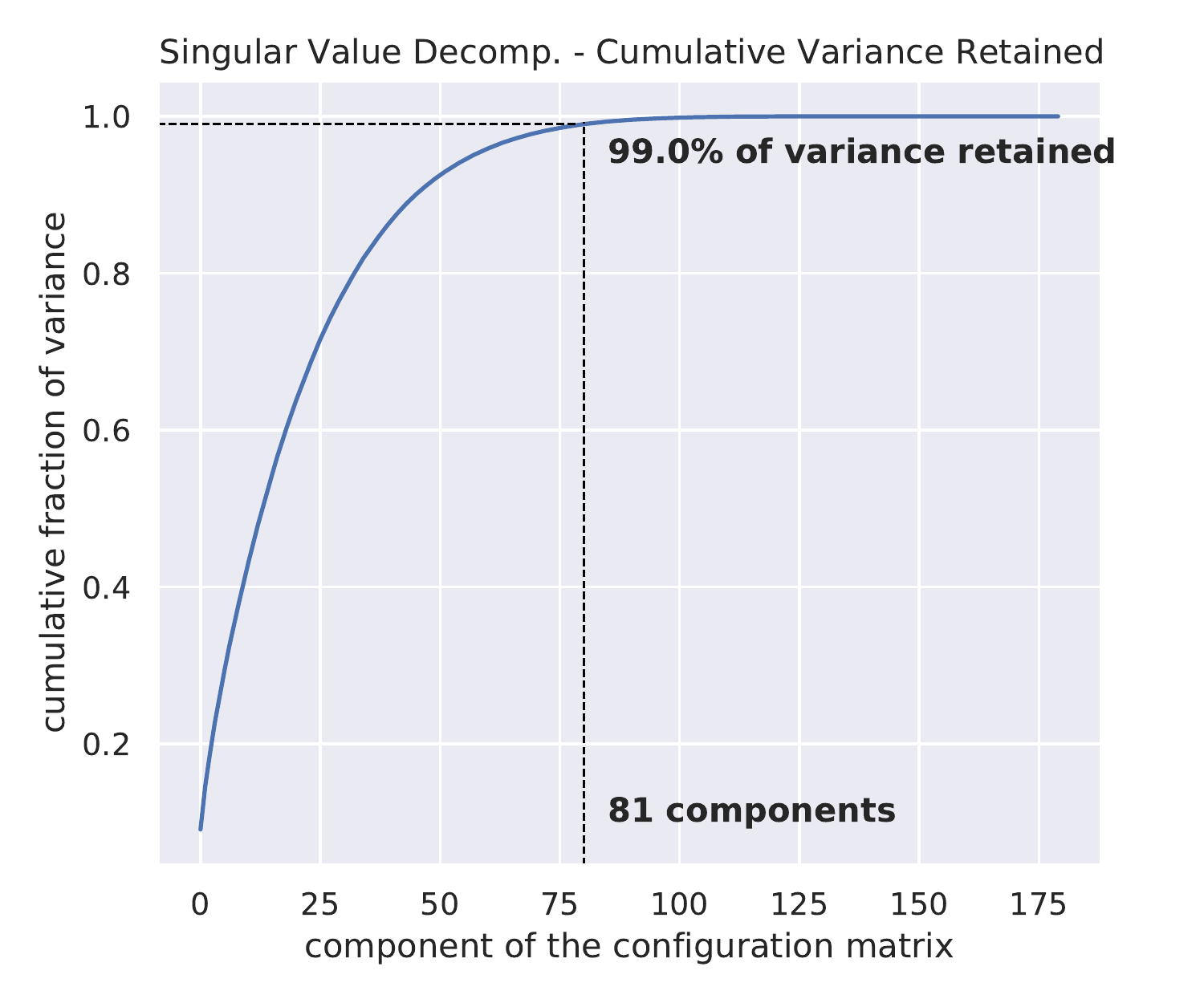}
		\caption{original dataset}
	\end{subfigure}
	\quad
	\begin{subfigure}[c]{.45\linewidth}
		\centering
		\includegraphics[width=\textwidth, trim={6in 0 0 0}, clip]{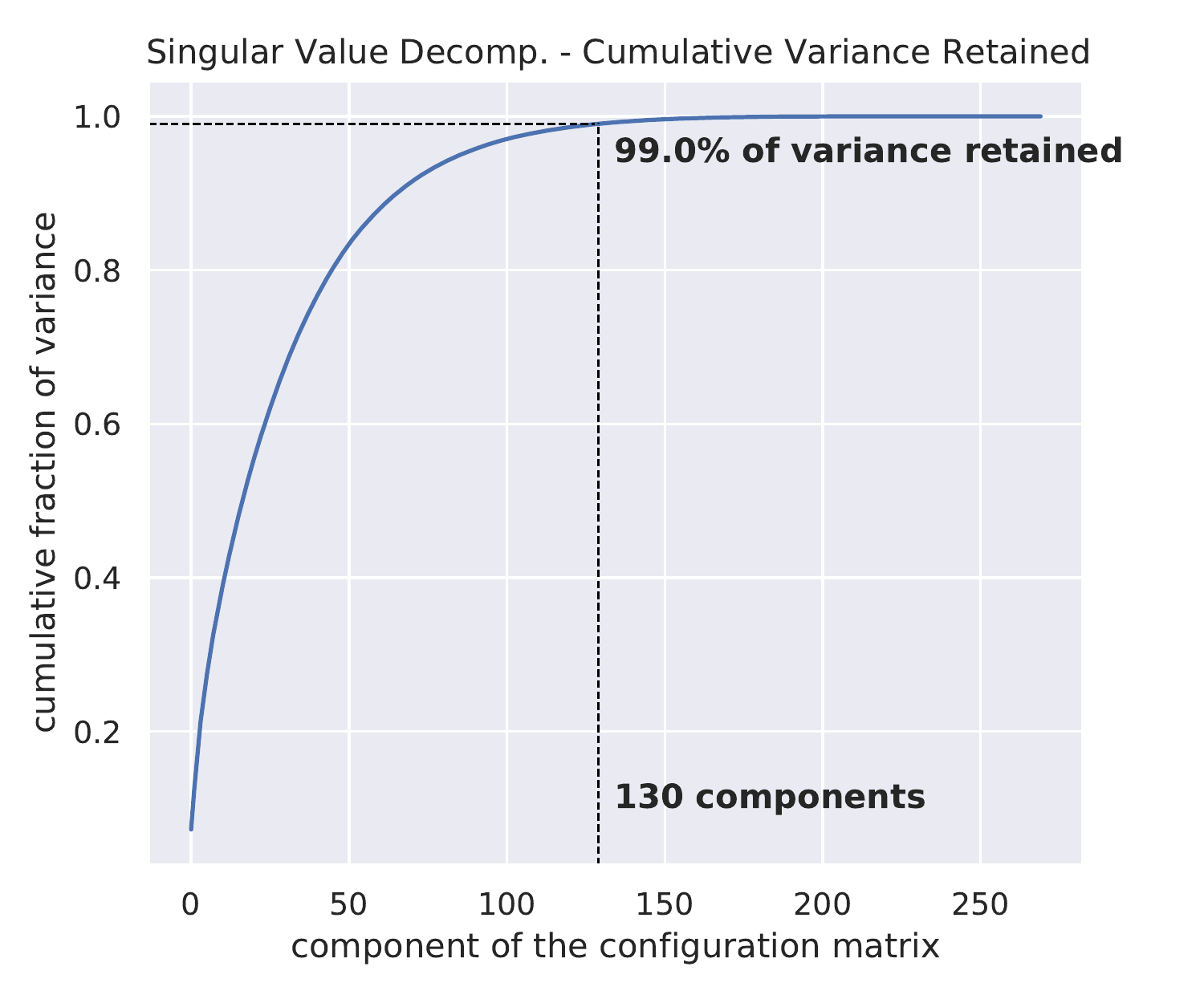}
		\caption{favourable dataset}
	\end{subfigure}
	
	\caption{Cumulative retained variance of the principal components of the configuration matrix in the original and favourable dataset.}
	\label{fig:eda:svd}
\end{figure}

Any transformation of the input data which has some mathematical meaning can be a useful feature.
We have established the following list of possibly useful quantities (most of them are already used to characterise CICY in the literature~\cite{Hubsch:1992:CalabiYauManifoldsBestiary}):
\begin{itemize}
	\item the number of projective spaces (rows), $m = $ \texttt{num\_cp};

	\item the number of equations (columns), $k = $ \texttt{num\_eqs};

	\item the number of $\P^1$, $f = $ \texttt{num\_cp\_1};

	\item the number of $\P^2$, \texttt{num\_cp\_2};

	\item the number of $\P^n$ with $n \neq 1$, $F = $ \texttt{num\_cp\_neq1};

	\item the excess number $N_{ex} = \sum\limits_{r=1}^F (n_r + f + m - 2k) =$ \texttt{num\_ex};

	\item the dimension of the cohomology group $H^0$ of the ambient space, \texttt{dim\_h0\_amb};

	\item the Frobenius norm of the matrix, \texttt{norm\_matrix};

	\item the list of the projective space dimensions \texttt{dim\_cp} and statistics thereof (min, max, median, mean);

	\item the list of the equation degrees \texttt{deg\_eqs} and statistics thereof (min, max, median, mean);

	\item $k$-means clustering on the components of the configuration matrix (with a number of clusters going from 2 to 15);\footnotemark{}
	\footnotetext{%
	    The algorithm determines the centroids of conglomerates of data called \textit{clusters} using an iterative process which computes the distance of each sample from the center of the cluster.
	    It then assigns the label of the cluster to the nearest samples.
		We used the class \texttt{cluster.KMeans} in \texttt{scikit-learn}.
	}%

	\item principal components of the configuration matrix derived using a principal components analysis (PCA) with 99\% of the variance retained (see \Cref{fig:eda:svd}).
\end{itemize}

\subsubsection{Selection}

\paragraph{Correlations}

\begin{figure}[htp]
	\centering

	\begin{subfigure}[c]{.45\linewidth}
		\centering
		\includegraphics[width=\textwidth]{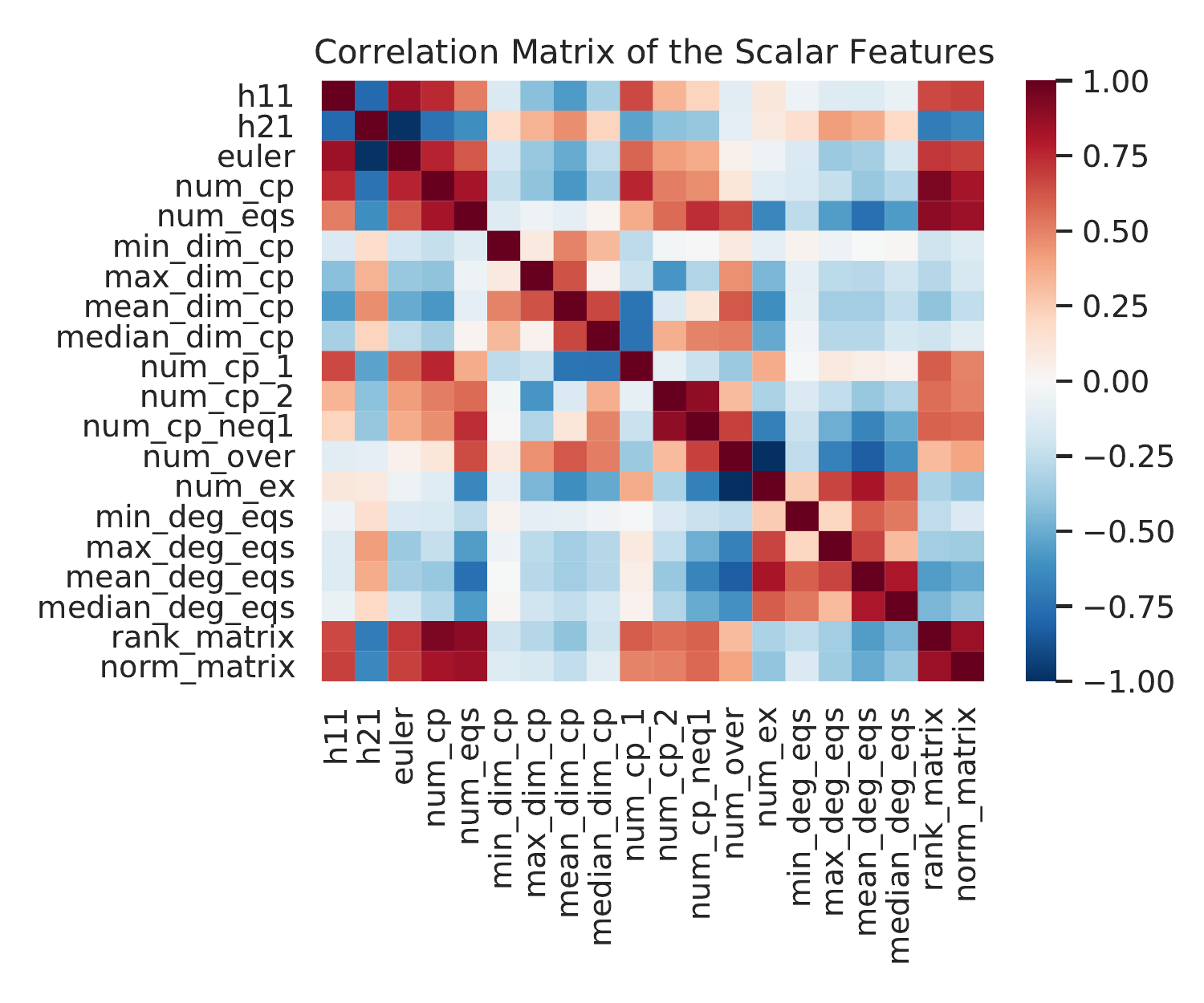}
		\caption{original dataset}
	\end{subfigure}
	\quad
	\begin{subfigure}[c]{.45\linewidth}
		\centering
		\includegraphics[width=\textwidth]{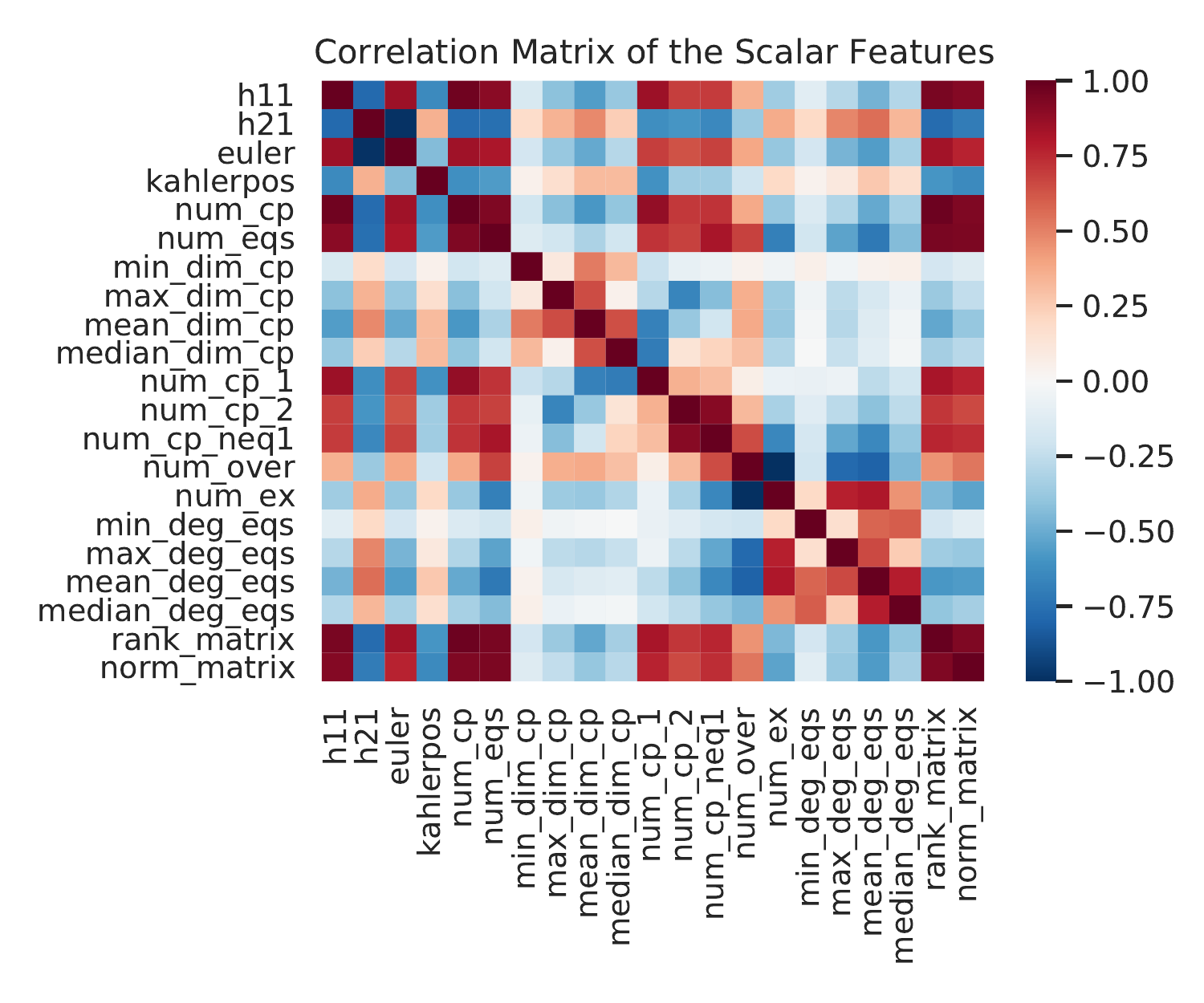}
		\caption{favourable dataset}
		\label{}
	\end{subfigure}

	\caption{Correlations between the engineered scalar features and the labels.}
	\label{fig:eda:corr}
\end{figure}

To get a first general idea, it is useful to take a look at the correlation matrix of the features and the labels.\footnotemark{}
\footnotetext{%
	The correlation is defined as the ratio between the covariance of two variables $\sigma(x, y) = \sum_{i} (x_i - \bar{x})(y_i - \bar{y})$ and the product of the standard deviations $\sigma(x)\sigma(y)$ (in this case $\bar{x}$ and $\bar{y}$ are the sample means).
}%
The correlation matrices for the scalar variables are displayed in \Cref{fig:eda:corr} for the original and favourable datasets (this excludes the configuration matrix).

As we can see, some engineered features are strongly correlated, especially in the favourable dataset.
In particular $h^{1,1}$ (respectively $h^{2,1}$) correlates (respectively anti-correlates) strongly with the number of projective spaces $m$ and with the norm and rank of the matrix.
This gives a first hint that these variables could help improve predictions by feeding them to the algorithm along with the matrix.
On the other hand, finer information on the number of projective spaces and equations do not correlate with the Hodge numbers.

From this analysis, in particular from \Cref{fig:eda:corr}, we find that the values of $h^{1,1}$ and $h^{2,1}$ are also correlated.
This motivates the simultaneous learning of both Hodge numbers since it can increase chances for the neural network to learn more universal features.
In fact, this is something that often happens in practice: counter-intuitively, it has been found that multi-tasking enhances the ability to generalize~\cite{Thrun:1995:LearningNthThing, Caruana:1997:MultitaskLearning, Baxter:2000:ModelInductiveBias, Maurer:2016:BenefitMultitaskRepresentation, Ndirango:2019:GeneralizationMultitaskDeep}.

\paragraph{Feature importance}

A second option is to sort the features by order of importance.
This can be done using a decision tree which is capable to determine the weight of each variable towards making a prediction.
One advantage over correlations is that the algorithm is non-linear and can thus determine subtler relations between the features and labels.
To avoid biasing the results using only one decision tree, we trained a random forest of trees (using \texttt{ensemble.RandomForestRegressor} in \texttt{scikit-learn}).
It consists in a large number of decision trees which are trained on different random subsets of the training dataset and averaged over the outputs (see also \Cref{sec:ml:trees,sec:app:trees}).
The algorithm determines the importance of the different features to make predictions as a by-product of the learning process, because the most relevant features tend to be found at the first branches since they are the most important to make the prediction.
The importance of a variable is a number between $0$ and $1$, and the sum over all of them must be $1$.
Since a random forest contains many trees, the robustness of the variable ranking usually improves with respect to a single tree (\Cref{sec:app:trees}).
Moreover, as the main objective is to obtain a qualitative preliminary understanding of the features, there is no need for fine tuning at this stage and we use the default parameters (in particular, $100$ decision trees).
We computed feature importance for both datasets and for two different set of variables: one containing the engineered features and the configuration matrix, and one with the engineered features and the PCA components.
In the following figures, we show several comparisons of the importance of the features, dividing the figures into scalars, vectors and configuration matrix (or its PCA), and clusters.
The sum of importance of all features equals $1$.

\begin{figure}[htp]
	\centering

	\begin{subfigure}[c]{0.45\linewidth}
		\centering
		\includegraphics[width=\textwidth, trim={0 0 6in 0}, clip]{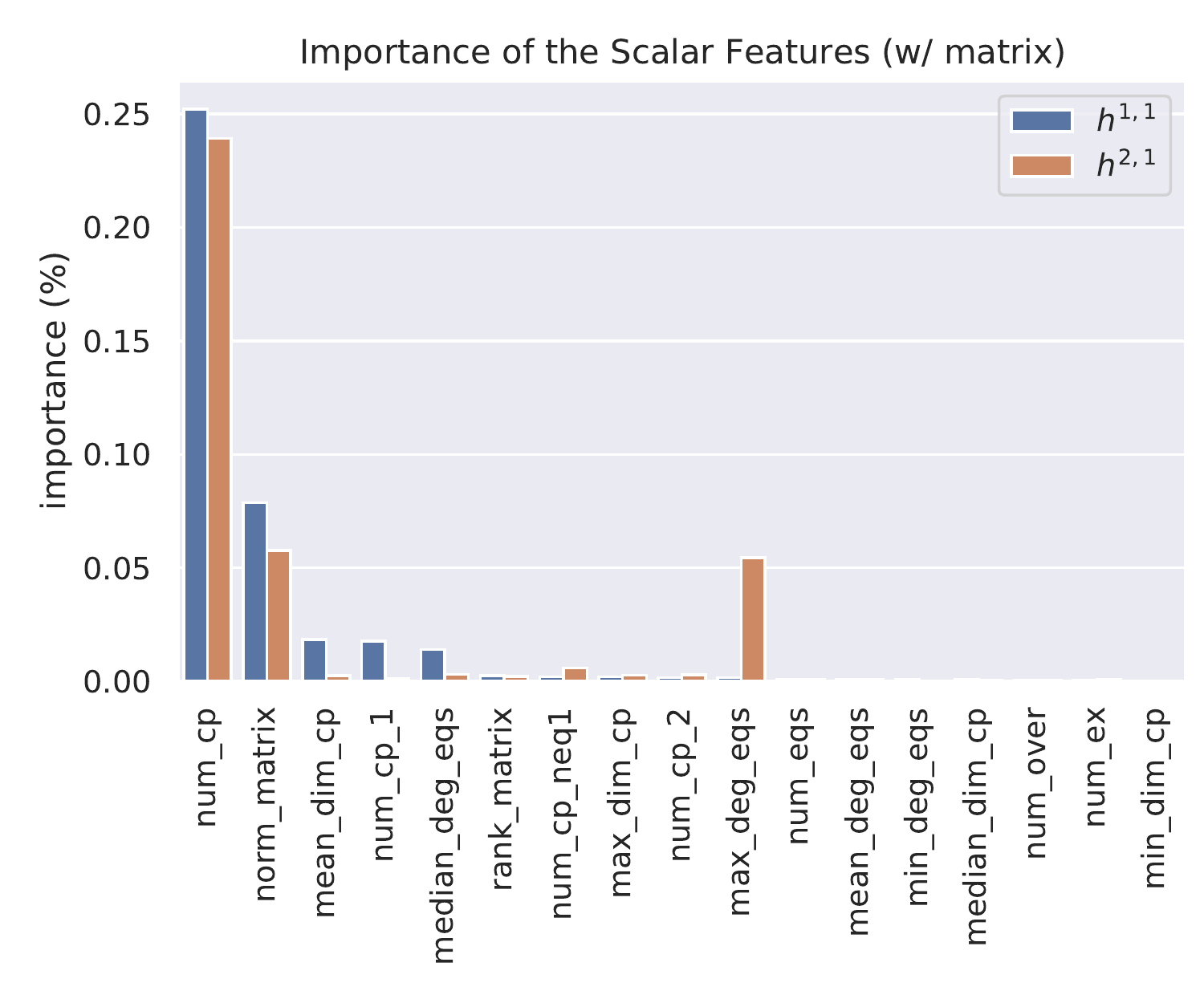}
		\caption{original dataset}
	\end{subfigure}
    \quad
	\begin{subfigure}[c]{0.45\linewidth}
		\centering
		\includegraphics[width=\textwidth, trim={0 0 6in 0}, clip]{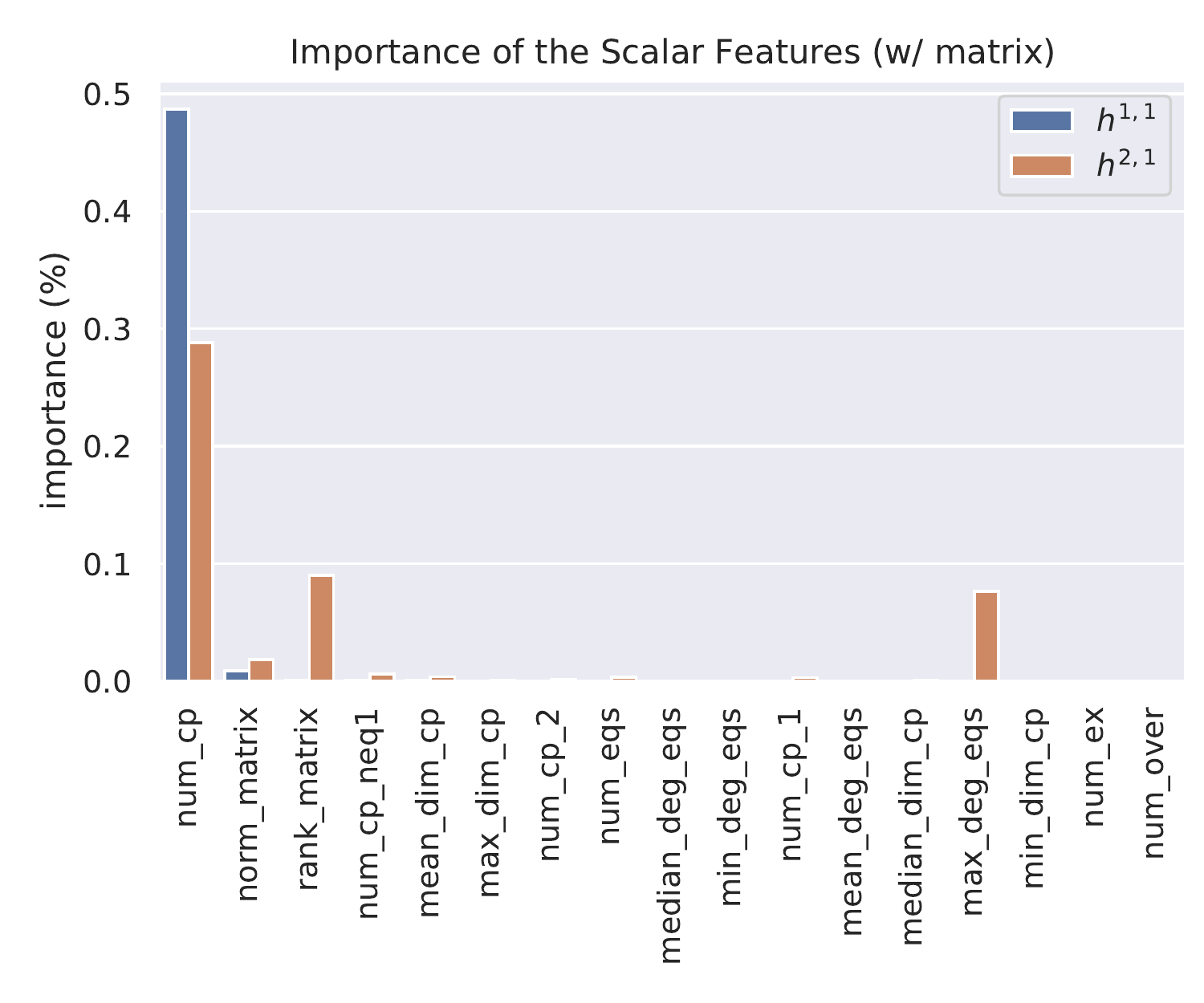}
		\caption{favourable dataset}
	\end{subfigure}

	\caption{Importance of the scalar features in the datasets.
	The same computation involving the PCA of the configuration matrix shows very marginal differences in this case: the importance of the scalar features is mostly unchanged, especially for the higher ranked variables.}
	\label{fig:eda:scalars}
\end{figure}

In \Cref{fig:eda:scalars}, we show the ranking of the scalar features in the two datasets (differences between the set using the configuration matrix and the other using the PCA are marginal and are not shown to avoid redundant plots).
As already mentioned, we find again that the number of projective spaces is the most important feature by far.
It is followed by the matrix norm in the original dataset, and by the matrix rank for $h^{2,1}$ in the favourable dataset, but in a lesser measure.
Finally, it points out that the other features have a negligible impact on the determination of the labels and may as well be ignored during training.

\begin{figure}[htp]
    \centering
    
	\begin{subfigure}[c]{\textwidth}
		\centering
		\includegraphics[width=\textwidth]{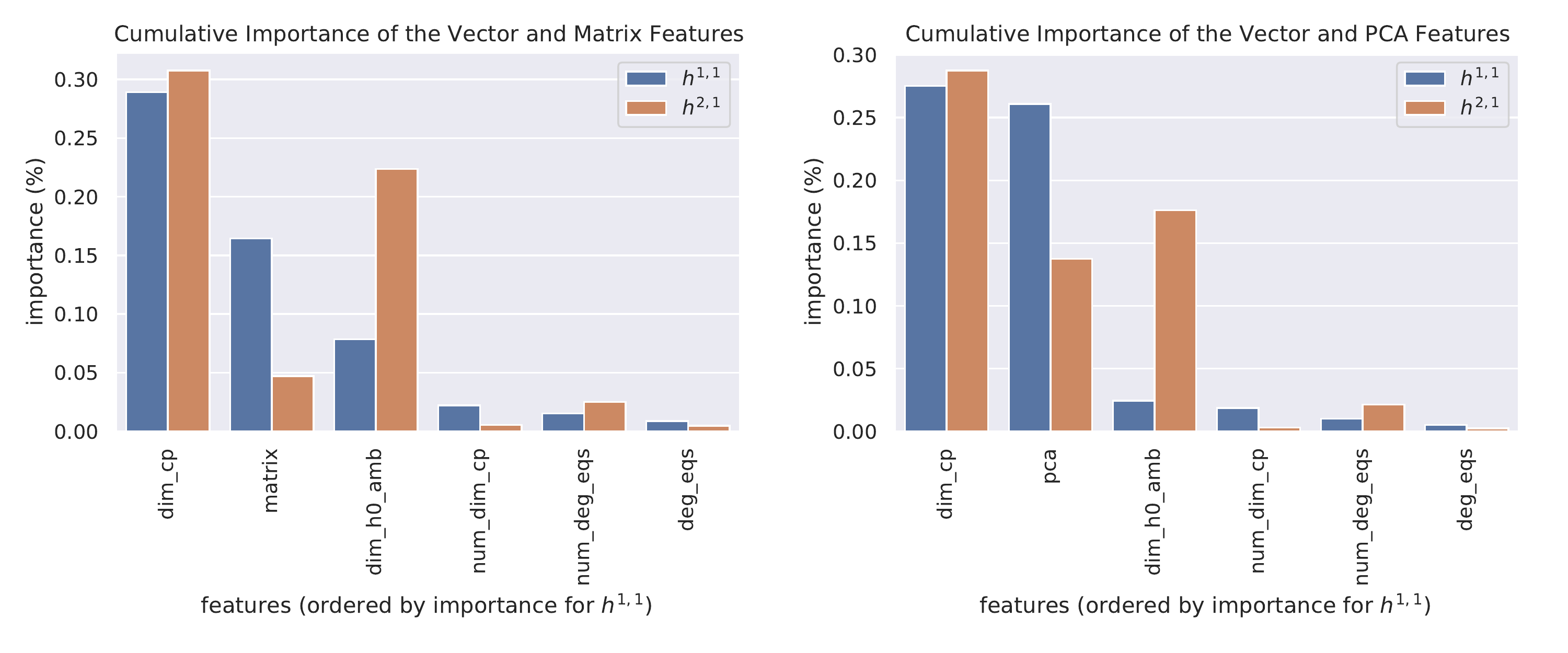}
		\caption{Original dataset}
	\end{subfigure}
	
	\begin{subfigure}[c]{\textwidth}
		\centering
		\includegraphics[width=\textwidth]{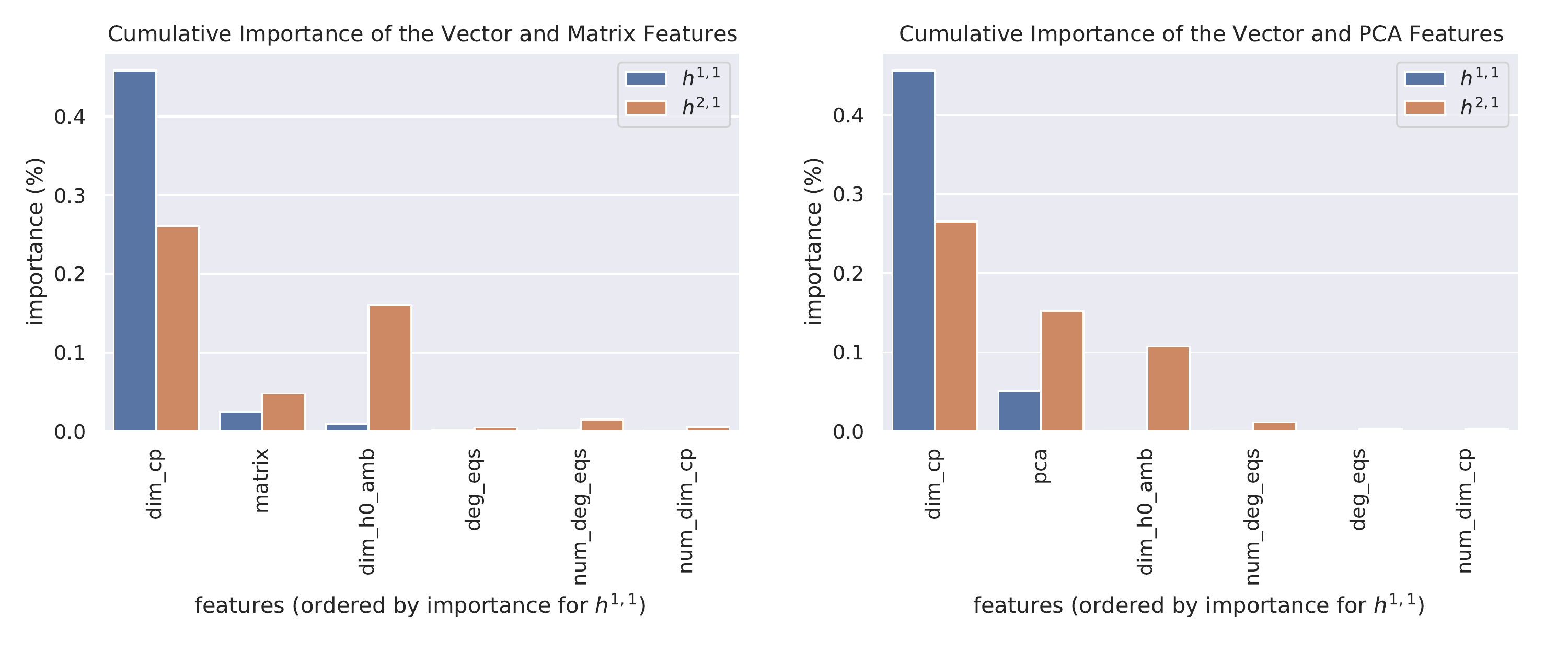}
		\caption{Favourable dataset}
	\end{subfigure}
	
	\caption{Importance of the vector features the configuration matrix (or its principal components) in the datasets: notice how the PCA plays a much more important role with respect to the full configuration matrix.}
	\label{fig:eda:tensor}
\end{figure}

The same analysis can be repeated for the vector features and the configuration matrix component by component.
In \Cref{fig:eda:tensor}, we show the cumulative importance of the features (i.e.\ the sum of the importance of each component).
We can appreciate that the list of the projective space dimensions plays a major role in the determination of the labels in both datasets.
In the case of $h^{2,1}$, we also have a large contribution from the dimensions of the cohomology group \texttt{dim\_h0\_amb}, as can be expected from algebraic topology~\cite{Hubsch:1992:CalabiYauManifoldsBestiary}.

\begin{figure}[htp]
	\centering
	
	\begin{subfigure}[c]{0.45\linewidth}
		\centering
		\includegraphics[width=\textwidth, trim={0 0 6in 0}, clip]{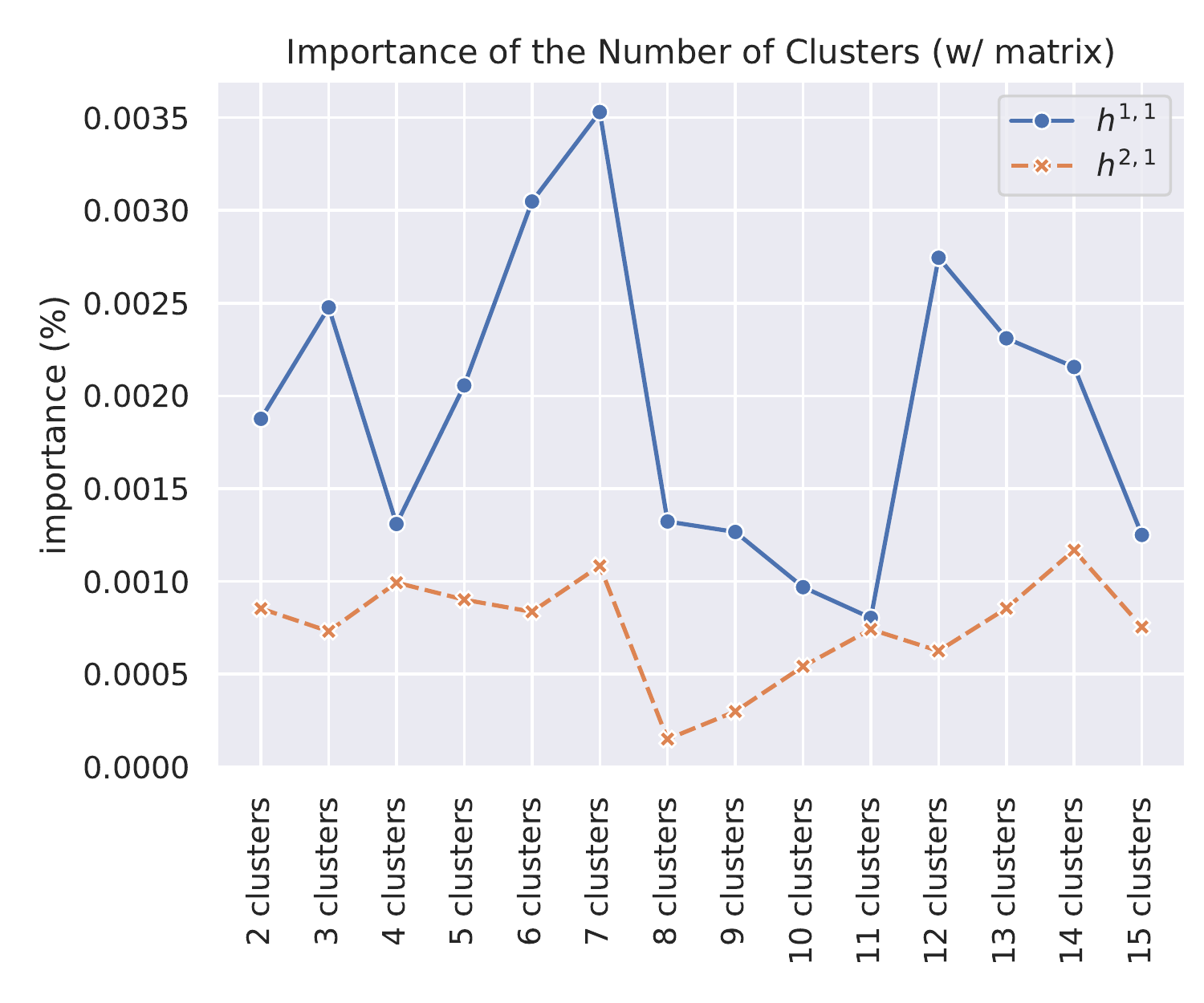}
		\caption{Original dataset}
	\end{subfigure}
	\qquad
	\begin{subfigure}[c]{0.45\linewidth}
		\centering
		\includegraphics[width=\textwidth, trim={0 0 6in 0}, clip]{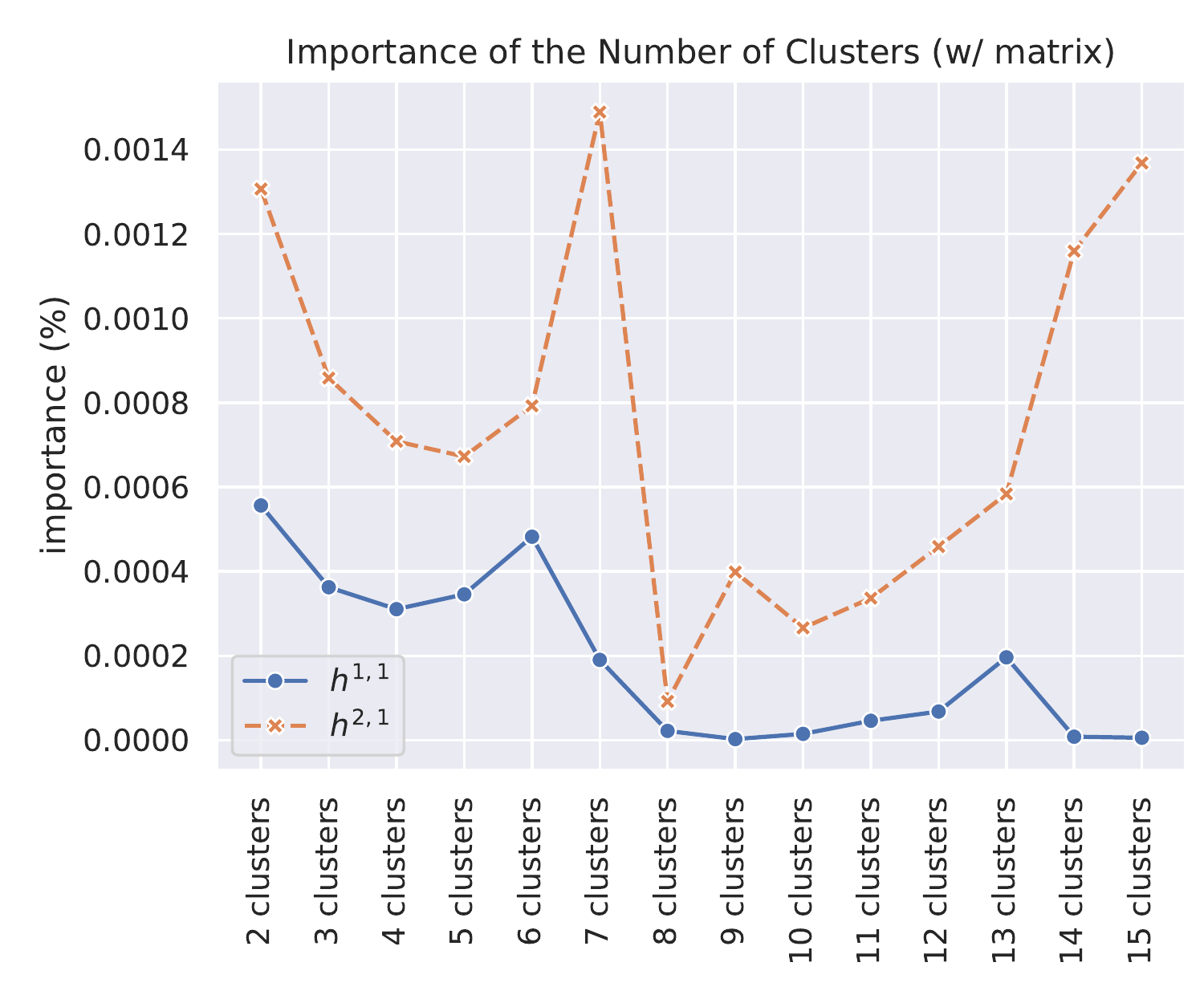}
		\caption{Favourable dataset}
	\end{subfigure}
	
	\caption{Incidence of the numbers of clusters on the variable ranking.
	Also in this case the difference between using the configuration matrix or its PCA is marginal and actually the clusters have even lower importance when using the latter.
	We therefore avoid presenting nugatory information and show only the importance of clusters when using the configuration matrix.}
	\label{fig:eda:cluster}
\end{figure}

In \Cref{fig:eda:cluster}, we finally show the importance associated to the number of clusters used during the EDA: no matter how many clusters we use, their relevance is definitely marginal compared to all other features used in the variable ranking (scalars, vectors, and the configuration matrix or its PCA) for both datasets.

\begin{figure}[htp]
	\centering
	\begin{minipage}[t]{\textwidth}
		\centering
		\includegraphics[width=\textwidth, trim={0 10in 0 0}, clip]{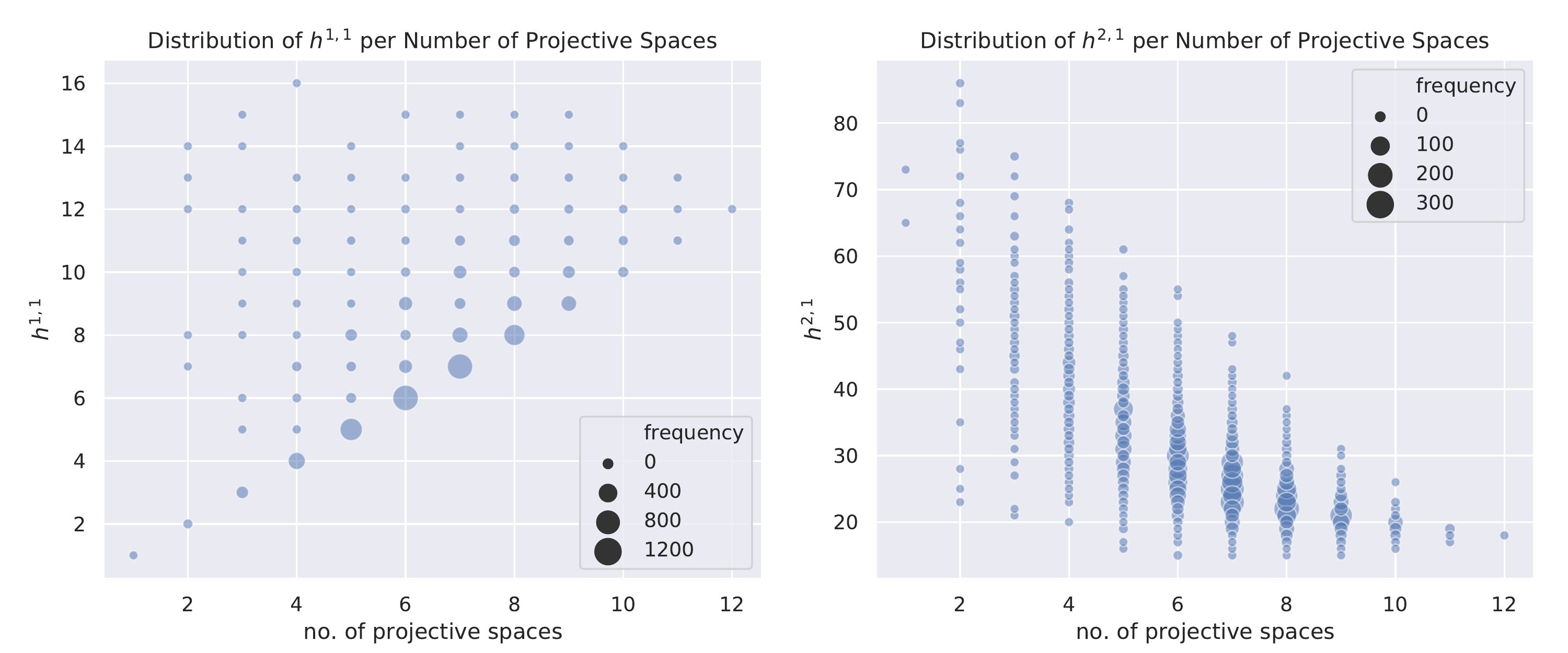}
		\caption*{Original dataset}
	\end{minipage}
	\begin{minipage}[t]{\textwidth}
		\centering
		\includegraphics[width=\textwidth, trim={0 10in 0 0}, clip]{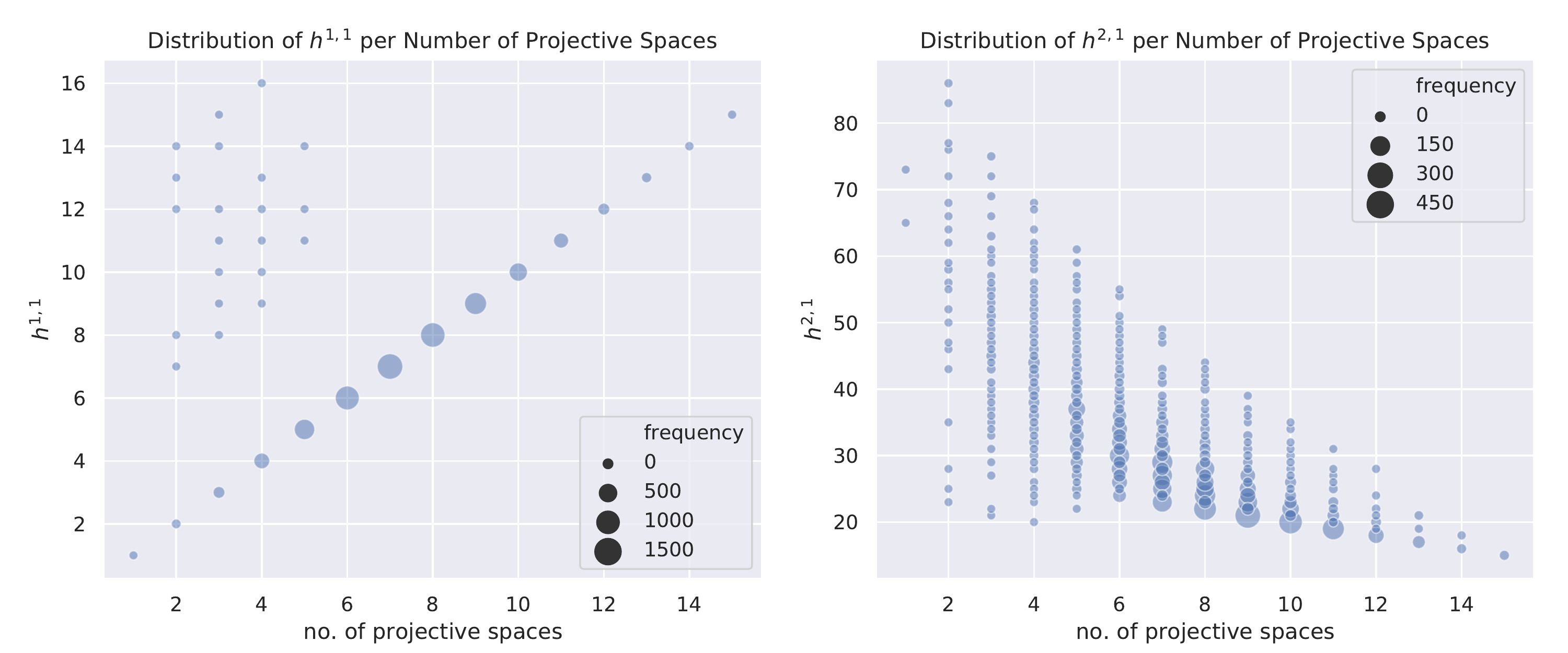}
		\caption*{Favourable dataset}
	\end{minipage}
	\caption{Distribution of the labels with respect to the number of projective spaces.}
	\label{fig:eda:distr}
\end{figure}

\paragraph{Conclusion}

It seems therefore that the number of projective spaces plays a relevant role in the determination of $h^{1,1}$ and $h^{2,1}$ as well as the list of dimensions of the projective spaces.
In order to validate this observation, in  \Cref{fig:eda:distr} we present a scatter plot of the Hodge number distributions versus the number of projective spaces: it shows that there is indeed a linear dependence in $m$ for $h^{1,1}$, especially in the favourable dataset.
In fact, the only exceptions to this pattern in the latter case are the manifolds which do not have a favourable embedding~\cite{Anderson:2017:FibrationsCICYThreefolds}.
Hence, a simple data analysis hints naturally towards this mathematical result.

Finally, we found other features which may be relevant and are worth to be included in the algorithm: the matrix rank and norm, the list of projective space dimensions and of the associated cohomology dimensions.
However, we want to emphasize one caveat to this analysis: correlations look only for linear relations, and the random forest has not been optimized or could just be not powerful enough to make good predictions.
This means that feature selection just gives a hint but it may be necessary to adapt.

\subsubsection{Removing Outliers}
\label{sec:data:eda:outliers}

\begin{figure}[htp]
	\centering
	
	\begin{subfigure}[c]{0.45\linewidth}
		\centering
		\includegraphics[width=\textwidth, trim={0 0 6in 0}, clip]{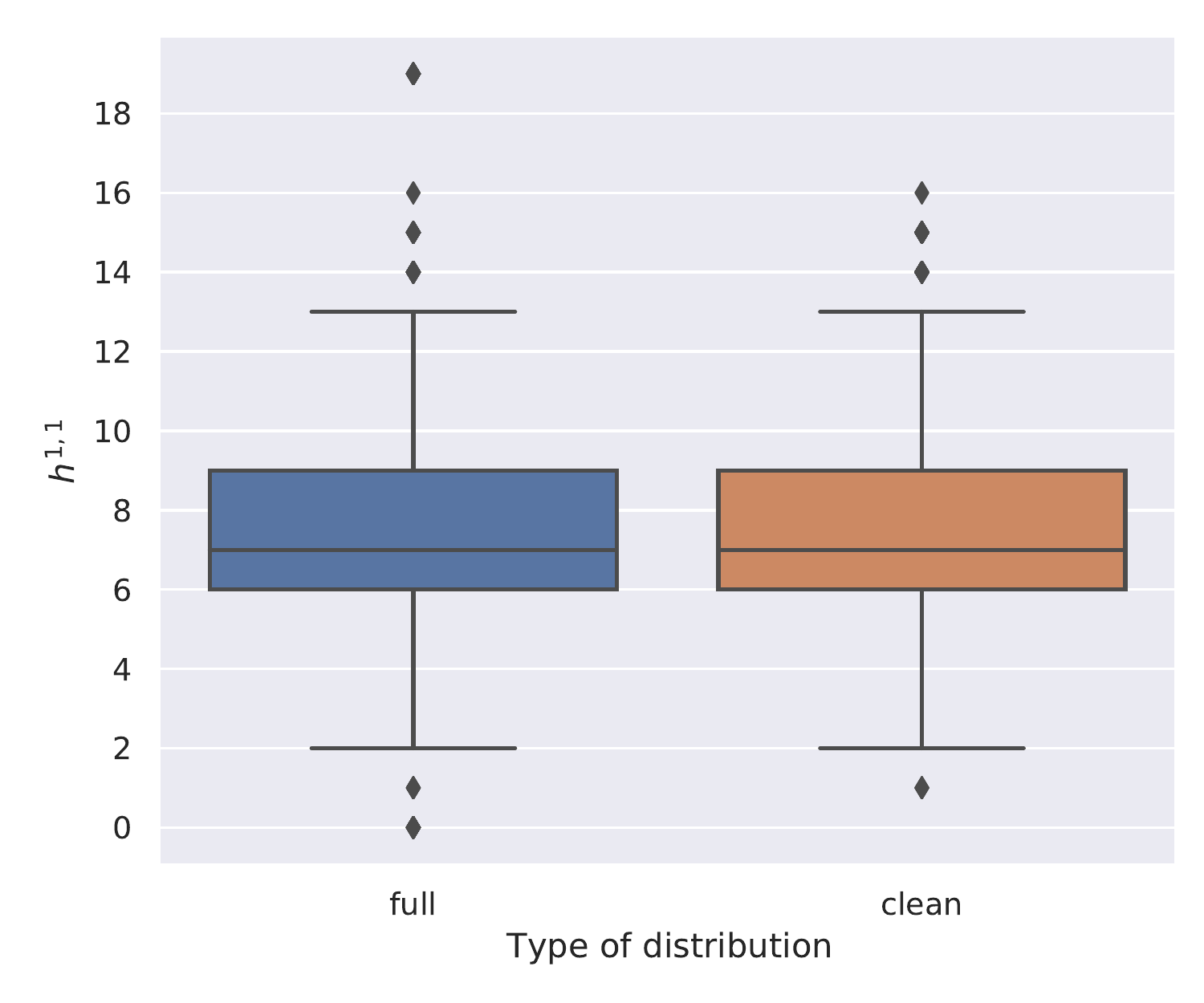}
		\caption{$h^{1,1}$}
	\end{subfigure}
	\qquad
	\begin{subfigure}[c]{0.45\linewidth}
		\centering
		\includegraphics[width=\textwidth, trim={6in 0 0 0}, clip]{images/label-distribution-compare_orig}
		\caption{$h^{2,1}$}
	\end{subfigure}
	
	\caption{%
		Summary of the statistics for the distributions of both Hodge numbers.
		The coloured box shows the three quartiles of the distributions, with the internal horizontal line corresponding to the median.
		The ``whiskers'' cover the interquartile range, i.e.\ $1.5$ times the distance between the first and third quartiles, from the lower and upper limits of the boxes.
		Isolated points show the remaining outliers.
		which we however choose to keep to avoid excessively pruning the dataset.
	}
	\label{fig:data:distr}
\end{figure}

The Hodge number distributions (\Cref{fig:data:hist-hodge,fig:data:distr}) display few outliers which lie outside the tail of the main distributions.
Such outliers may negatively impact the learning process and drive down the accuracy: it makes sense to remove them from the training set.

It is easy to see that the $22$ outlying manifolds with $h^{1,1} = h^{2,1} = 0$ are product spaces, recognisable from their block-diagonal matrix.
Moreover, we will also remove outliers with $h^{1,1} = 19$ and $h^{2,1} > 86$, which represent $15$ and $2$ samples.
In total, this represents $39$ samples, or $\SI{0.49}{\percent}$ of the total data.

To simplify the overall presentation and because the dataset is complete, we will mainly focus on the pruned subset of the data obtained by removing outliers, even from the test set.\footnotemark{}
\footnotetext{%
	There is no obligation to use a ML algorithm to label outliers in the training set, it is perfectly fine to decide which data to include or not, even based on targets.
	However, for a real-world application, outliers in the test set should be labeled by some process based only on the input features.
	Flagging possible outliers may improve the predictions by helping the machine understand that such samples require more caution.
}%
This implies that Hodge numbers lie in the ranges $1 \le h^{1,1} \le 16$ and $15 \le h^{2,1} \le 86$.
Except when stated otherwise, accuracy is indicated for this pruned dataset.
Obviously, the very small percentage of outliers makes the effect of removing them from the test set negligible when stating accuracy.

\section{Machine Learning Analysis}
\label{sec:ml}

In this section, we compare the performances of different ML algorithms: linear regression, SVM, random forests, gradient boosted trees and neural networks.
Before reporting the results for each algorithm, we detail the feature selection (\Cref{sec:ml:selection}) and the evaluation strategy (\Cref{sec:ml:strategy}).
We obtain the best results in \Cref{sec:ml:nn:inception} where we present a neural network inspired by the Inception model~\cite{Szegedy:2014:GoingDeeperConvolutions, Szegedy:2015:RethinkingInceptionArchitecture, Szegedy:2016:Inceptionv4InceptionResNetImpact}.
We provide some details on the different algorithms in \Cref{app:ml-algo} and refer the reader to the literature~\cite{Goodfellow:2016:DeepLearning, Chollet:2017:DeepLearningPython, Geron:2019:HandsOnMachineLearning, Coursera:HowWinData, Skiena:2017:DataScience, Mehta:2019:HighbiasLowvarianceIntroduction, Carleo:2019:MachineLearningPhysical, Ruehle:2020:DataScienceApplications} for more details.

\subsection{Feature Extraction}
\label{sec:ml:selection}

In \Cref{sec:data}, the EDA showed that several engineered features are promising for predicting the Hodge numbers.
In what follows, we will compare the performances of various algorithms using different subsets of features:
\begin{itemize}
	\item only the configuration matrix (no feature engineering);

	\item only the number of projective spaces $m$;

	\item only a subset of engineered features and not the configuration matrix nor its PCA;

	\item a subset of engineered features and the PCA of the matrix.
\end{itemize}

Following the EDA and feature engineering, we finally select the features we use in the analysis by choosing the highest ranked features.
We will therefore keep the number of projective spaces (\texttt{num\_cp} in the dataset) and the list of the dimension of the projective spaces (\texttt{dim\_cp}) for both $h^{1,1}$ and $h^{2,1}$).
We will also include the dimension of the cohomology group of the ambient space \texttt{dim\_h0\_amb} but only for $h^{2,1}$.

\subsection{Analysis Strategy}
\label{sec:ml:strategy}

For the ML analysis, we split the dataset into training and test sets: we fit the algorithms on the first and then show the predictions on the test set, which will not be touched until the algorithms are ready.

\paragraph{Test split and validation}

The training set is made of \SI{90}{\percent} of the samples for training, which leaves the remaining \SI{10}{\percent} in the test set (i.e.\ $785$ manifolds out of the $7851$ in the set).\footnotemark{}
\footnotetext{%
	Remember that we have removed outliers, see \Cref{sec:data:eda:outliers}.
	Scores quoted in this paper are slightly different from~\cite{Erbin:2020:InceptionCICY} because, in that paper, outliers are kept in the test set.
}%

For most algorithms, we use \emph{leave-one-out} cross-validation on the training set as evaluation of the algorithm: we subdivide the training set in $9$ subsets, each of them containing \SI{10}{\percent} of the \emph{total} amount of samples, then, we train the algorithm on $8$ of them and evaluate it on the $9$th.
We then repeat the procedure changing the evaluation fold until the algorithm has been trained and evaluated on all of them.
The performance measure in validation is given by the average over all the left out folds.

When training neural networks, we will however use a single \emph{holdout validation} set made of \SI{10}{\percent} of the \emph{total} samples.

\paragraph{Predictions and metrics}

Since we are interested in predicting exactly the Hodge numbers, the appropriate metric measuring the success of the predictions is the accuracy (for each Hodge number separately):
\begin{equation}
	\text{accuracy}
		= \frac{1}{N}
			\sum_{i=1}^N \delta\big(y_i^{\text{true}} - y_i^{\text{pred}} \big),
\end{equation}
where $N$ is the number of samples.
In the paper, accuracy of the predictions on the test set is rounded to the nearest integer.

Since the Hodge numbers are integers, the problem of predicting them looks like a classification task.
However, as argued in the introduction, we prefer to use a regression approach.
Indeed, regression does not require to specify the data boundaries and allows to extrapolate beyond them, contrary to a classification approach where the categories are fixed at the beginning.\footnotemark{}
\footnotetext{%
	A natural way to transform the problem in a regression task is to \emph{normalize} the Hodge numbers, for example by shifting by the mean value and diving by the standard deviation.
	Under this transformation, the Hodge numbers are mapped to real numbers.
	While normalizing often improve ML algorithms, we found that the impact was mild or even negative.
}%

Most algorithms need a differentiable loss function since the optimization of parameters (such as neural networks weights) uses some variant of gradient descent.
For this reason, the accuracy cannot be used and the models are trained by minimizing the mean squared error (MSE), which is simply the squared $\ell_2$-norm between of the difference between the predictions and the real values.
There will however be also a restricted number of cases in which we will use either the mean absolute error (MAE), which is the $\ell_1$-norm of the same difference, or a weighted linear combination of MSE and MAE (also known as \textit{Huber} loss): we will point them out at the right time.
When predicting both Hodge numbers together, the total loss is the sum of each individual loss with equal weight: $h^{1,1}$ is simpler to learn so it is useful to put emphasis on learning $h^{2,1}$, but the magnitudes of the latter are higher, such that the associated loss is naturally bigger (since we did not normalize the data).

Since predictions are real numbers, we need to turn them into integers.
In general, rounding to the nearest integer gives the best result, but we found algorithms (such as linear regression) for which flooring to the integer below works better.
The optimal choice of the integer function is found for each algorithm as part of the hyperparameter optimization (described below).
The accuracy is computed after the rounding stage.

Learning curves for some models are displayed.
They show how the performances of a model improves by using more training data, for fixed hyperparameters.
To obtain it, we train models using from \SI{10}{\percent} to \SI{90}{\percent} of all the data (``training ratio'') and evaluate the accuracy on the remaining data.\footnotemark{}
\footnotetext{%
	Statistics are not provided due to the limitations of our available computational resources.
	However, we check manually on few examples that the reported results are typical.
}%

To avoid redundant information and to avoid cluttering the paper with graphs, the results for models predicting separately the Hodge numbers for the test set are reported in tables, while the results for the models predicting both numbers together are reported in the learning curves.
For the same reason, the latter are not displayed for the favourable dataset.

\paragraph{Visualisation of the performance}

Complementary to the predictions and the accuracy results, we also provide different visualisations of the performance of the models in the form of univariate plots (histograms) and multivariate distributions (scatter plots).

The usual assumption behind the statistical inference of a distribution is that the difference between the observed data and the predicted values can be modelled by a random variable called \textit{residual}~\cite{Lista:2017:StatisticalMethods,Coursera:DataScience}.\footnotemark{}
\footnotetext{The difference between the non observable \textit{true} value of the model and the observed data is known as \textit{statistical error}.
The difference between residuals and errors is subtle but the two definitions have different interpretations in the context of the regression analysis: in a sense, residuals are an estimate of the errors.}
As such we expect that its values can be sampled from a normal distribution with a constant variance (i.e.\ constant width), since it should not depend on specific observations, and centered around zero, since the regression algorithm tries to minimise the squared difference between observed and predicted values.
Histograms of the residual errors should therefore exhibit such properties graphically.

Another interesting kind of visual realisation of the residuals is to show their distribution against the variables used for the regression model: in the case of a simple regression model in one variable, it is customary to plot the residuals as a function of the independent variable, but in a multivariable regression analysis (such as the case at hand) the choice usually falls on the values predicted by the fit (not the observed data).
We shall therefore plot the residuals as functions of the predicted values.\footnotemark{}
\footnotetext{We will use the same strategy also for the fit using just the number of projective spaces in order to provide a way to compare the plots across different models.}
Given the assumption of the random distribution of the residuals, they should not present strong correlations with the predictions and should not exhibit trends.
In general the presence of correlated residuals is an indication of an incomplete or incorrect model which cannot explain the variance of the predicted data, meaning that the model is either not suitable for predictions or that we should add information (that is, add features) to it.

\paragraph{Hyperparameter optimisation}

One of the key steps in a ML analysis is the optimisation of the \emph{hyperparameters} of the algorithm.
These are internal parameters of each estimator (such as the number of trees in a random forest or the amount of regularisation in a linear model): they are not modified during the training of the model, but they directly influence it in terms of performance and outcome.

Hyperparameter optimization is performed by training many models with different hyperparameters, and keeping those which perform best according to some metric on the validation set(s).
As it does not need to be differentiable, we use the accuracy as a scoring function to evaluate the models.
There is however subtle issue because it is not clear how to combine the accuracy of $h^{1,1}$ and $h^{2,1}$ to get a single metric.
For this reason, we will perform the analysis on both Hodge numbers separately.
Then, we can design a single model computing both Hodge numbers simultaneously by making a compromise by hand between the hyperparameters found for the two models computing the Hodge numbers separately.

The optimization is implemented using the API from \texttt{scikit-learn}, using the function \texttt{metrics.make\_scorer} and the accuracy as a custom scoring function.
There are several approaches to perform this search automatically, in particular: grid search, random search, genetic evolution, and Bayes optimization.

Grid and random search are natively implemented in \texttt{scikit-learn}.
The first takes a list of possible discrete values of the hyperparameters and will evaluate the algorithm over all possible combinations.
The second samples values in both discrete sets and continuous intervals according to some probability distributions, repeating the process a fixed number of times.
The grid search method is particularly useful for discrete hyperparameters, less refined searches or for a small number of combinations, while the second method can be used to explore the hyperparameter space on a larger scale~\cite{Bergstra:2012:RandomSearchHyperparameter}.
Genetic algorithms are based on improving the choice of hyperparameters over \emph{generations} that successively select only the most promising values: in general, they require a lot of tuning and are easily influenced by the fact that the replication process can also lead to worse results totally at random~\cite{Rudolph:1994:GeneticAlgorithms}.
They are however effective when dealing with very deep or complex neural networks.

Bayes optimisation~\cite{Snoek:2012:PracticalBayesianOptimization, Shahriari:2016:TakingHumanOut} is a very well established mathematical procedure to find the stationary points of a function without knowing its analytical form~\cite{Mockus:1975:BayesianMethodsSeeking}.
It relies on assigning a \emph{prior} probability to a given parameter and then multiply it by the probability distribution (or \emph{likelihood}) of the scoring function to compute the probability of finding a better results given a set of hyperparameters.
This has proven to be very effective in our case and we adopted this solution as it does not require fine tuning and leads to better results for models which are not deep neural networks.
We choose to use \texttt{scikit-optimize}~\cite{Head:Scikitoptimize} whose method \texttt{BayesSearchCV} has a very well implemented Python interface compatible with \texttt{scikit-learn}.
We will in general perform $50$ iterations of the Bayes search algorithm, unless otherwise specified.

\subsection{Linear Models}

Linear models attempt to describe the labels as a linear combinations of the input features while keeping the coefficients at order one (\Cref{sec:app:linreg}).
However, non-linearity can still be introduced by engineering features which are non-linear in terms of the original data.

From the results of \Cref{sec:data:eda}, we made a hypothesis on the linear dependence of $h^{1,1}$ on the number of projective spaces $m$.
As a first approach, we can try to fit a linear model to the data as a baseline computation and to test whether there is actual linear correlation between the two quantities.
We will consider different linear models, including their regularised versions.

\paragraph{Parameters}

The linear regression is performed with the class \lstinline!linear_model.ElasticNet! from \lstinline!scikit-learn!.
The hyperparameters involved in this case are: the amount of regularisation $\alpha$, the relative ratio (\texttt{l1\_ratio}) between the $\ell_1$ and $\ell_2$ regularization losses, and the fit of the intercept.

By performing the hyperparameter optimization, we found that $\ell_2$ regularization has a minor impact and can be removed, which corresponds to setting the relative ratio to $1$ (this is equivalent to using \texttt{linear\_model.Lasso}).

In \Cref{tab:hyp:lin} we show the choices of the hyperparameters for the different models we built using the $\ell_1$ regularised linear regression.

For the original dataset, we floored the predictions to the integers below, while in the favourable we rounded to the next integer.
This choice for the original dataset makes sense: the majority of the samples lie on the line $h^{1,1} = m$, but there are still many samples with $h^{1,1} > m$ (see \Cref{fig:eda:distr}).
As a consequence, the ML prediction pulls the line up, which can only damage the accuracy.
Choosing the floor function is a way to counteract this effect.
Note that accuracy for $h^{2,1}$ is only slightly affected by the choice of rounding, so we just choose the same one as $h^{1,1}$ for simplification.

\begin{table}[htp]
\centering
\resizebox{\textwidth}{!}{%
\begin{tabular}{@{}lccccccccc@{}}
\toprule
                                                          &           & \multicolumn{2}{c}{\textbf{matrix}}         & \multicolumn{2}{c}{\textbf{num\_cp}} & \multicolumn{2}{c}{\textbf{eng. feat.}}     & \multicolumn{2}{c}{\textbf{PCA}}            \\ \midrule
                                                          &           & \textit{old}         & \textit{fav.}        & \textit{old}  & \textit{fav.}        & \textit{old}         & \textit{fav.}        & \textit{old}         & \textit{fav.}        \\ \midrule
\multirow{2}{*}{$\alpha$}                                 & $h^{1,1}$ & $2.0 \times 10^{-6}$ & $3.0 \times 10^{-5}$ & 0.10          & $2.0 \times 10^{-6}$ & 0.05                 & 0.05                 & 0.07                 & 0.08                 \\
                                                          & $h^{2,1}$ & $1.0 \times 10^{-6}$ & $1.0 \times 10^{-5}$ & 0.1           & $1.0 \times 10^{-6}$ & $3.0 \times 10^{-4}$ & $1.2 \times 10^{-3}$ & $2.0 \times 10^{-6}$ & $1.2 \times 10^{-3}$ \\ \midrule
\multirow{2}{*}{\texttt{fit\_intercept}} & $h^{1,1}$ & False                & False                & True          & False                & True                 & True                 & False                & True                 \\
                                                          & $h^{2,1}$ & True                 & True                 & True          & True                 & True                 & False                & True                 & False                \\ \midrule
\multirow{2}{*}{\texttt{normalize}}      & $h^{1,1}$ & ---                  & ---                  & False         & ---                  & False                & False                & ---                  & False                \\
                                                          & $h^{2,1}$ & False                & True                 & False         & False                & False                & ---                  & True                 & ---                  \\ \bottomrule
\end{tabular}%
}
\caption{Hyperparameter choices of the $\ell_1$ regression model used. In addition to the known hyperparameters $\alpha$ and \texttt{fit\_intercept}, we also include the \texttt{normalize} parameter which indicates whether the samples have been centered and scaled by their $\ell_2$ norm before the fit: it is ignored when the intercept is ignored.}
\label{tab:hyp:lin}
\end{table}

\paragraph{Results}

In \Cref{tab:res:lin}, we show the accuracy for the best hyperparameters.
For $h^{1,1}$, the most precise predictions are given by the number of projective spaces which actually confirms the hypothesis of a strong linear dependence of $h^{1,1}$ on the number of projective spaces.
In fact, this gives close to $100\%$ accuracy for the favourable dataset, which shows that there is no need for more advanced ML algorithms.
Moreover, adding more engineered features \emph{decreases} the accuracy in most cases where regularization is not appropriate.
The accuracy for $h^{2,1}$ remains low but including engineered features definitely improves it.

\begin{figure}[htp]
  \centering
  \includegraphics[width=\textwidth]{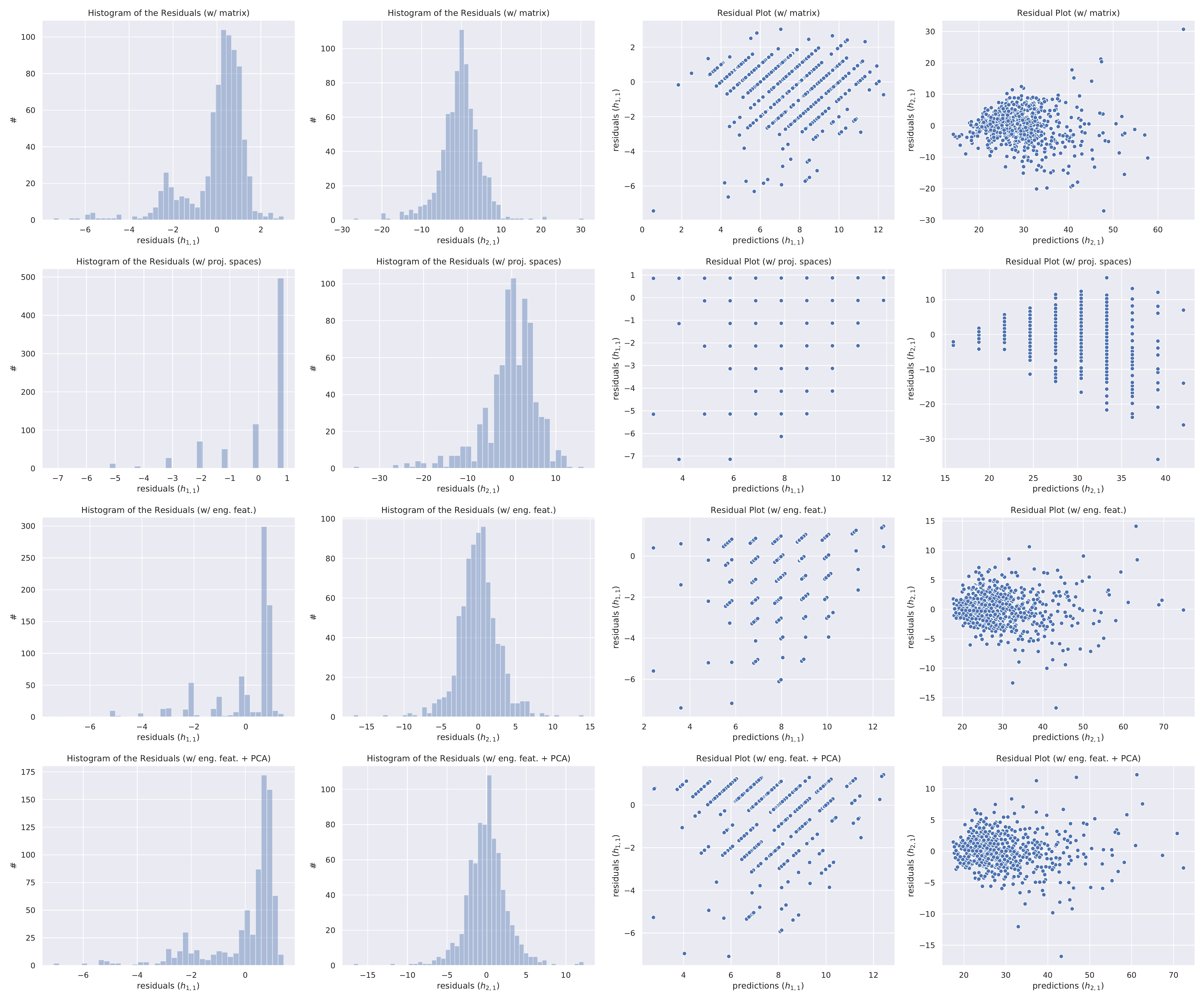}
  \caption{%
  		Plots of the residual error for the $\ell_1$ regularised linear model: rows show the different scenarios (fit with only the matrix, with only the number of projective spaces, with the engineered features, with the engineered features and the PCA).
  		Plots refer to the test split of the original dataset.
  }
  \label{fig:res:lin}
\end{figure}

In \Cref{fig:res:lin}, we show the plots of the residual errors of the model on the original dataset.
For the $\ell_1$ regularised linear model, the univariate plots show that the errors seem to follow normal distributions peaked at $0$ as they generally should: in the case of $h^{1,1}$, the width is also quite contained.
The scatter plots instead show that, in general, there is no correlation between a particular sector of the predictions and the error made by the model, thus the variance of the residuals is in general randomly distributed over the predictions.
Only the case of the fit of the number of projective spaces seems to show a slight correlation for $h^{2,1}$, signalling that the model using only one feature might be actually incomplete: in fact it is better to include also other engineered features.

\begin{figure}[htp]
	\centering

	\begin{subfigure}[c]{0.45\linewidth}
		\centering
		\includegraphics[width=\textwidth]{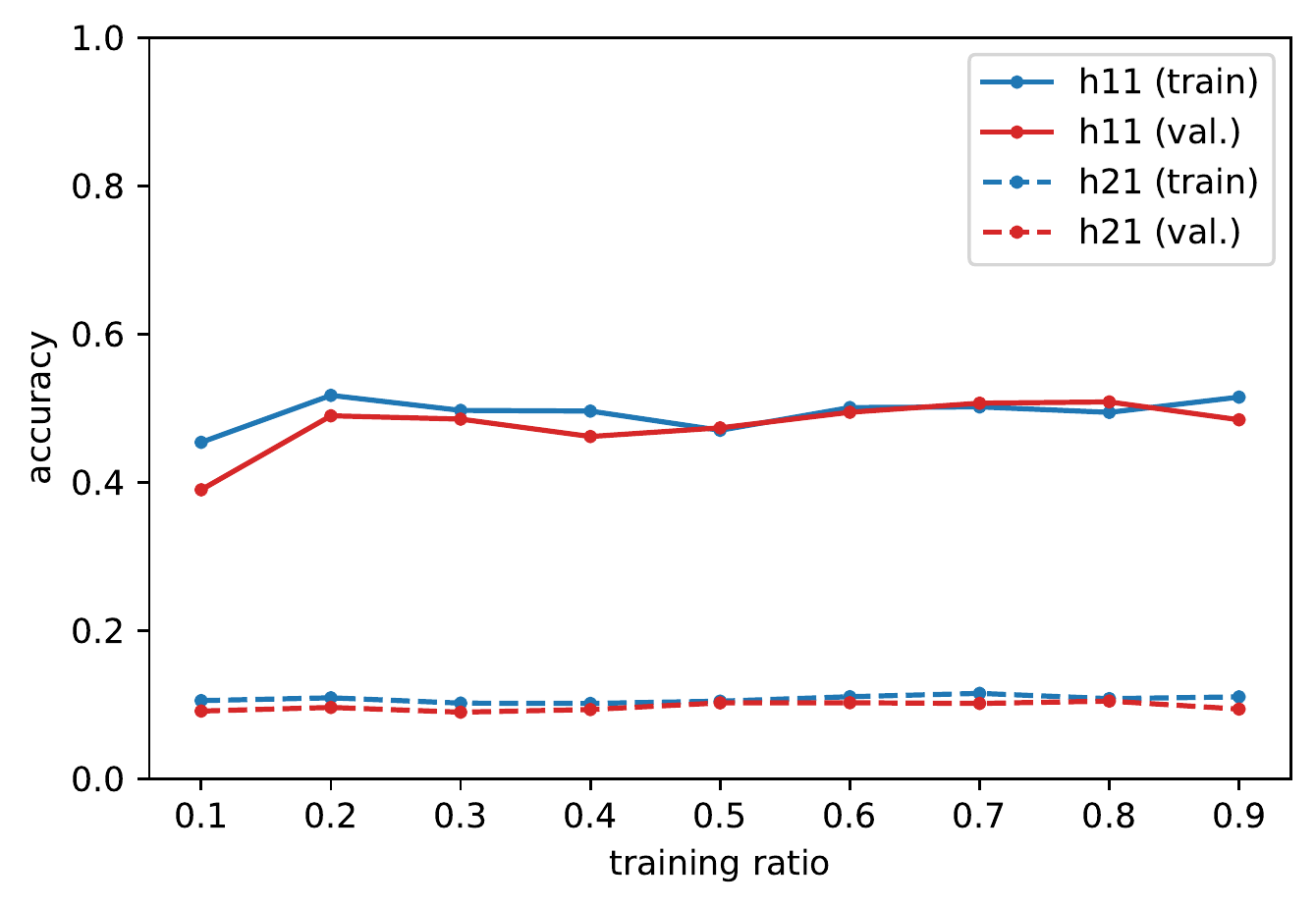}
		\caption{input: \lstinline!matrix!, $\alpha = \num{2e-4}$}
	\end{subfigure}
	\qquad
	\begin{subfigure}[c]{0.45\linewidth}
		\centering
		\includegraphics[width=\textwidth]{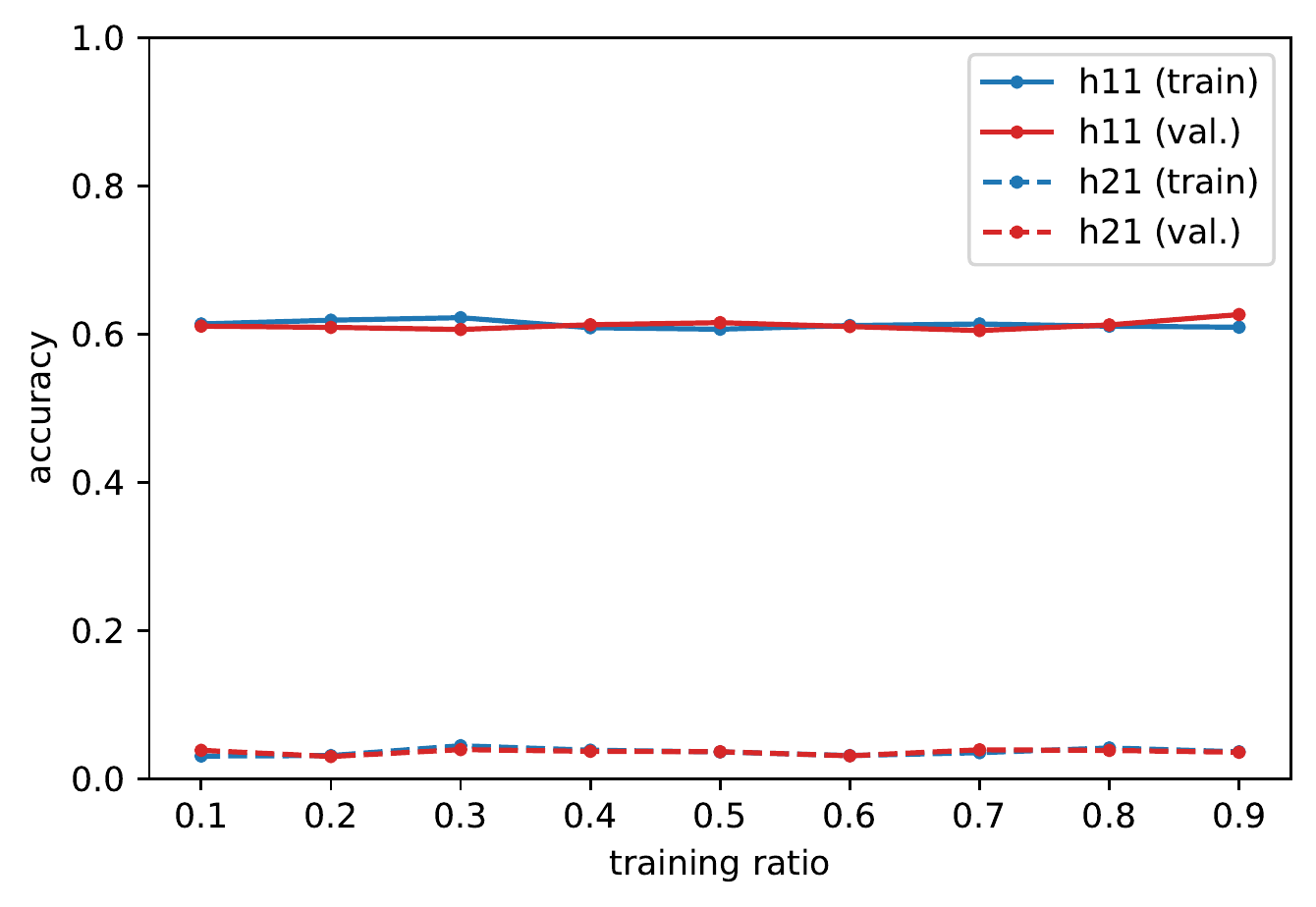}
		\caption{input: \lstinline!num_cp!, $\alpha = 1$}
	\end{subfigure}

	\caption{Learning curves for the linear regression (original dataset), including outliers and using a single model for both Hodge numbers.}
	\label{fig:lc:lin}
\end{figure}

\begin{figure}[htp]
	\centering

	\begin{subfigure}[c]{0.45\linewidth}
		\centering
		\includegraphics[width=\textwidth]{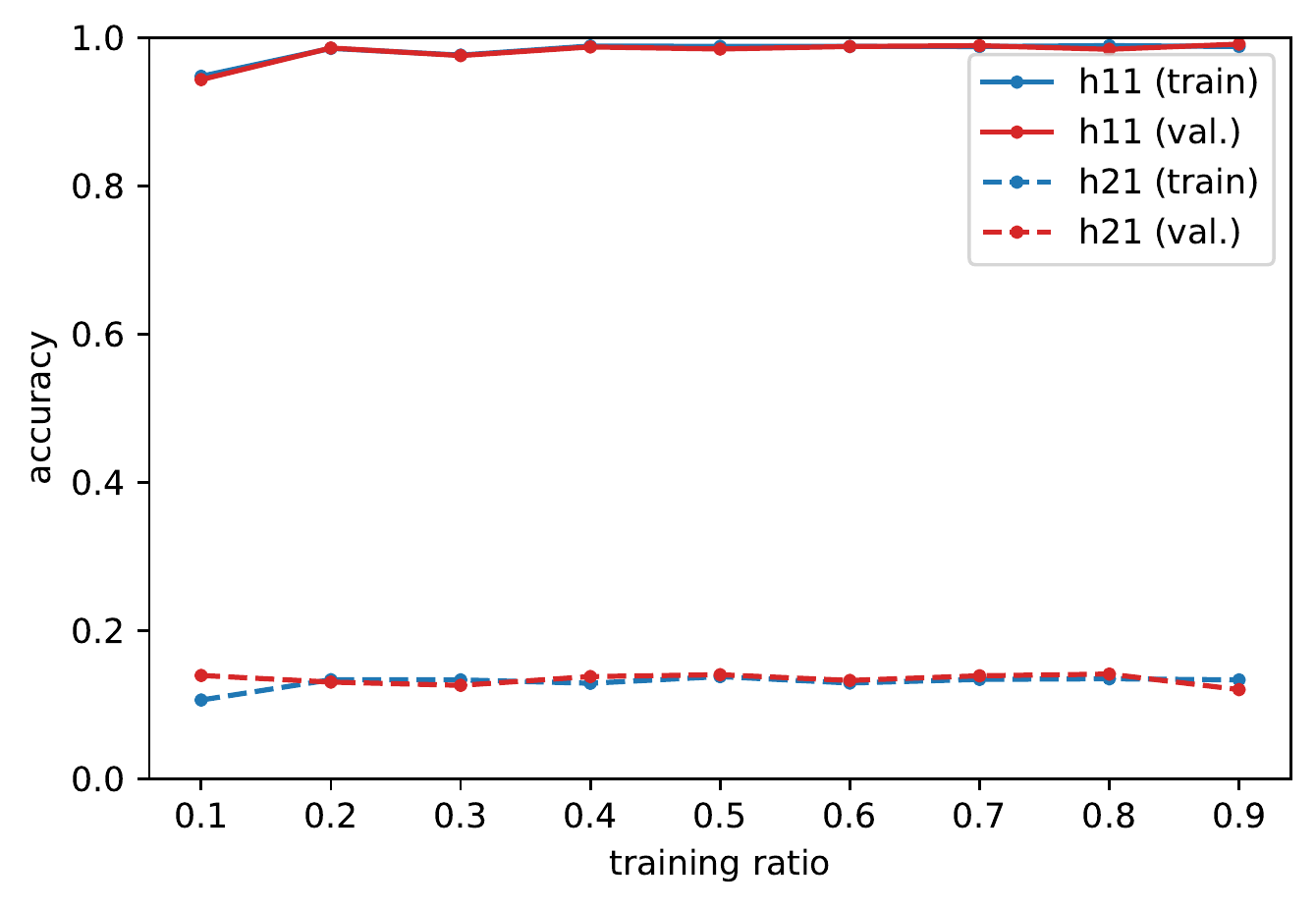}
		\caption{input: \lstinline!num_cp!, $\alpha = 1$}
	\end{subfigure}

	\caption{Learning curves for the linear regression (favourable dataset), including outliers and using a single model for both Hodge numbers.}
	\label{fig:lc:lin-fav}
\end{figure}

The learning curves (\Cref{fig:lc:lin}) clearly shows that the model underfits.
Moreover, we also noticed that the models are only marginally affected by the number of samples used for training.
In particular, this provides a very strong baseline for $h^{1,1}$.
For comparison, we also give the learning curve for the favourable dataset in \Cref{fig:lc:lin-fav}: this shows that a linear regression is completely sufficient to determine $h^{1,1}$ in that case.

\begin{table}[htp]
  \centering
  \begin{tabular}{@{}cccccc@{}}
    \toprule
                            &           & \textbf{matrix} & \textbf{num\_cp} & \textbf{eng. feat.} & \textbf{PCA} \\ \midrule
      \multirow{2}{*}
      {\emph{original}}   & $h^{1,1}$ & 51\%            & 63\%             & 63\%                & 64\%         \\
                          & $h^{2,1}$ & 11\%            & 8\%              & 21\%                & 21\%         \\ \midrule
      \multirow{2}{*}
      {\emph{favourable}} & $h^{1,1}$ & 95\%            & 100\%            & 100\%               & 100\%        \\
                          & $h^{2,1}$ & 14\%            & 15\%             & 19\%                & 19\%         \\ \bottomrule
  \end{tabular}
  \caption{Best accuracy of the linear model using $\ell_1$ regularisation on the test split.}
  \label{tab:res:lin}
\end{table}

\subsection{Support Vector Machines}
\label{sec:res:svr}

Support Vector Machines (SVM) are a family of algorithms which use a \emph{kernel trick} to map the space of input data vectors into a higher dimensional space where samples can be accurately separated and fitted to an appropriate curve (\Cref{sec:app:svr}).

In this analysis, we show two such kernels, namely a linear kernel (also known as \emph{no kernel} since no transformations are involved) and a Gaussian kernel (known as \texttt{rbf} in ML literature, from \emph{radial basis function}).

\subsubsection{Linear Kernel}

For this model we use the class \texttt{svm.LinearSVR} in \texttt{scikit-learn}.

\paragraph{Parameters}

In \Cref{tab:hyp:linsvr} we show the choices of the hyperparameters used for the model.
As we show in \Cref{sec:app:svr} parameters $C$ and $\epsilon$ are related to the penalty assigned to the samples lying outside the no-penalty boundary (the loss in this case is computed according to the $\ell_1$ or $\ell_2$ norm of the distance from the boundary as specified by the \texttt{loss} hyperparameter).
Other parameters are related to the use of the intercept to improve the prediction.

We rounded the predictions to the floor for the original dataset and to the next integer for the favourable dataset.

\begin{table}[htp]
\centering
\resizebox{\textwidth}{!}{%
\begin{tabular}{@{}lccccccccc@{}}
\toprule
                                             &           & \multicolumn{2}{c}{\textbf{matrix}}                           & \multicolumn{2}{c}{\textbf{num\_cp}}                                            & \multicolumn{2}{c}{\textbf{eng. feat.}}                       & \multicolumn{2}{c}{\textbf{PCA}}                                       \\ \midrule
                                             &           & \textit{old}                  & \textit{fav.}                 & \textit{old}                           & \textit{fav.}                          & \textit{old}                  & \textit{fav.}                 & \textit{old}                           & \textit{fav.}                 \\ \midrule
\multirow{2}{*}{\texttt{C}}                  & $h^{1,1}$ & 0.13                          & 24                            & 0.001                                  & 0.0010                                 & 0.13                          & 0.001                         & 0.007                                  & 0.4                           \\
                                             & $h^{2,1}$ & 0.30                          & 100                           & 0.05                                  & 0.0016                                 & 0.5                          & 0.4                         & 1.5                                  & 0.4                           \\ \midrule
\multirow{2}{*}{$\epsilon$}                  & $h^{1,1}$ & 0.7                           & 0.3                           & 0.4                                    & 0.00                                   & 0.9                           & 0.0                           & 0.5                                    & 0.0                           \\
                                             & $h^{2,1}$ & 0.0                           & 0.0                           & 10                                   & 0.03                                   & 0.0                           & 0.0                           & 0.0                                    & 0.6                           \\ \midrule
\multirow{2}{*}{\texttt{fit\_intercept}}     & $h^{1,1}$ & True                          & False                         & True                                   & False                                  & True                          & False                         & False                                  & False                         \\
                                             & $h^{2,1}$ & True                          & False                         & True                                   & True                                   & True                          & True                          & True                                   & False                         \\ \midrule
\multirow{2}{*}{\texttt{intercept\_scaling}} & $h^{1,1}$ & 0.13                          & ---                           & 100                                    & ---                                    & 0.01                          & ---                           & ---                                    & ---                           \\
                                             & $h^{2,1}$ & 100                           & ---                           & 13                                     & 92                                     & 100                        & 0.01                          & 100                                    & ---                           \\ \midrule
\multirow{2}{*}{\texttt{loss}}               & $h^{1,1}$ & $|\epsilon|$ & $|\epsilon|$ & $|\epsilon|$          & $||\epsilon||^2$ & $|\epsilon|$ & $|\epsilon|$ & $|\epsilon|$ & $|\epsilon|$ \\
                                             & $h^{2,1}$ & $|\epsilon|$ & $|\epsilon|$ & $||\epsilon||^2$ & $|\epsilon|$          & $|\epsilon|$ & $|\epsilon|$ & $|\epsilon|$          & $|\epsilon|$ \\ \bottomrule
\end{tabular}%
}
\caption{Hyperparameter choices of the linear SVR regression. The parameter \texttt{intercept\_scaling} is clearly only relevant when the intercept is used. The different losses used simply distinguish between the $\ell_1$ norm of the $\epsilon$-dependent boundary where no penalty is assigned and its $\ell_2$ norm.}
\label{tab:hyp:linsvr}
\end{table}

\paragraph{Results}

In \Cref{tab:res:linsvr}, we show the accuracy on the test set for the linear kernel.
As we can see the performance of the algorithm strongly resembles a linear model in terms of the accuracy reached.

It is interesting to notice that the contributions of the PCA do not improve the predictions using just the engineered features: it seems that the latter work better than using the configuration matrix or its principal components.

\begin{figure}[htp]
	\centering
	\includegraphics[width=0.9\textwidth]{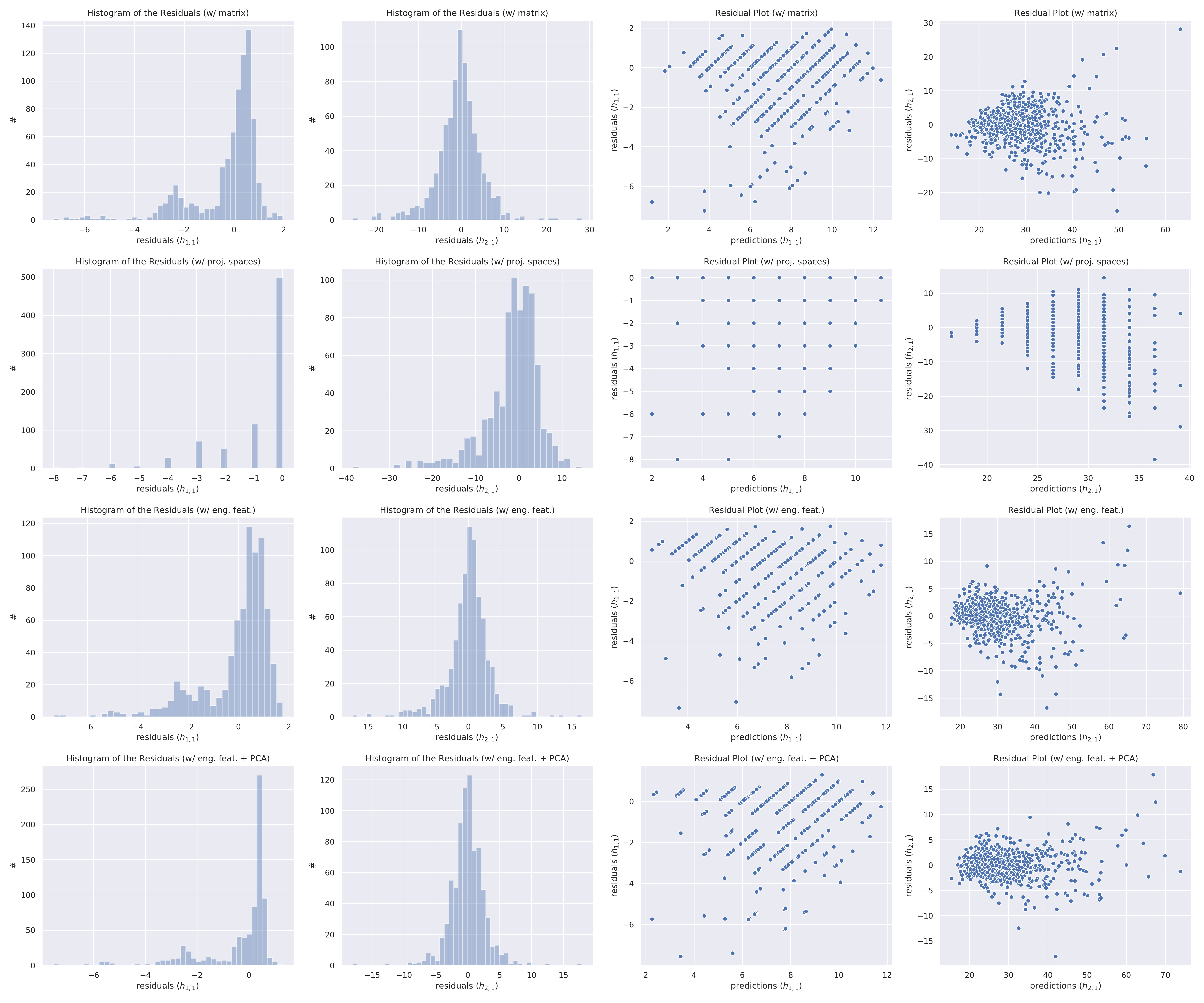}
	\caption{Plots of the residual errors for the SVM with linear kernel.}
	\label{fig:res:linsvr}
\end{figure}

The residual plots in \Cref{fig:res:linsvr} confirm what we already said about the linear models with regularisation: the model with only the number of projective spaces shows a tendency to heteroscedasticity\footnotemark{} which can be balanced by adding more engineered feature, also helping in having more precise predictions (translated into peaked univariate distributions).
\footnotetext{%
	That is the tendency to have a correlation between the predictions and the residuals: theoretically speaking, there should not be any, since we suppose the residuals to be independent on the model and normally sampled.
}%
In all cases, we notice that the model slightly overestimates the real values (residuals are computed as the difference between the prediction and the real value) as the second, small peaks in the histograms for $h^{1,1}$ suggest: this may also explain why flooring the predictions produces the highest accuracy.

As in general for linear models, the influence of the number of samples used for training is marginal also in this case: we only noticed a decrease in accuracy when also including the PCA or directly the matrix.

\begin{table}[htp]
\centering
\begin{tabular}{@{}cccccc@{}}
  \toprule
                          &           & \textbf{matrix} & \textbf{num\_cp} & \textbf{eng. feat.} & \textbf{PCA} \\ \midrule
    \multirow{2}{*}
    {\emph{original}}   & $h^{1,1}$ & 61\%            & 63\%             & 65\%                & 62\%         \\
                          & $h^{2,1}$ & 11\%            & 9\%              & 21\%                & 20\%         \\ \midrule
    \multirow{2}{*}
    {\emph{favourable}} & $h^{1,1}$ & 96\%            & 100\%            & 100\%               & 100\%        \\
                          & $h^{2,1}$ & 14\%            & 14\%             & 19\%                & 20\%         \\ \bottomrule
	\end{tabular}
	\caption{Accuracy of the linear SVM on the test split.}
	\label{tab:res:linsvr}
\end{table}

\subsubsection{Gaussian Kernel}

We then consider SVM using a Gaussian function as kernel.
The choice of the function can heavily influence the outcome of the predictions since they map the samples into a much higher dimensional space and create highly non-linear combinations of the features before fitting the algorithm.
In general, this can help in the presence of ``obscure'' features which badly correlate one another.
In our case, we can hope to leverage the already good correlations we found in the EDA with the kernel trick.
The implementation is done with the class \texttt{svm.SVR} from \texttt{scikit-learn}.

\paragraph{Parameters}

As we show in \Cref{sec:app:svr}, this particular choice of kernel leads to profoundly different behaviour with respect to linear models: we will round the predictions to the next integer in both datasets since the loss function strongly penalises unaligned samples.

In \Cref{tab:hyp:svrrbf}, we show the choices of the hyperparameters for the models using the Gaussian kernel.
As usual the hyperparameter \texttt{C} is connected to the penalty assigned to the samples outside the soft margin boundary (see \Cref{sec:app:svr}) delimited by the $\epsilon$.
Given the presence of a non linear kernel we have to introduce an additional hyperparameter $\gamma$ which controls the width of the Gaussian function used for the support vectors.

\begin{table}[htp]
\centering
\begin{tabular}{@{}lccccccccc@{}}
\toprule
                            &           & \multicolumn{2}{c}{\textbf{matrix}} & \multicolumn{2}{c}{\textbf{num\_cp}} & \multicolumn{2}{c}{\textbf{eng. feat.}} & \multicolumn{2}{c}{\textbf{PCA}} \\ \midrule
                            &           & \textit{old}     & \textit{fav.}    & \textit{old}     & \textit{fav.}     & \textit{old}       & \textit{fav.}      & \textit{old}   & \textit{fav.}   \\ \midrule
\multirow{2}{*}{\texttt{C}} & $h^{1,1}$ & 14               & 1000             & 170              & 36                & 3                  & 40                 & 1.0            & 1000            \\
                            & $h^{2,1}$ & 40               & 1000             & 1.0              & 1.0               & 84                 & 62                 & 45             & 40              \\ \midrule
\multirow{2}{*}{$\epsilon$} & $h^{1,1}$ & 0.01             & 0.01             & 0.45             & 0.03              & 0.05               & 0.3                & 0.02           & 0.01            \\
                            & $h^{2,1}$ & 0.01             & 0.01             & 0.01             & 0.09              & 0.29               & 0.10               & 0.20           & 0.09            \\ \midrule
\multirow{2}{*}{$\gamma$}   & $h^{1,1}$ & 0.03             & 0.002            & 0.110            & 0.009             & 0.07               & 0.003              & 0.02           & 0.001           \\
                            & $h^{2,1}$ & 0.06             & 0.100            & 0.013            & 1000              & 0.016              & 0.005              & 0.013          & 0.006           \\ \bottomrule
\end{tabular}%
\caption{Hyperparameter choices of the SVR regression with Gaussian kernel.}
\label{tab:hyp:svrrbf}
\end{table}

\paragraph{Results}

In \Cref{tab:res:svrrbf}, we show the accuracy of the predictions on the test sets.
In the favourable dataset, we can immediately appreciate the strong linear dependence of $h^{1,1}$ on the number of projective spaces: even though there are a few non favourable embeddings in the dataset, the kernel trick is able to map them in a better representation and improve the accuracy.
The predictions for the original dataset have also improved and are the best results we found using shallow learning.
The predictions using only the configuration matrix matches~\cite{Bull:2018:MachineLearningCICY}, but we can slightly improve the accuracy by using a combination of engineered features and PCA.

\begin{figure}[htp]
	\centering
	\includegraphics[width=\textwidth]{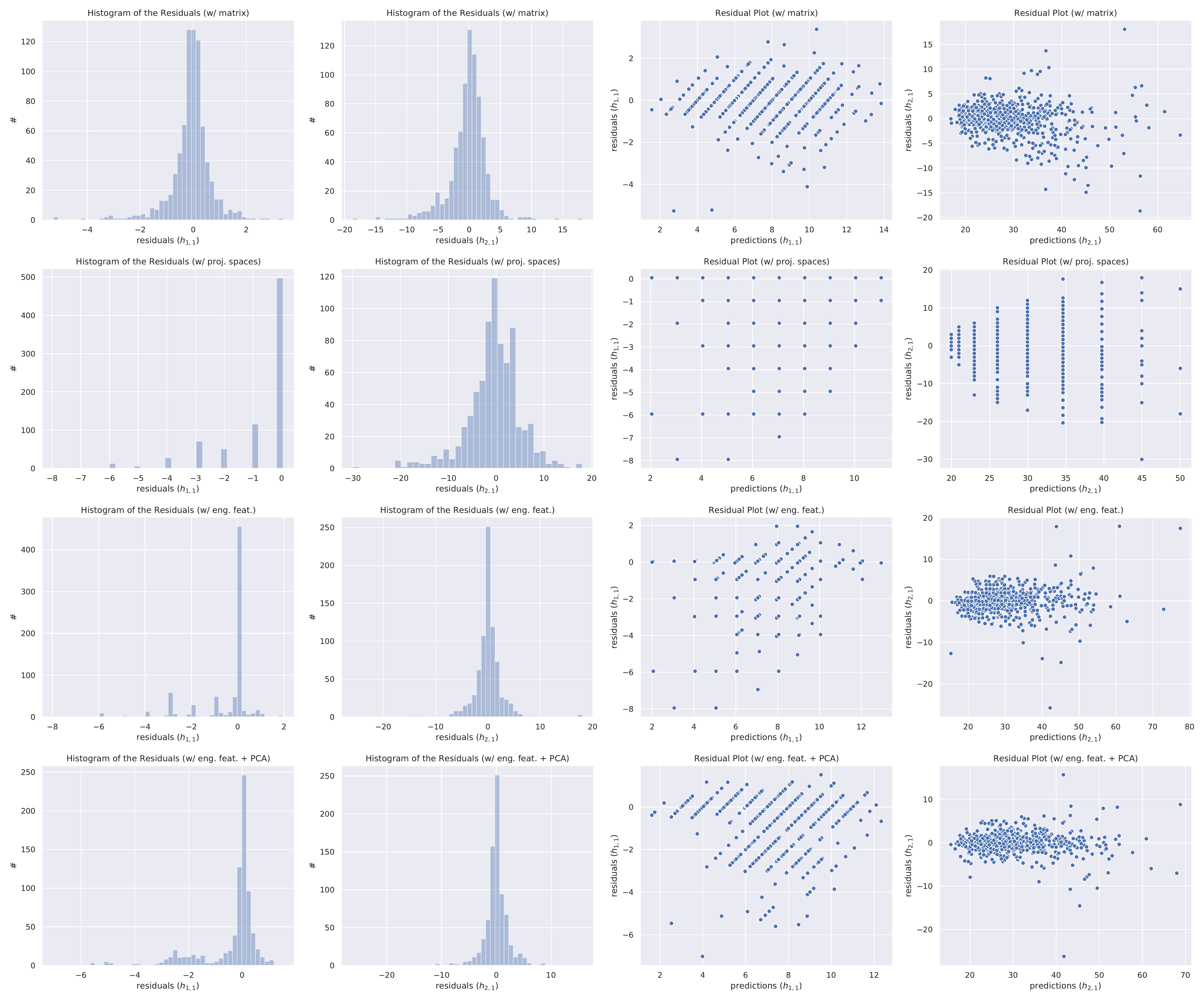}
	\caption{Plots of the residual errors for the SVM with Gaussian kernel.}
	\label{fig:res:svrrbf}
\end{figure}

In \Cref{fig:res:svrrbf}, we show the residual plots and their histograms for the original dataset: residuals follow peaked distributions which, in this case, do not present a second smaller peak (thus we need to round to the next integer the predictions) and good variate distribution over the predictions.

\begin{figure}[htp]
	\centering

	\begin{subfigure}[c]{0.45\linewidth}
		\centering
		\includegraphics[width=\textwidth]{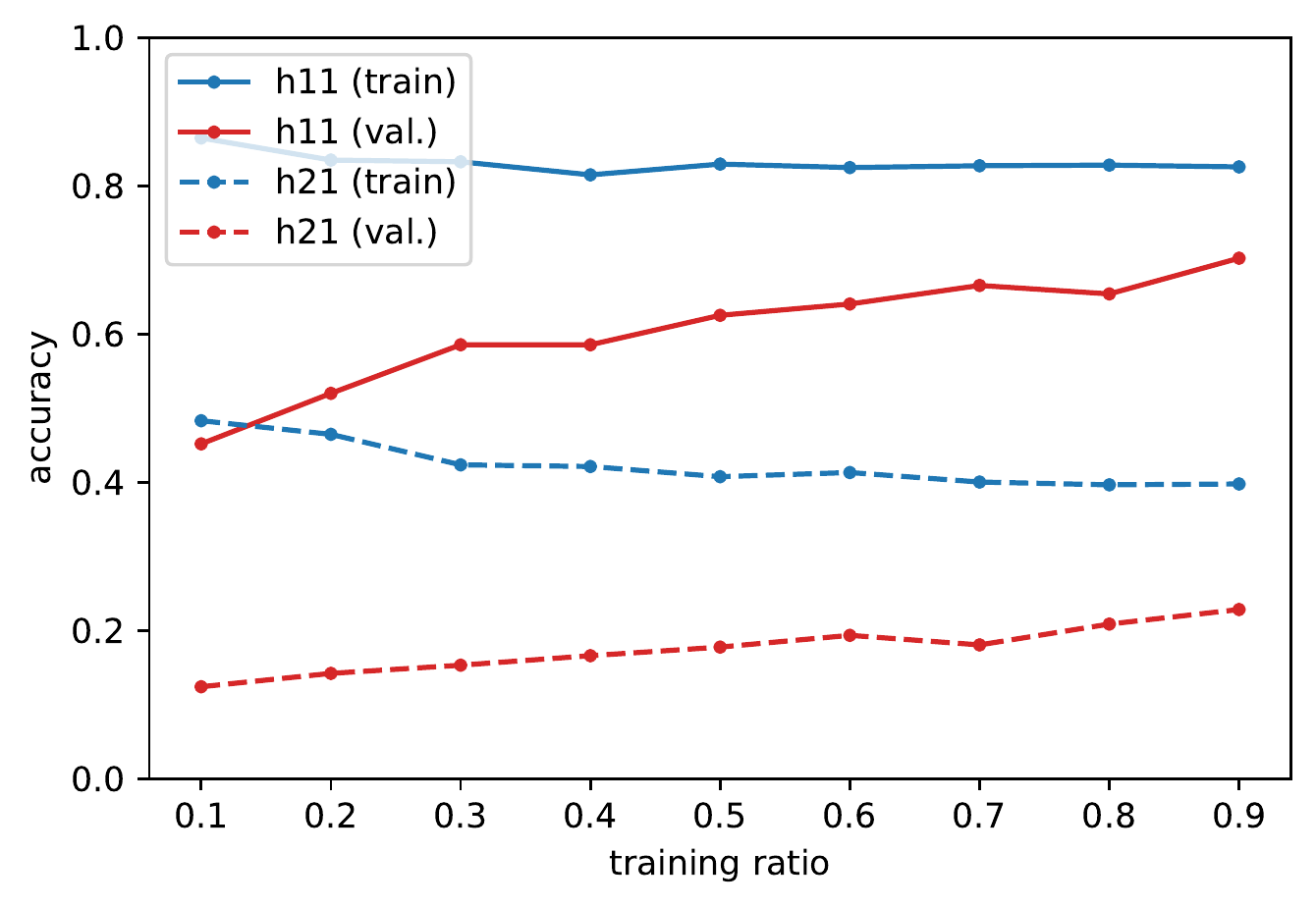}
		\caption{input: \lstinline!matrix!, $C = 15, \gamma = 0.03, \epsilon = 0.1$}
	\end{subfigure}
	\qquad
	\begin{subfigure}[c]{0.45\linewidth}
		\centering
		\includegraphics[width=\textwidth]{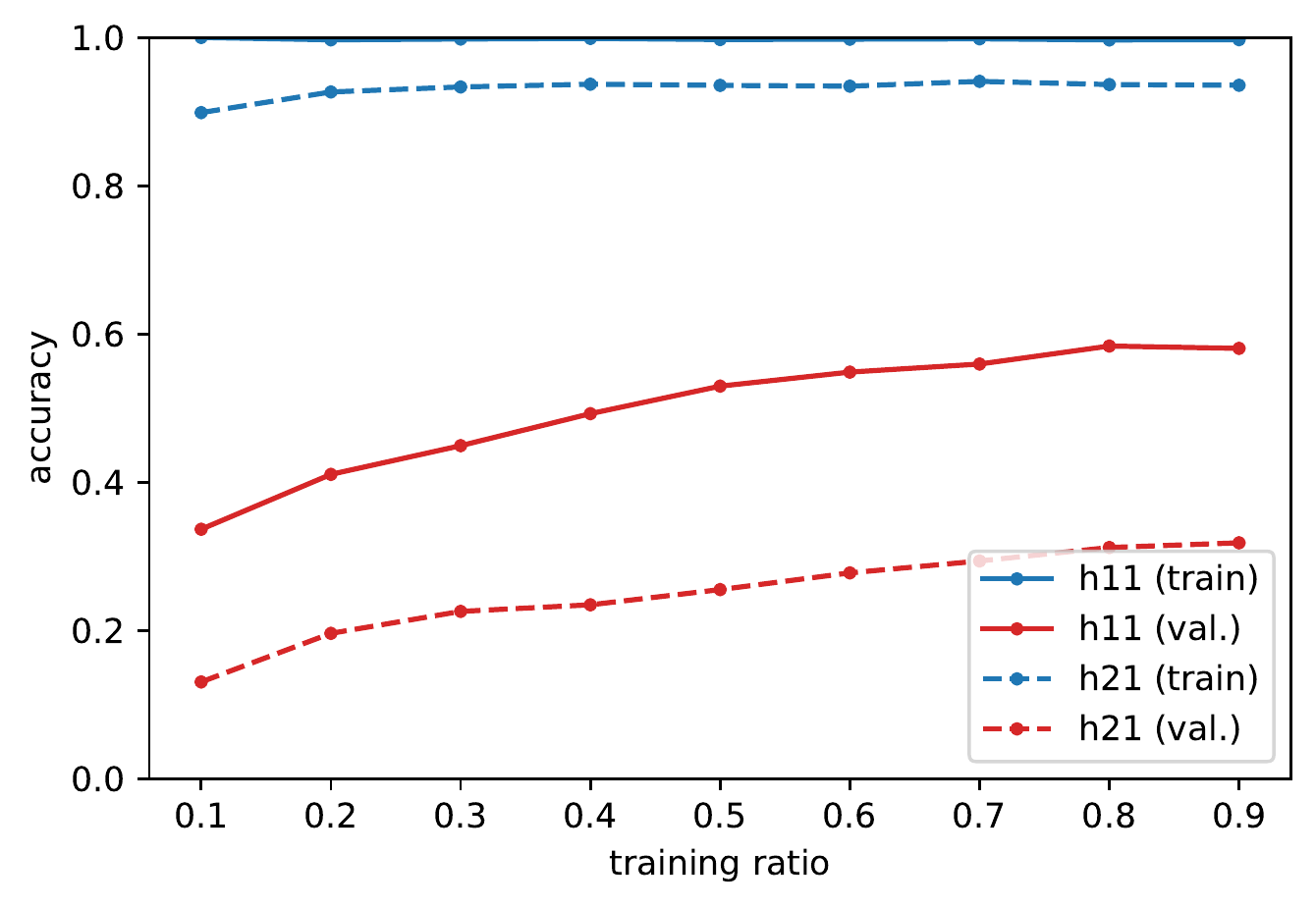}
		\caption{input: all, $C = 10, \gamma = 0.03, \epsilon = 0.1$}
	\end{subfigure}

	\caption{Learning curves for the SVM with Gaussian kernel (original dataset), using a single model for both Hodge numbers.}
	\label{fig:lc:svrrbf}
\end{figure}

\begin{figure}[htp]
	\centering

	\begin{subfigure}[c]{0.45\linewidth}
		\centering
		\includegraphics[width=\textwidth]{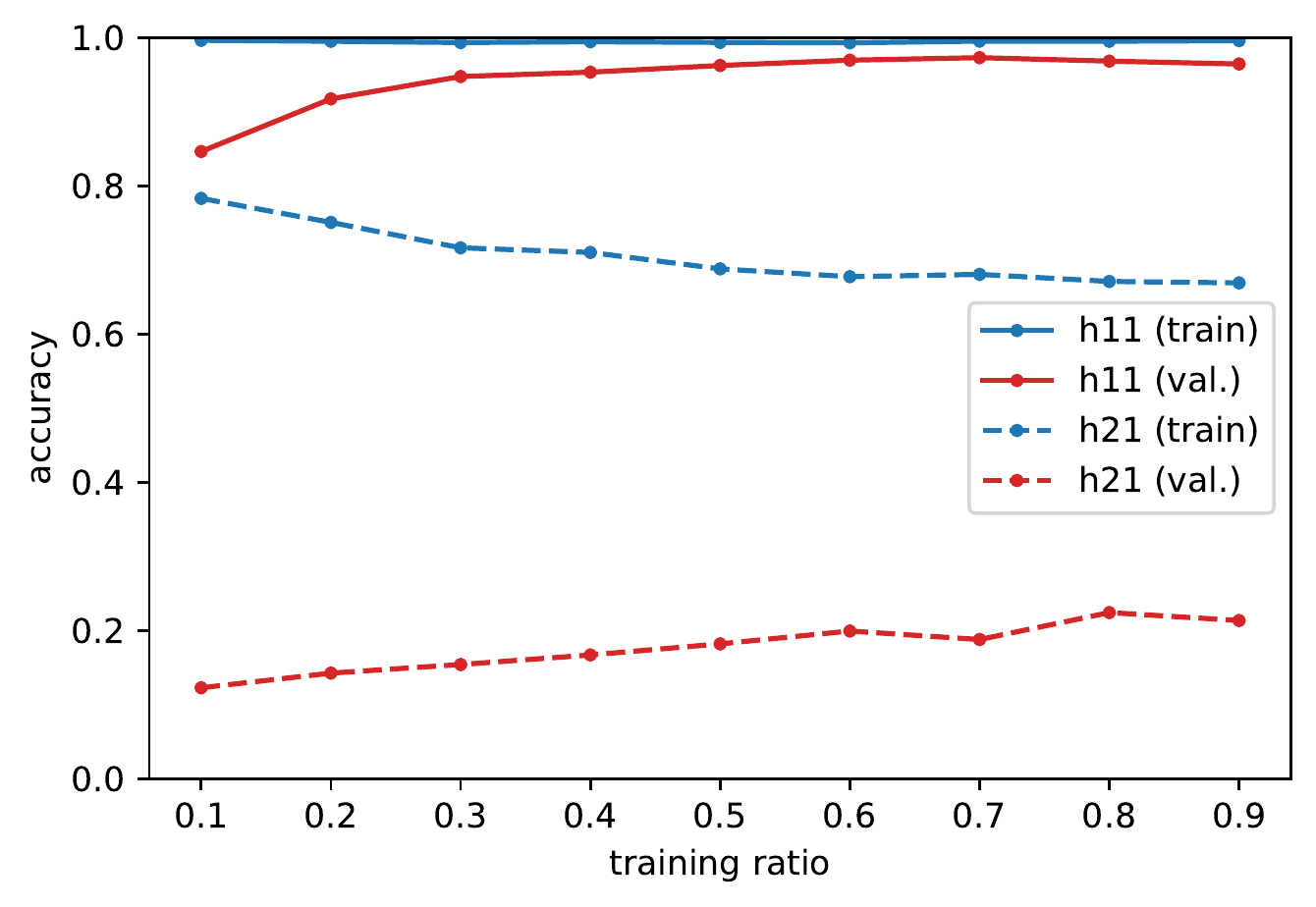}
		\caption{input: \lstinline!matrix!, $C = 20, \gamma = \mathtt{scale}, \epsilon = 0.1$}
	\end{subfigure}
	\qquad
	\begin{subfigure}[c]{0.45\linewidth}
		\centering
		\includegraphics[width=\textwidth]{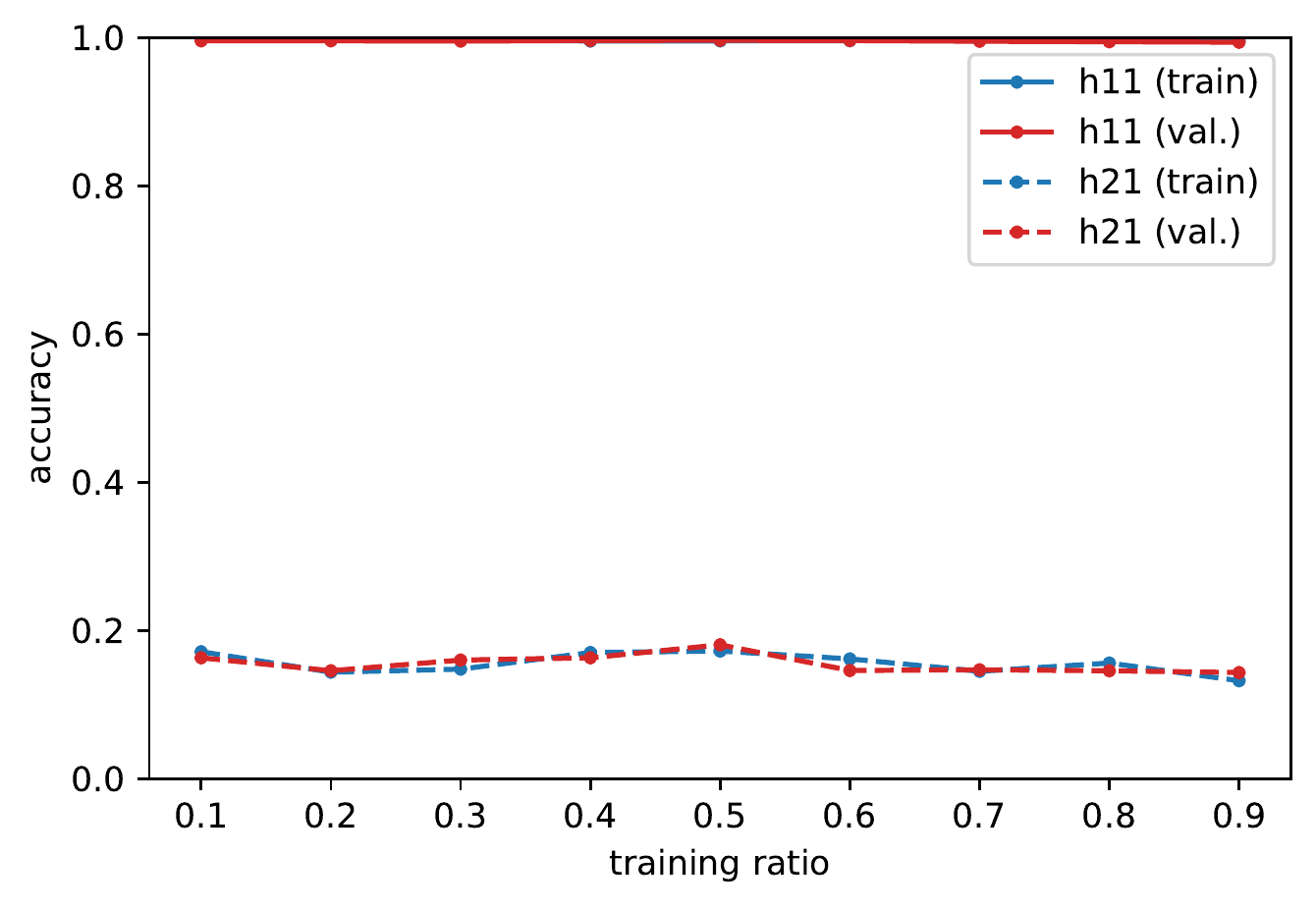}
		\caption{input: all, $C = 20, \gamma = \mathtt{scale}, \epsilon = 0.1$}
	\end{subfigure}

	\caption{Learning curves for the SVM with Gaussian kernel (favourable dataset), using a single model for both Hodge numbers.}
	\label{fig:lc:svrrbf-fav}
\end{figure}

The Gaussian kernel is also more influenced by the size of the training set.
Using 50\% of the samples as training set we witnessed a drop in accuracy of 3\% while using engineered features and the PCA, and around 1\% to 2\% in all other cases.
The learning curves (\Cref{fig:lc:svrrbf}) show that the accuracy improves by using more data.
Interestingly, it shows that using all engineered features leads to an overfit on the training data since both Hodge numbers reach almost $100\%$, while this is not the case for $h^{2,1}$.
For comparison, we also display in \Cref{fig:lc:svrrbf-fav} the learning curve for the favourable dataset: this shows that predicting $h^{1,1}$ accurately works out-of-the-box.

\begin{table}[htp]
\centering
\begin{tabular}{@{}cccccc@{}}
  \toprule
                          &           & \textbf{matrix} & \textbf{num\_cp} & \textbf{eng. feat.} & \textbf{PCA} \\ \midrule
    \multirow{2}{*}
    {\emph{original}}   & $h^{1,1}$ & 70\%            & 63\%             & 66\%                & 72\%         \\
                          & $h^{2,1}$ & 22\%            & 10\%             & 36\%                & 34\%         \\ \midrule
    \multirow{2}{*}
    {\emph{favourable}} & $h^{1,1}$ & 99\%            & 100\%            & 100\%               & 100\%        \\
                          & $h^{2,1}$ & 22\%            & 17\%             & 32\%                & 33\%         \\ \bottomrule
\end{tabular}
\caption{Accuracy of the Gaussian SVM on the test split.}
\label{tab:res:svrrbf}
\end{table}

\subsection{Decision Trees}
\label{sec:ml:trees}

We now consider two algorithms based on decision trees: random forests and gradient boosted trees.
Decision trees are powerful algorithms which implement a simple decision rule (in the style of an \emph{if\dots then\dots else\dots} statement) to classify or assign a value to the predictions.
However, they have a tendency to adapt too well to the training set and to not be robust enough against small changes in the training data.
We consider a generalisation of this algorithm used for \emph{ensemble learning}: this is a technique in ML which uses multiple estimators (they can be the same or different) to improve the performances.
We will present the results of \emph{random forests} of trees which increase the bias compared to a single decision tree, and \emph{gradient boosted} decision trees, which can use smaller trees to decrease the variance and learn better representations of the input data by iterating their decision functions and use information on the previous runs to improve (see \Cref{sec:app:trees} for a more in-depth description).

\subsubsection{Random Forests}

The random forest algorithm is implemented with Scikit's \lstinline!ensemble.RandomForestRegressor!.

\paragraph{Parameters}

Hyperparameter tuning for decision trees can in general be quite challenging.
From the general theory on random forests (\Cref{sec:app:trees}), we can try and look for particular shapes of the trees: this ensemble learning technique usually prefers a small number of fully grown trees.
We performed only 25 iterations of the optimisation process due to the very long time taken to train all the decision trees.

In \Cref{tab:hyp:rndfor}, we show the hyperparameters used for the predictions.
As we can see from \texttt{n\_estimator}, random forests are usually built with a small number of fully grown (specified by \texttt{max\_depth} and \texttt{max\_leaf\_nodes}) trees (not always the case, though).
In order to avoid overfit we also tried to increase the number of samples necessary to split a branch or create a leaf node using \texttt{min\_samples\_leaf} and \texttt{min\_samples\_split} (introducing also a weight on the samples in the leaf nodes specified by \texttt{min\_weight\_fraction\_leaf} to balance the tree).
Finally the \texttt{criterion} chosen by the optimisation reflects the choice of the trees to measure the impurity of the predictions using either the mean squared error (\texttt{mse}) or the mean absolute error (\texttt{mae}) of the predictions (see \Cref{sec:app:trees}).

\begin{table}[htp]
\centering
\resizebox{\textwidth}{!}{%
\begin{tabular}{@{}lccccccccc@{}}
\toprule
                                                      &           & \multicolumn{2}{c}{\textbf{matrix}}  & \multicolumn{2}{c}{\textbf{num\_cp}}        & \multicolumn{2}{c}{\textbf{eng. feat.}} & \multicolumn{2}{c}{\textbf{PCA}} \\ \midrule
                                                      &           & \textit{old}         & \textit{fav.} & \textit{old}         & \textit{fav.}        & \textit{old}   & \textit{fav.}          & \textit{old}   & \textit{fav.}   \\ \midrule
\multirow{2}{*}{\texttt{criterion}}                            & $h^{1,1}$ & \texttt{mse}         & \texttt{mse}  & \texttt{mae}         & \texttt{mae}         & \texttt{mae}   & \texttt{mse}           & \texttt{mae}   & \texttt{mae}    \\
                                                      & $h^{2,1}$ & \texttt{mae}         & \texttt{mae}  & \texttt{mae}         & \texttt{mae}         & \texttt{mae}   & \texttt{mae}           & \texttt{mae}   & \texttt{mae}    \\ \midrule
\multirow{2}{*}{\texttt{max\_depth}}                  & $h^{1,1}$ & 100                  & 100           & 100                  & 30                   & 90             & 30                     & 30             & 60              \\
                                                      & $h^{2,1}$ & 90                   & 100           & 90                   & 75                   & 100            & 100                    & 100            & 60              \\ \midrule
\multirow{2}{*}{\texttt{max\_leaf\_nodes}}            & $h^{1,1}$ & 100                  & 80            & 90                   & 20                   & 20             & 35                     & 90             & 90              \\
                                                      & $h^{2,1}$ & 90                   & 100           & 100                  & 75                   & 100            & 60                     & 100            & 100             \\ \midrule
\multirow{2}{*}{\texttt{min\_samples\_leaf}}          & $h^{1,1}$ & 1                    & 1             & 1                    & 15                   & 1              & 15                     & 1              & 1               \\
                                                      & $h^{2,1}$ & 3                    & 1             & 4                    & 70                   & 1              & 70                     & 30             & 1               \\ \midrule
\multirow{2}{*}{\texttt{min\_samples\_split}}         & $h^{1,1}$ & 2                    & 30            & 20                   & 35                   & 10             & 10                     & 100            & 100             \\
                                                      & $h^{2,1}$ & 30                   & 2             & 50                   & 45                   & 2              & 100                    & 2              & 100             \\ \midrule
\multirow{2}{*}{\texttt{min\_weight\_fraction\_leaf}} & $h^{1,1}$ & 0.0                  & 0.0           & 0.0                  & $1.7 \times 10^{-3}$ & 0.0            & 0.009   & 0.0            & 0.0             \\
                                                      & $h^{2,1}$ & $3.0 \times 10^{-4}$ & 0.0           & $1.0 \times 10^{-4}$ & 0.13                 & 0.0            & 0.0                    & 0.0            & 0.0             \\ \midrule
\multirow{2}{*}{\texttt{n\_estimators}}               & $h^{1,1}$ & 10                   & 100           & 45                   & 120                  & 155            & 300                    & 10             & 300             \\
                                                      & $h^{2,1}$ & 190                  & 10            & 160                  & 300                  & 10             & 10                     & 10             & 300             \\ \bottomrule
\end{tabular}%
}
\caption{Hyperparameter choices of the random forest regression.}
\label{tab:hyp:rndfor}
\end{table}

\paragraph{Results}

In \Cref{tab:res:rndfor}, we summarise the accuracy reached using random forests of decision trees as estimators.
As we already expected, the contribution of the number of projective spaces helps the algorithm to generate better predictions.
In general, it seems that the engineered features alone can already provide a good basis for predictions.
In the case of $h^{2,1}$, the introduction of the principal components of the configuration matrix also increases the prediction capabilities.
As in most other cases, we used the floor function for the predictions on the original dataset and the rounding to next integer for the favourable one.

\begin{figure}[htp]
	\centering
	\includegraphics[width=0.9\textwidth]{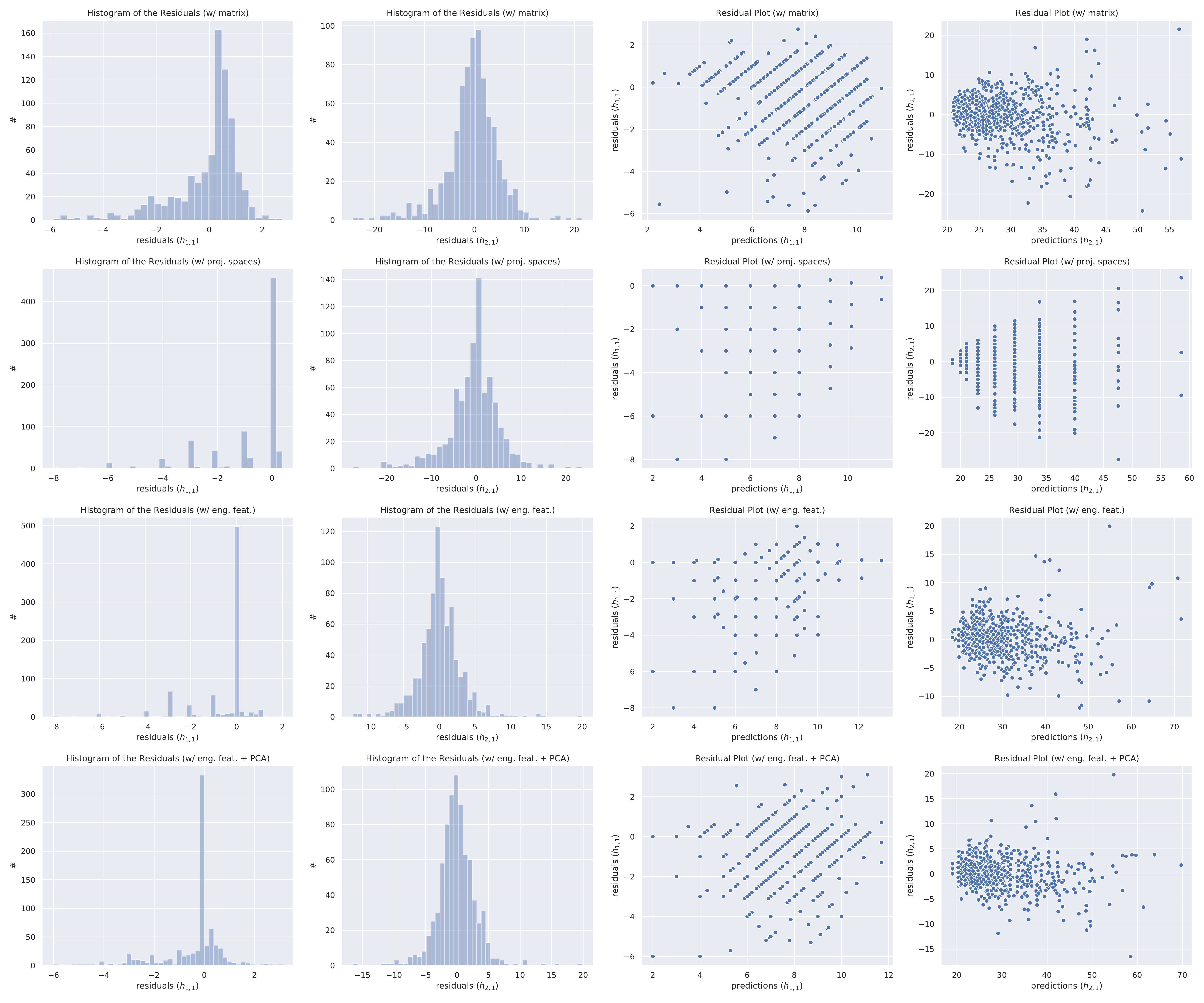}
	\caption{Plots of the residual errors for the random forests.}
	\label{fig:res:rndfor}
\end{figure}

\begin{figure}[htp]
	\centering

	\begin{subfigure}[c]{0.45\linewidth}
		\centering
		\includegraphics[width=\textwidth]{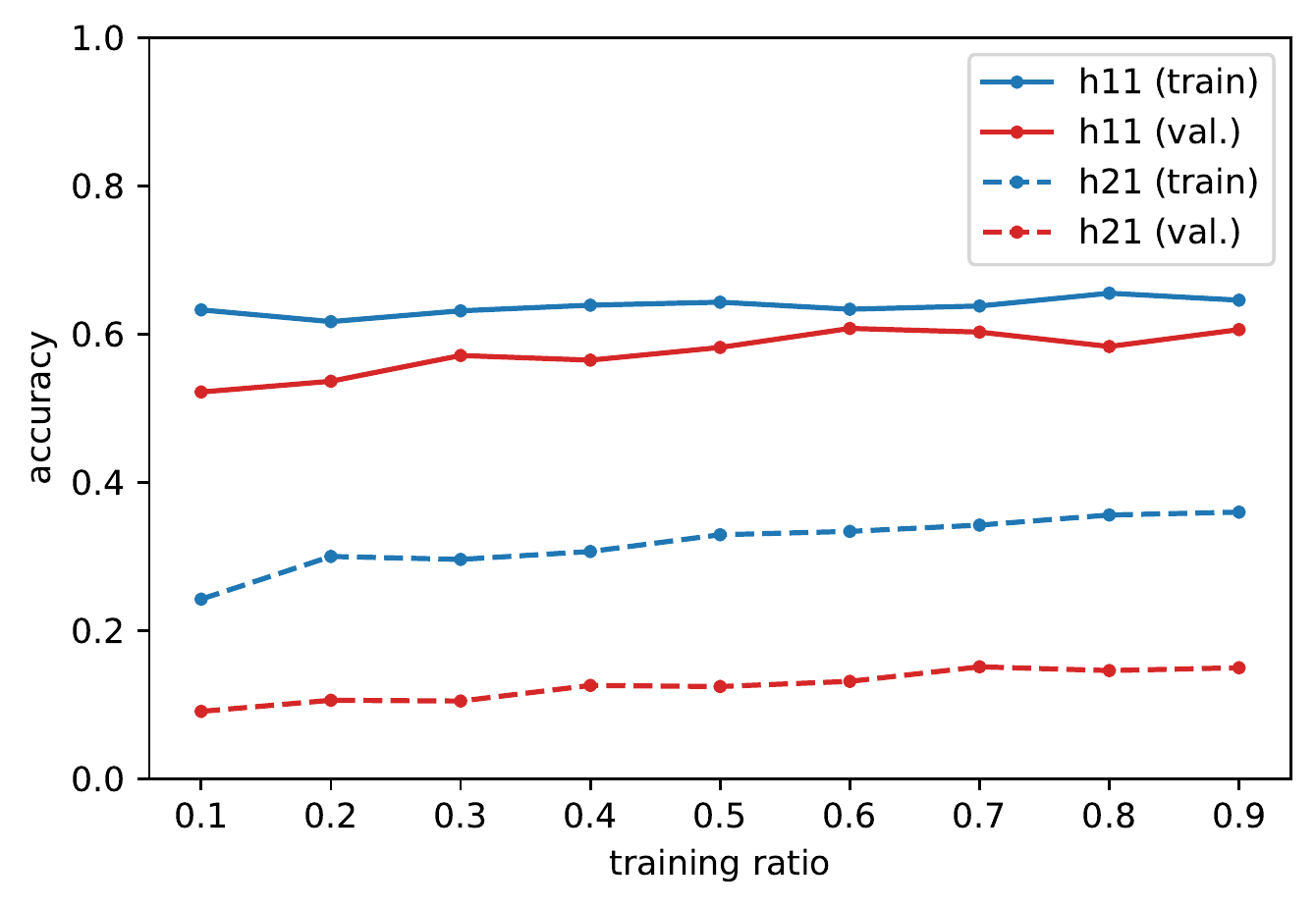}
		\caption{input: \lstinline!matrix!, default parameters}
	\end{subfigure}
	\qquad
	\begin{subfigure}[c]{0.45\linewidth}
		\centering
		\includegraphics[width=\textwidth]{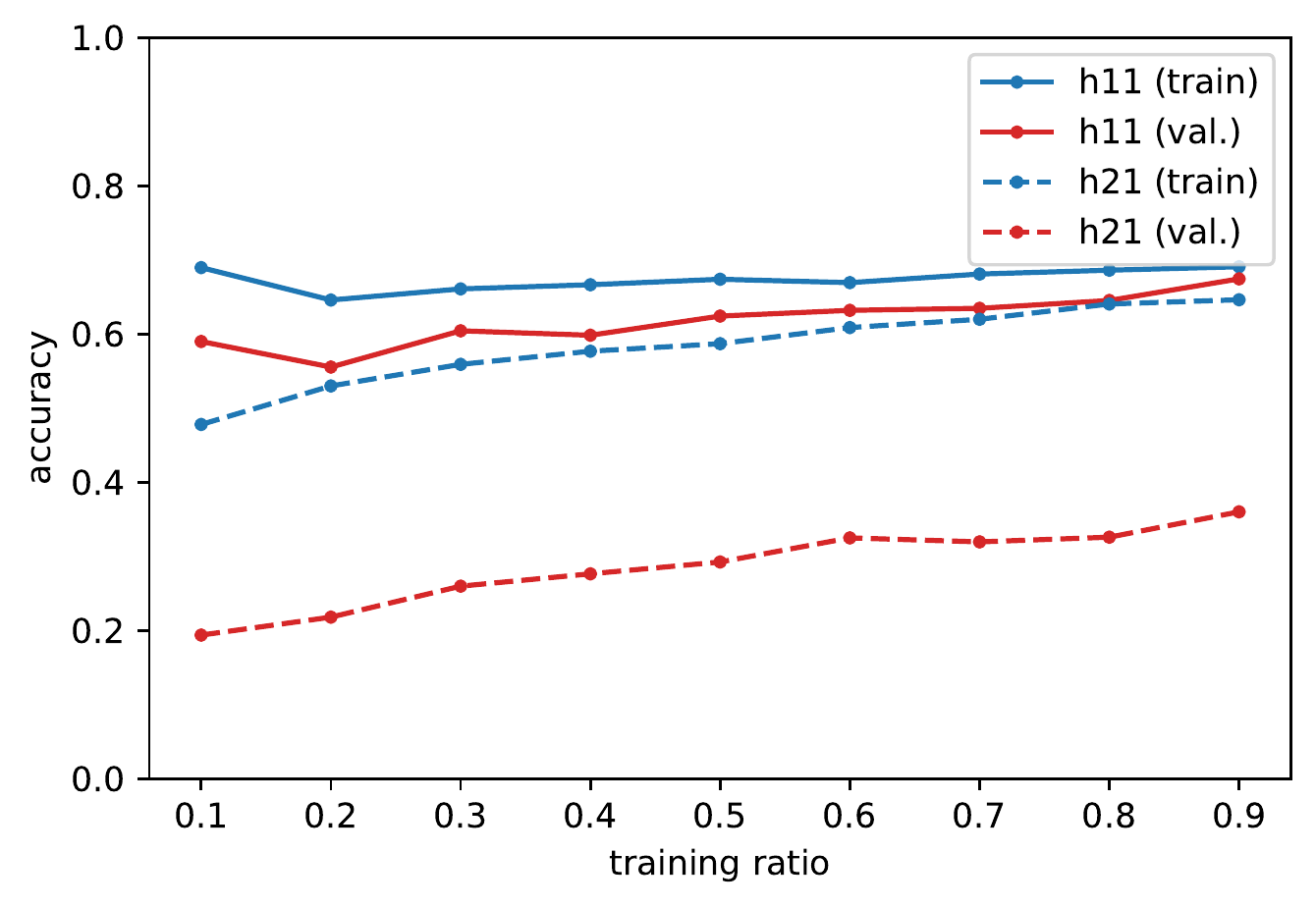}
		\caption{input: all, default parameters}
	\end{subfigure}

	\caption{Learning curves for the random forest (original dataset), including outliers and using a single model for both Hodge numbers.}
	\label{fig:lc:rndfor}
\end{figure}

As usual, in \Cref{fig:res:rndfor} we show the histograms of the distribution of the residual errors and the scatter plots of the residuals.
While the distributions of the errors are slightly wider than the SVM algorithms, the scatter plots of the residual show a strong heteroscedasticity in the case of the fit using the number of projective spaces: though quite accurate, the model is strongly incomplete.
The inclusion of the other engineered features definitely helps and also leads to better predictions.
Learning curves are displayed in \Cref{fig:lc:rndfor}.

\begin{table}[htp]
\centering
\begin{tabular}{@{}cccccc@{}}
  \toprule
                          &           & \textbf{matrix} & \textbf{num\_cp} & \textbf{eng. feat.} & \textbf{PCA} \\ \midrule
    \multirow{2}{*}
    {\emph{original}}   & $h^{1,1}$ & 55\%            & 63\%             & 66\%                & 64\%         \\
                          & $h^{2,1}$ & 12\%            & 9\%              & 17\%                & 18\%         \\ \midrule
    \multirow{2}{*}
    {\emph{favourable}} & $h^{1,1}$ & 89\%            & 99\%             & 98\%                & 98\%         \\
                          & $h^{2,1}$ & 14\%            & 17\%             & 22\%                & 27\%         \\ \bottomrule
\end{tabular}
\caption{Accuracy of the random forests on the test split.}
\label{tab:res:rndfor}
\end{table}

\subsubsection{Gradient Boosted Trees}

We used the class \lstinline!ensemble.GradientBoostingRegressor! from Scikit in order to implement the gradient boosted trees.

\paragraph{Parameters}

Hyperparameter optimisation has been performed using 25 iterations of the Bayes search algorithm since by comparison the gradient boosting algorithms took the longest learning time.
We show the chosen hyperparameters in \Cref{tab:hyp:grdbst}.

With respect to the random forests, for the gradient boosting we also need to introduce the \texttt{learning\_rate} (or \emph{shrinking parameter}) which controls the gradient descent of the optimisation which is driven by the choice of the \texttt{loss} parameters (\texttt{ls} is the ordinary least squares loss, \texttt{lad} is the least absolute deviation and \texttt{huber} is a combination of the previous two losses weighted by the hyperparameter $\alpha$).
We also introduce the \texttt{subsample} hyperparameter which chooses a fraction of the samples to be fed into the algorithm at each iteration.
This procedure has both a regularisation effect on the trees, which should not adapt too much to the training set, and speeds up the training (at least by a very small amount).

\begin{table}[htp]
\centering
\resizebox{\textwidth}{!}{%
\begin{tabular}{@{}lccccccccc@{}}
\toprule
                                                      &           & \multicolumn{2}{c}{\textbf{matrix}} & \multicolumn{2}{c}{\textbf{num\_cp}}   & \multicolumn{2}{c}{\textbf{eng. feat.}}         & \multicolumn{2}{c}{\textbf{PCA}} \\ \midrule
                                                      &           & \textit{old}     & \textit{fav.}    & \textit{old}           & \textit{fav.} & \textit{old}           & \textit{fav.}          & \textit{old}   & \textit{fav.}   \\ \midrule
\multirow{2}{*}{$\alpha$}                             & $h^{1,1}$ & 0.4              & ---              & ---                    & ---           & ---                    & ---                    & ---            & ---             \\
                                                      & $h^{2,1}$ & ---              & 0.11             & ---                    & ---           & 0.99                   & ---                    & ---            & ---             \\ \midrule
\multirow{2}{*}{\texttt{criterion}}                   & $h^{1,1}$ & \texttt{mae}     & \texttt{mae}     & \texttt{friedman\_mse} & \texttt{mae}  & \texttt{friedman\_mse} & \texttt{friedman\_mse} & \texttt{mae}   & \texttt{mae}    \\
                                                      & $h^{2,1}$ & \texttt{mae}     & \texttt{mae}     & \texttt{friedman\_mse} & \texttt{mae}  & \texttt{mae}           & \texttt{mae}           & \texttt{mae}   & \texttt{mae}    \\ \midrule
\multirow{2}{*}{\texttt{learning\_rate}}              & $h^{1,1}$ & 0.3              & 0.04             & 0.6                    & 0.03          & 0.15                   & 0.5                    & 0.04           & 0.03            \\
                                                      & $h^{2,1}$ & 0.6              & 0.5              & 0.3                    & 0.5           & 0.04                   & 0.02                   & 0.03           & 0.07            \\ \midrule
\multirow{2}{*}{\texttt{loss}}                        & $h^{1,1}$ & huber            & ls               & lad                    & ls            & ls                     & lad                    & ls             & ls              \\
                                                      & $h^{2,1}$ & ls               & huber            & ls                     & ls            & huber                  & ls                     & ls             & lad             \\ \midrule
\multirow{2}{*}{\texttt{max\_depth}}                  & $h^{1,1}$ & 100              & 100              & 15                     & 60            & 2                      & 100                    & 55             & 2               \\
                                                      & $h^{2,1}$ & 85               & 100              & 100                    & 30            & 35                     & 60                     & 15             & 2               \\ \midrule
\multirow{2}{*}{\texttt{min\_samples\_split}}         & $h^{1,1}$ & 2                & 30               & 20                     & 35            & 10                     & 10                     & 100            & 100             \\
                                                      & $h^{2,1}$ & 30               & 2                & 50                     & 45            & 2                      & 100                    & 2              & 100             \\ \midrule
\multirow{2}{*}{\texttt{min\_weight\_fraction\_leaf}} & $h^{1,1}$ & 0.03             & 0.0              & 0.0                    & 0.2           & 0.2                    & 0.0                    & 0.06           & 0.0             \\
                                                      & $h^{2,1}$ & 0.0              & 0.0              & 0.16                   & 0.004         & 0.0                    & 0.0                    & 0.0            & 0.0             \\ \midrule
\multirow{2}{*}{\texttt{n\_estimators}}               & $h^{1,1}$ & 90               & 240              & 120                    & 220           & 100                    & 130                    & 180            & 290             \\
                                                      & $h^{2,1}$ & 100              & 300              & 10                     & 20            & 200                    & 300                    & 300            & 300             \\ \midrule
\multirow{2}{*}{\texttt{subsample}}                   & $h^{1,1}$ & 0.8              & 0.8              & 0.9                    & 0.6           & 0.1                    & 0.1                    & 1.0            & 0.9             \\
                                                      & $h^{2,1}$ & 0.7              & 1.0              & 0.1                    & 0.9           & 0.1                    & 0.9                    & 0.1            & 0.2             \\ \bottomrule
\end{tabular}%
}
\caption{Hyperparameter choices of the gradient boosted decision trees.}
\label{tab:hyp:grdbst}
\end{table}

\paragraph{Results}

We show the results of gradient boosting in \Cref{tab:res:grdbst}.
As usual, the linear dependence of $h^{1,1}$ on the number of projective spaces is evident and in this case also produces the best accuracy result (using the floor function for the original dataset and rounding to the next integer for the favourable dataset) for $h^{1,1}$.
$h^{2,1}$ is once again strongly helped by the presence of the redundant features.

\begin{figure}[htp]
	\centering
	\includegraphics[width=0.9\textwidth]{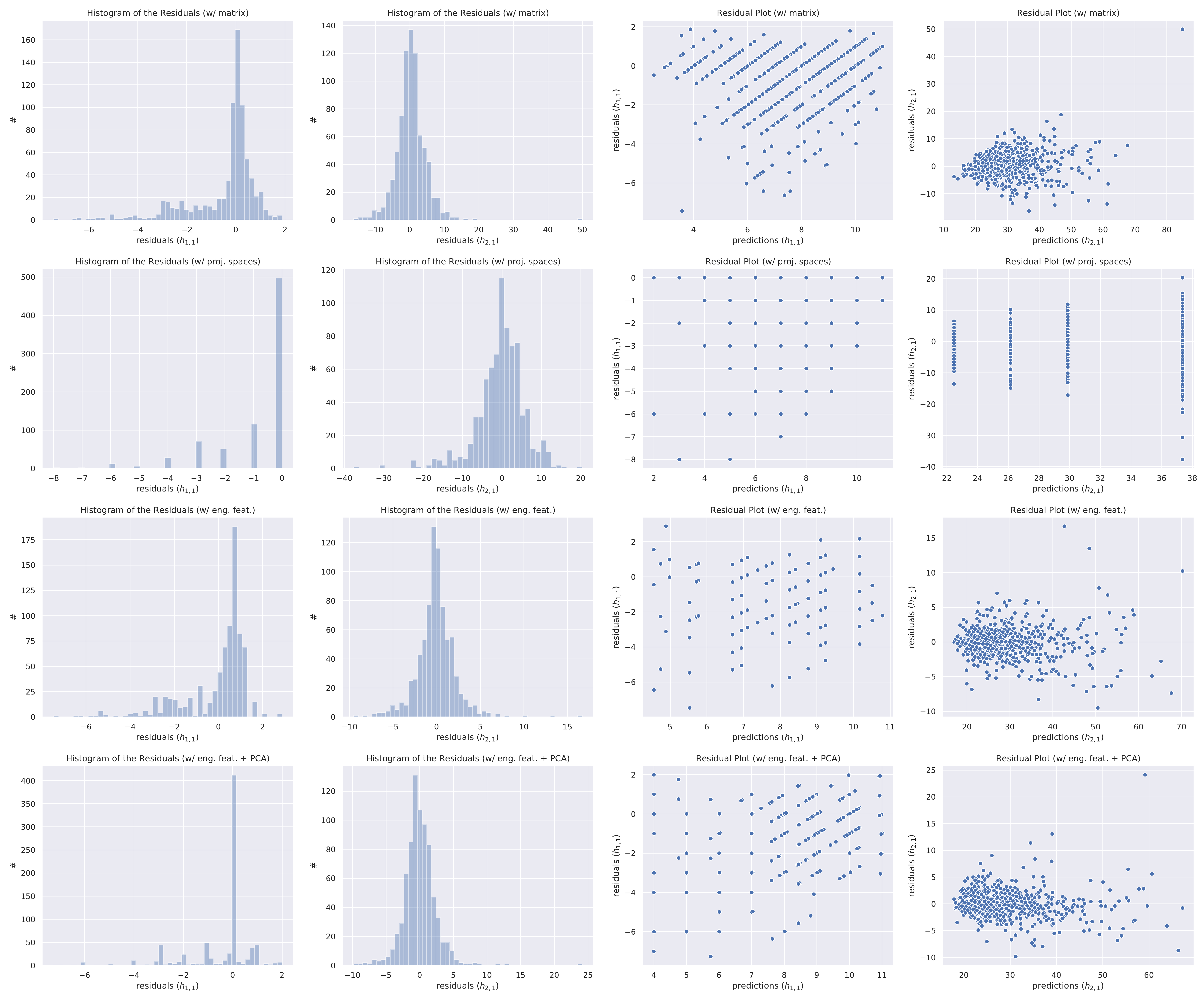}
	\caption{Plots of the residual errors for the gradient boosted trees.}
	\label{fig:res:grdbst}
\end{figure}

In \Cref{fig:res:grdbst}, we finally show the histograms and the scatter plots of the residual errors for the original dataset showing that also in this case the choice of the floor function can be justified and that the addition of the engineered features certainly improves the overall variance of the residuals.

\begin{table}[htp]
\centering
\begin{tabular}{@{}cccccc@{}}
  \toprule
                          &           & \textbf{matrix} & \textbf{num\_cp} & \textbf{eng. feat.} & \textbf{PCA} \\ \midrule
    \multirow{2}{*}
    {\emph{original}}   & $h^{1,1}$ & 50\%            & 63\%             & 61\%                & 58\%         \\
                          & $h^{2,1}$ & 14\%            & 9\%              & 23\%                & 21\%         \\ \midrule
    \multirow{2}{*}
    {\emph{favourable}} & $h^{1,1}$ & 97\%            & 100\%            & 99\%                & 99\%         \\
                          & $h^{2,1}$ & 17\%            & 16\%             & 35\%                & 22\%         \\ \bottomrule
\end{tabular}
\caption{Accuracy of the gradient boosting on the test split.}
\label{tab:res:grdbst}
\end{table}

\subsection{Neural Networks}

In this section we approach the problem of predicting the Hodge numbers using artificial neural networks (ANN), which we briefly review in \Cref{sec:app:nn}.
We use Google's \emph{Tensorflow} framework and \emph{Keras}, its high-level API, to implement the architectures and train the networks~\cite{Package:Tensorflow, Package:Keras}.
We explore different architectures and discuss the results.

Differently from the previous algorithms, we do not perform a cross-validation scoring but we simply retain \SI{10}{\percent} of the total set as a holdout validation set (also referred to as \emph{development} set) due to the computation power available.
Thus, we use \SI{80}{\percent} of the samples for training, \SI{10}{\percent} for evaluation and \SI{10}{\percent} as a test set.
For the same reason, the optimisation of the algorithm has been performed manually.

We always use the Adam optimiser with default learning rate $\num{e-3}$ to perform the gradient descent and a fix batch size of $32$.
The network is trained for a large number of epochs to avoid missing possible local optima.
In order to avoid overshooting the minimum, we dynamically reduce the learning rate both using the \emph{Adam} optimiser, which implements learning rate decay, and through the callback \texttt{callbacks.ReduceLROnPlateau} in Keras, which scales the learning rate by a given factor when the monitored quantity (e.g.\ the validation loss) does not decrease): we choose to reduce it by $0.3$ when the validation loss does not improve for at least $75$ epochs.
Moreover, we stop training when the validation loss does not improve during $200$ epochs.
Clearly, we then keep only the weights of the neural networks which gave the best results.
Batch normalization layers are used with a momentum of $0. 99$.

Training and evaluation were performed on a \texttt{NVidia GeForce 940MX} laptop GPU with \SI{2}{\giga B} of RAM memory.

\subsubsection{Fully Connected Network}

First, we reproduce the analysis from~\cite{Bull:2018:MachineLearningCICY} for the prediction of $h^{1,1}$.

\paragraph{Model}

\begin{figure}[htp]
    \centering
    \begin{minipage}[t]{0.475\textwidth}
        \centering
        \includegraphics[width=\textwidth]{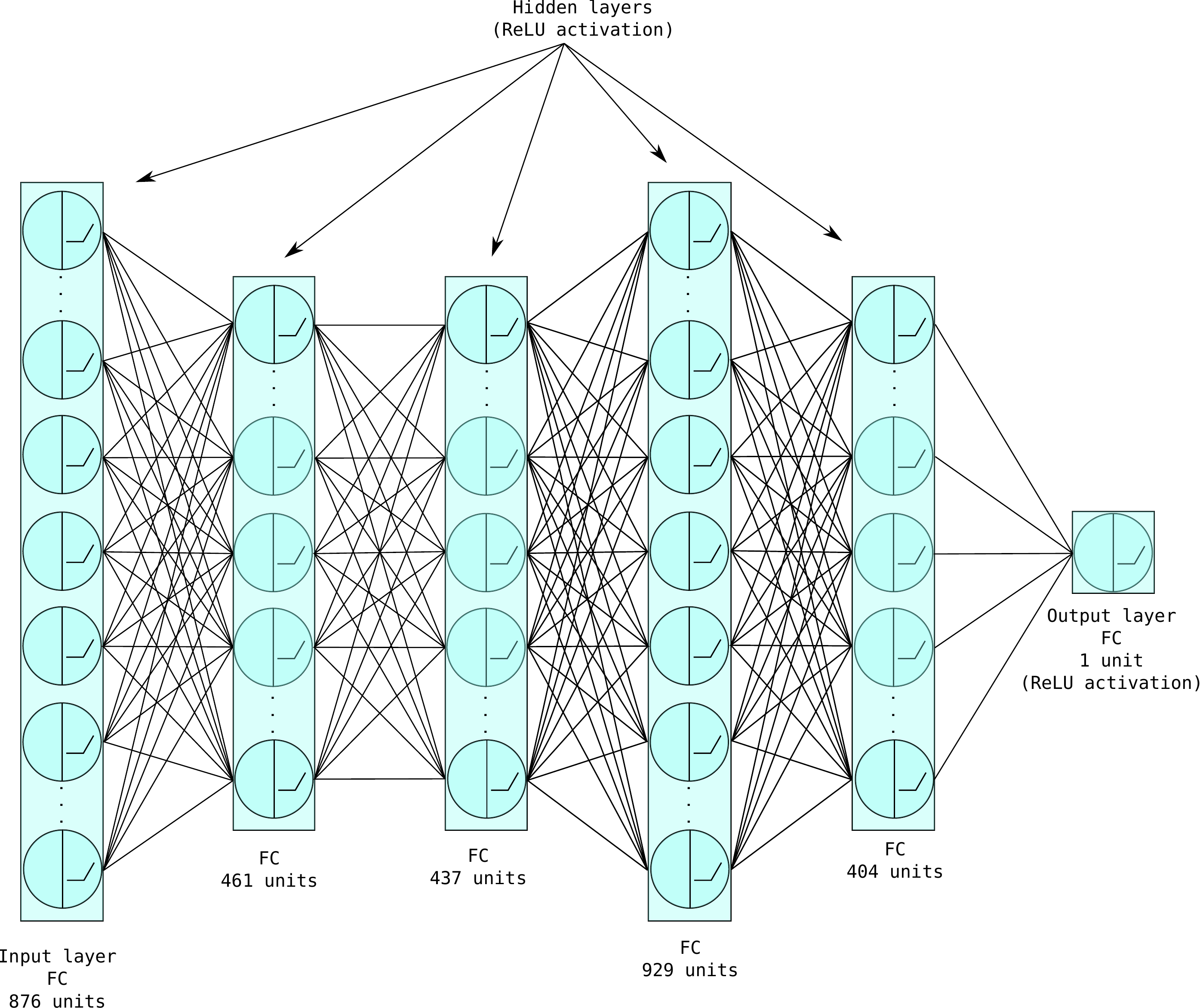}
        \caption{Architecture of the fully connected network to predict $h^{1,1}$.
        For simplicity we do not draw the dropout and batch normalisation layers present after every FC layer.}
        \label{fig:nn:dense}
    \end{minipage}
    \hfill
    \begin{minipage}[t]{0.475\textwidth}
        \centering
        \includegraphics[width=\textwidth, trim={0 0 6in 0}, clip]{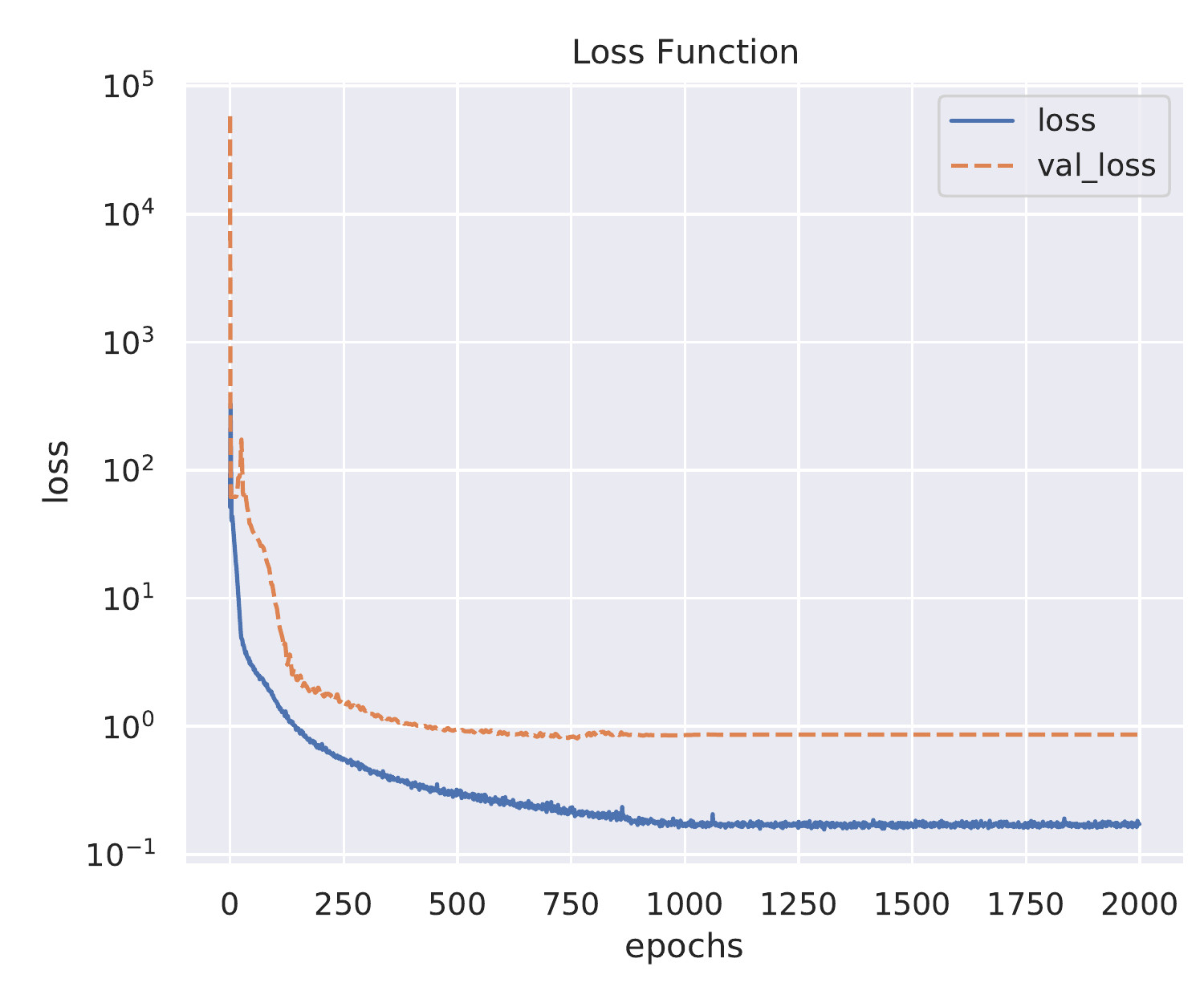}
        \caption{Loss function of the FC network in the original dataset.}
        \label{fig:nn:bull_et_al_loss}
    \end{minipage}
\end{figure}

The neural network presented in~\cite{Bull:2018:MachineLearningCICY} for the regression task contains $5$ hidden layers with $876$, $461$, $437$, $929$ and $404$ units (\Cref{fig:nn:dense}).
All layers (including the output layer) are followed by a ReLU activation and by a dropout layer with a rate of $\num{0.2072}$.
This network contains roughly $\num{1.58e6}$ parameters.

The other hyperparameters (like the optimiser, batch size, number of epochs, regularisation, etc.) are not mentioned.
In order to reproduce the results, we have filled the gap as follows:
\begin{itemize}
    \item Adam optimiser with batch size of $32$;
    
    \item maximal number epochs of $2000$ without early stopping;\footnote{It took around 20 minutes to train the model.}
    
    \item we implement learning rate reduction by $0.3$ after $75$ epochs without improvement of the validation loss;
    
    \item no $\ell_1$ or $\ell_2$ regularisation;
    
    \item a batch normalization layer~\cite{Ioffe:2015:BatchNormalizationAccelerating} after each fully connected layer.
\end{itemize}

\paragraph{Results}

We have first reproduced the results from~\cite{Bull:2018:MachineLearningCICY}, which are summarized in \Cref{tab:res:neuralnet-bull}.
The training process was very quick and the loss function is reported in \Cref{fig:nn:bull_et_al_loss}.
We obtain an accuracy of $77\%$ both on the development and the test set of the original dataset with $80\%$ of training data (see \Cref{tab:res:ann}).
Using the same network, we also achieved $97\%$ of accuracy in the favourable dataset.

\begin{table}[htb]
        \centering
        \begin{tabular}{@{}cccccc@{}}
            \toprule
            &
            \multicolumn{5}{c}{\textbf{training data}}
            \\
                &
                    $\num{10}\%$ &
                    $\num{30}\%$ &
                    $\num{50}\%$ &
                    $\num{70}\%$ &
                    $\num{90}\%$
                \\
                \midrule
                regression &
                        $\num{58}\%$ &
                        $\num{68}\%$ &
                        $\num{72}\%$ &
                        $\num{75}\%$ &
                        $\num{75}\%$
                \\
                classification &
                        $\num{68}\%$ &
                        $\num{78}\%$ &
                        $\num{82}\%$ &
                        $\num{85}\%$ &
                        $\num{88}\%$
                \\
                \bottomrule
        \end{tabular}
        \caption{Accuracy (approximate) for $h^{1,1}$ obtained in \cite[Figure~1]{Bull:2018:MachineLearningCICY}.}
        \label{tab:res:neuralnet-bull}
\end{table}

\subsubsection{Convolutional Network}

We then present a new purely convolutional network to predict $h^{1,1}$ and $h^{2,1}$, separately or together.
The advantage of such networks is that it requires a smaller number of parameters and is insensitive to the size of the inputs.
The latter point can be helpful to work without padding the matrices (of the same or different representations), but the use of a flatten layer removes this benefit.

\paragraph{Model}

The neural network has $4$ convolutional layers.
They are connected to the output layer with a intermediate flatten layer.
After each convolutional layer, we use the ReLU activation function and a batch normalisation layer (with momentum 0.99).
Convolutional layers use the padding option \lstinline!same! and a kernel of size $(5, 5)$ to be able to extract more meaningful representations of the input, treating the configuration matrix somewhat similarly to an object segmentation task~\cite{Peng:2017:LargeKernelMatters}.
The output layer is also followed by a ReLU activation in order to force the prediction to be a positive number.
We use a dropout layer only after the convolutional network (before the flatten layer), but we introduced a combination of $\ell_2$ and $\ell_1$ regularisation to reduce the variance.
The dropout rate is 0.2 in the original dataset and 0.4 for the favourable dataset, while $\ell_1$ and $\ell_2$ regularisation are set to $10^{-5}$.
We train the model using the \emph{Adam} optimiser with a starting learning rate of $10^{-3}$ and a mini-batch size of $32$.

The architecture is more similar in style to the old \emph{LeNet} presented for the first time in 1998 by Y.\ LeCun during the ImageNet competition.
In our implementation, however, we do not include the pooling operations and swap the usual order of batch normalisation and activation function by first putting the ReLU activation.

\begin{figure}[htp]
    \centering
    \includegraphics[width=0.75\textwidth]{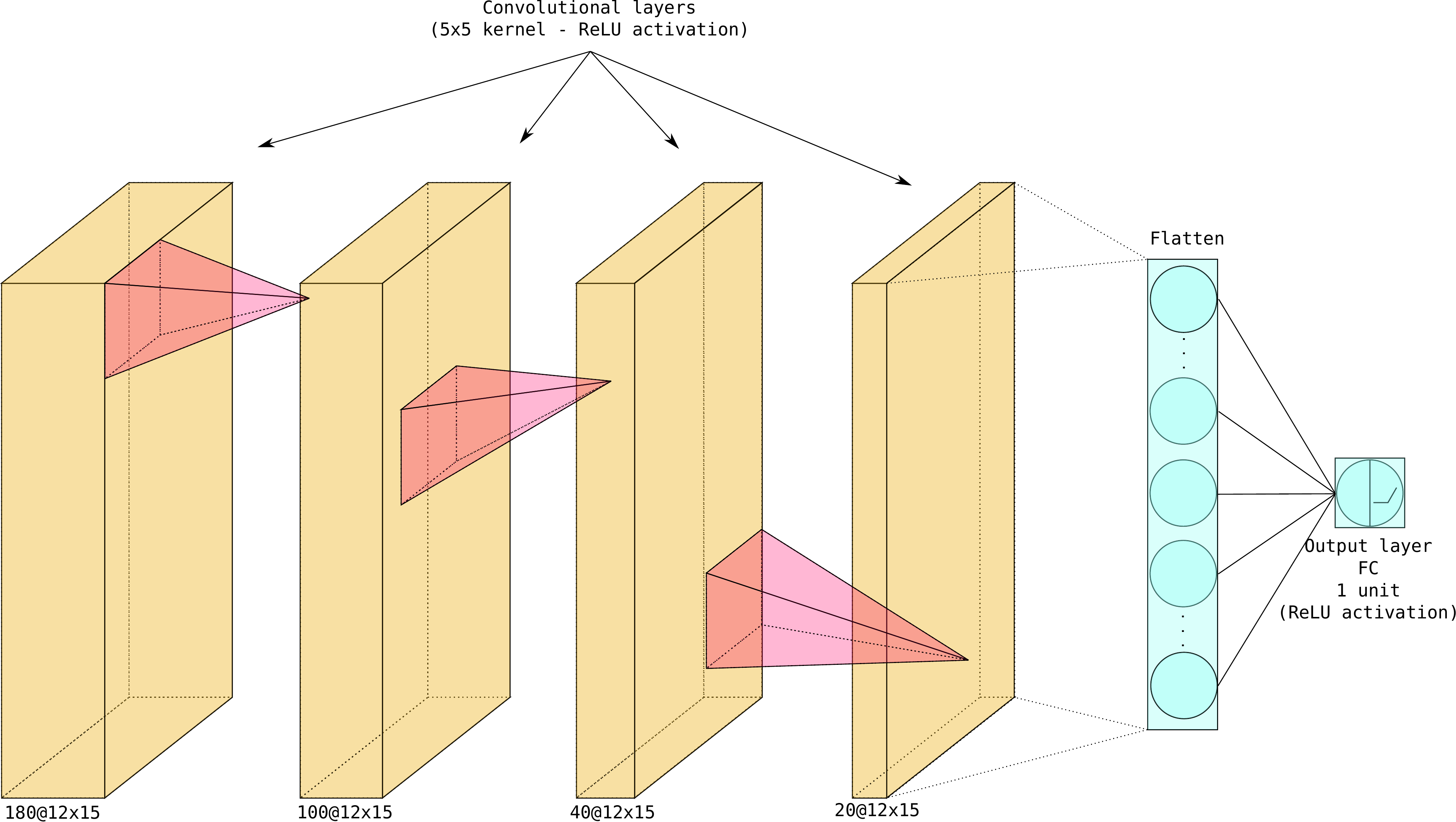}
    \caption{%
		Pure convolutional neural network for predicting $h^{1,1}$.
		It is made of $4$ modules composed by convolutional layer, ReLU activation, batch normalisation (in this order), followed by a dropout layer, a flatten layer and the output layer (in this order).
    }
    \label{fig:nn:lenet}
\end{figure}

In \Cref{fig:nn:lenet}, we show the model architecture in the case of the original dataset and of predicting $h^{1,1}$ alone.
The convolution layers have $180$, $100$, $40$ and $20$ units each.

\paragraph{Results}

With this setup, we were able to achieve an accuracy of 94\% on both the development and the test sets for the ``old'' database and 99\% for the favourable dataset in both validation and test sets (results are briefly summarised in \Cref{tab:res:ann}).
We thus improved the results of the densely connected network and proved that convolutional networks can be valuable assets when dealing with the extraction of a good representation of the input data: not only are CNNs very good at recognising patterns and rotationally invariant objects inside pictures or general matrices of data, but deep architectures are also capable of transforming the input using non linear transformations~\cite{Mallat:2016:UnderstandingDeepConvolutional} to create new patterns which can then be used for predictions.

\begin{figure}[htp]
    \centering
    \begin{subfigure}[c]{0.45\linewidth}
      \centering
      \includegraphics[width=\textwidth, trim={0 0 6in 0}, clip]{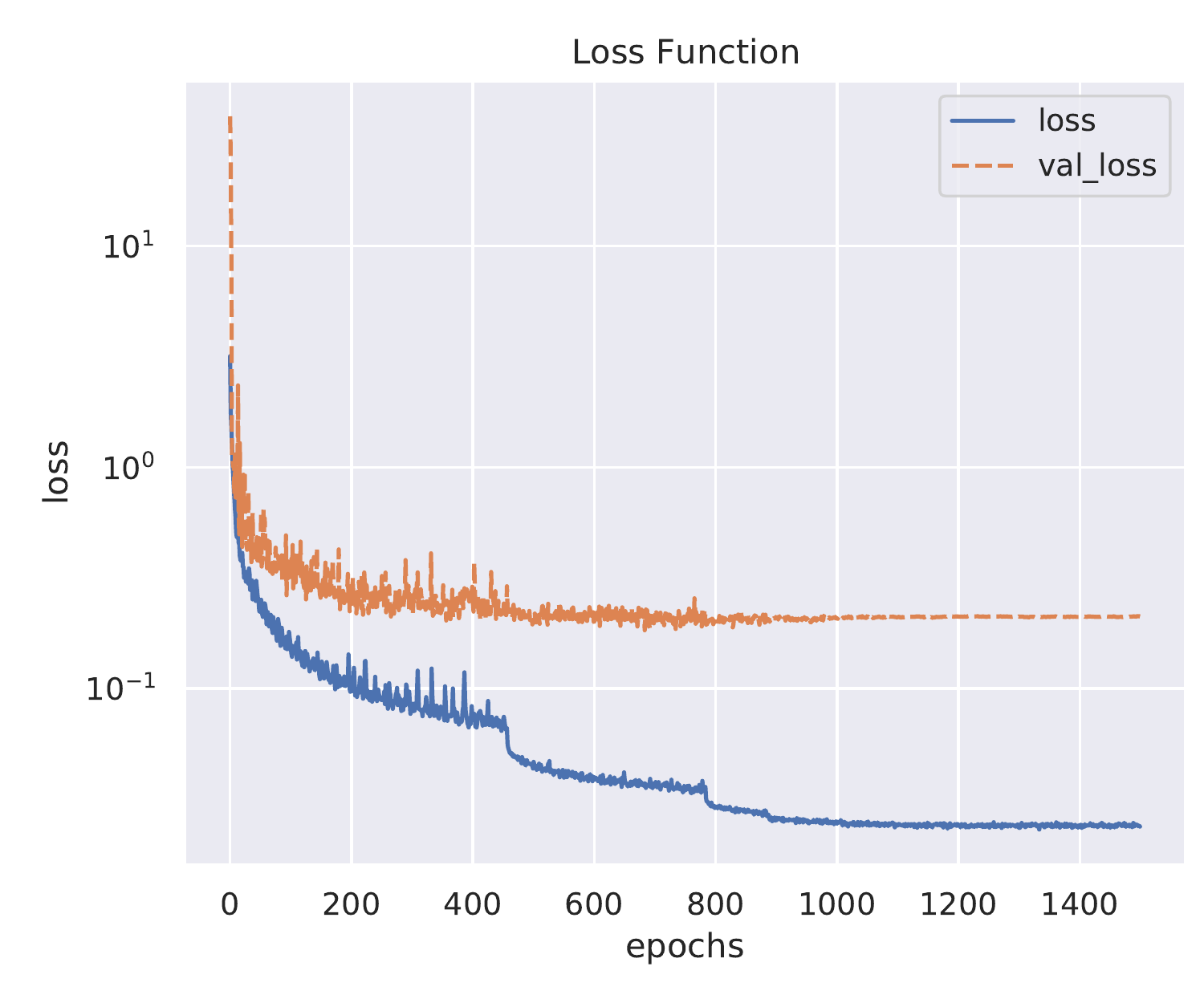}
      \caption{Loss function of $h^{1,1}$.}
    \end{subfigure}
    \quad
    \begin{subfigure}[c]{0.45\linewidth}
      \centering
      \includegraphics[width=\textwidth, trim={0 0 6in 0}, clip]{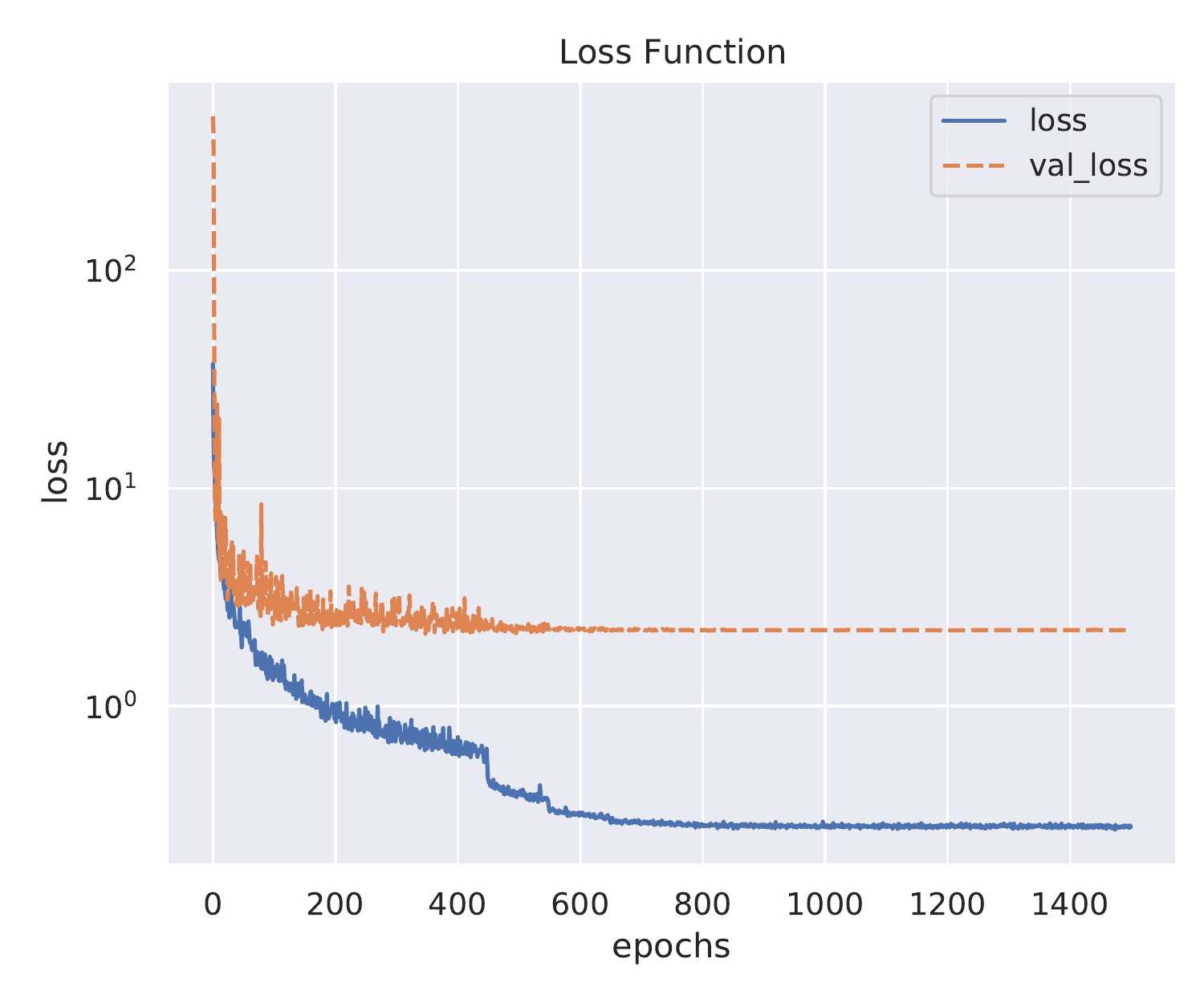}
      \caption{Loss function of $h^{2,1}$.}
    \end{subfigure}
    \caption{
		Loss function of the networks for the prediction of $h^{1,1}$ and $h^{2,1}$.
		We can see that the validation loss flattens out while the training loss keeps decreasing: we took care of the overfit by using the weights of the network when the validation loss reached its minimum.
		The use of mini-batch gradient descent also completely spoils the monotonicity of the loss functions which can therefore increase moving from one epoch to the other, while keeping the descending trend for most of its evolution.
	}
    \label{fig:cnn:class-ccnn}
\end{figure}

Even though the convolution operation is very time consuming, another advantage of CNN is the extremely reduced number of parameters with respect to FC networks.\footnotemark{}
\footnotetext{%
	It took around 4 hours of training (and no optimisation) for each Hodge number in each dataset.
}%
The architectures we used were in fact made of approximately $\num{5.8e5}$ parameters: way less than half the number of parameters used in the FC network.
Ultimately, this leads to a smaller number of training epochs necessary to achieve good predictions (see \Cref{fig:cnn:class-ccnn}).

Using this classic setup, we tried different architectures.
The network for the original dataset seems to work best in the presence of larger kernels, dropping by roughly $5\%$ in accuracy when a more ``classical'' $3 \times 3$ kernel is used.
We also tried to use to set the padding to \lstinline!valid!, reducing the input from a $12 \times 15$ matrix to a $1 \times 1$ feature map over the course of $5$ layers with $180$, $100$, $75$, $40$ and $20$ filters.
The advantage is the reduction of the number of parameters (namely $\sim \num{4.9e5}$) mainly due to the small FC network at the end, but accuracy dropped to $87\%$.
The favourable dataset seems instead to be more independent of the specific architecture, retaining accuracy also with smaller kernels.

The analysis for $h^{2,1}$ follows the same prescriptions.
For both the original and favourable dataset, we opted for 4 convolutional layers with 250, 150, 100 and 50 filters and no FC network for a total amount of $\num{2.1e6}$ parameters.

In this scenario we were able to achieve $36\%$ of accuracy in the development set and $40\%$ on the test set for $h^{2,1}$ in the ``old'' dataset and $31\%$ in both development and test sets in the favourable set (see \Cref{tab:res:ann}).

\begin{figure}[htp]
	\centering

	\includegraphics[width=0.6\textwidth]{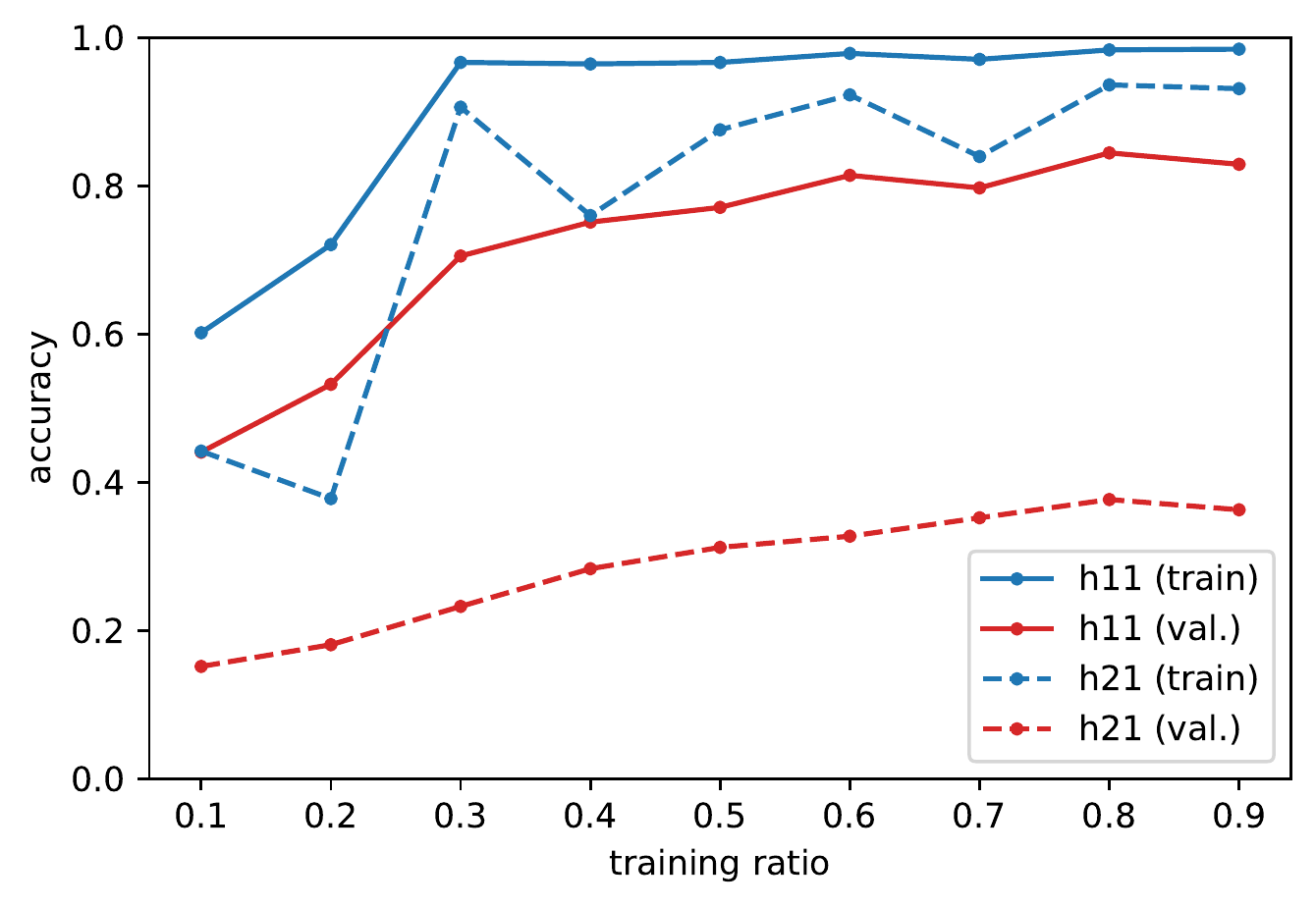}

	\caption{%
		Learning curves for the classic convolutional neural network (original dataset), using a single model for both Hodge numbers.
	}
	\label{fig:lc:class-ccnn}
\end{figure}

The learning curves for both Hodge numbers are given in \Cref{fig:lc:class-ccnn}.
This model uses the same architecture as the one for predicting $h^{1,1}$ only, which explains why it is less accurate as it needs to also adapt to compute $h^{2,1}$ -- a difficult task, as we have seen (see for example \Cref{fig:lc:inception}).

\subsubsection{Inception-like Neural Network}
\label{sec:ml:nn:inception}

In the effort to find a better architecture, we took inspiration from Google's winning CNN in the annual \href{https://image-net.org/challenges/LSVRC/}{\emph{ImageNet challenge}} in 2014~\cite{Szegedy:2014:GoingDeeperConvolutions, Szegedy:2015:RethinkingInceptionArchitecture, Szegedy:2016:Inceptionv4InceptionResNetImpact}.
The architecture presented uses \emph{inception} modules in which separate $3 \times 3$, $5 \times 5$ convolutions are performed side by side (together with \emph{max pooling} operations) before recombining the outputs.
The modules are then repeated until the output layer is reached.
This has two evident advantages: users can avoid taking a completely arbitrary decision on the type of convolution to use since the network will take care of it tuning the weights, and the number of parameters is extremely restricted as the network can learn complicated functions using fewer layers.
As a consequence the architecture of such models can be made very deep while keeping the number of parameters contained, thus being able to learn very difficult representations of the input and producing accurate predictions.
Moreover, while the training phase might become very long due to the complicated convolutional operations, the small number of parameters is such that predictions can be generated in a very small amount of time, making inception-like models extremely appropriate whenever quick predictions are necessary.
Another advantage of the architecture is the presence of different kernel sizes inside each module: the network automatically learns features at different scales and different positions, thus leveraging the advantages of a deep architecture with the ability to learn different representations at the same time and compare them.

\paragraph{Model}

\begin{figure}[htp]
    \centering
    \includegraphics[width=0.9\textwidth]{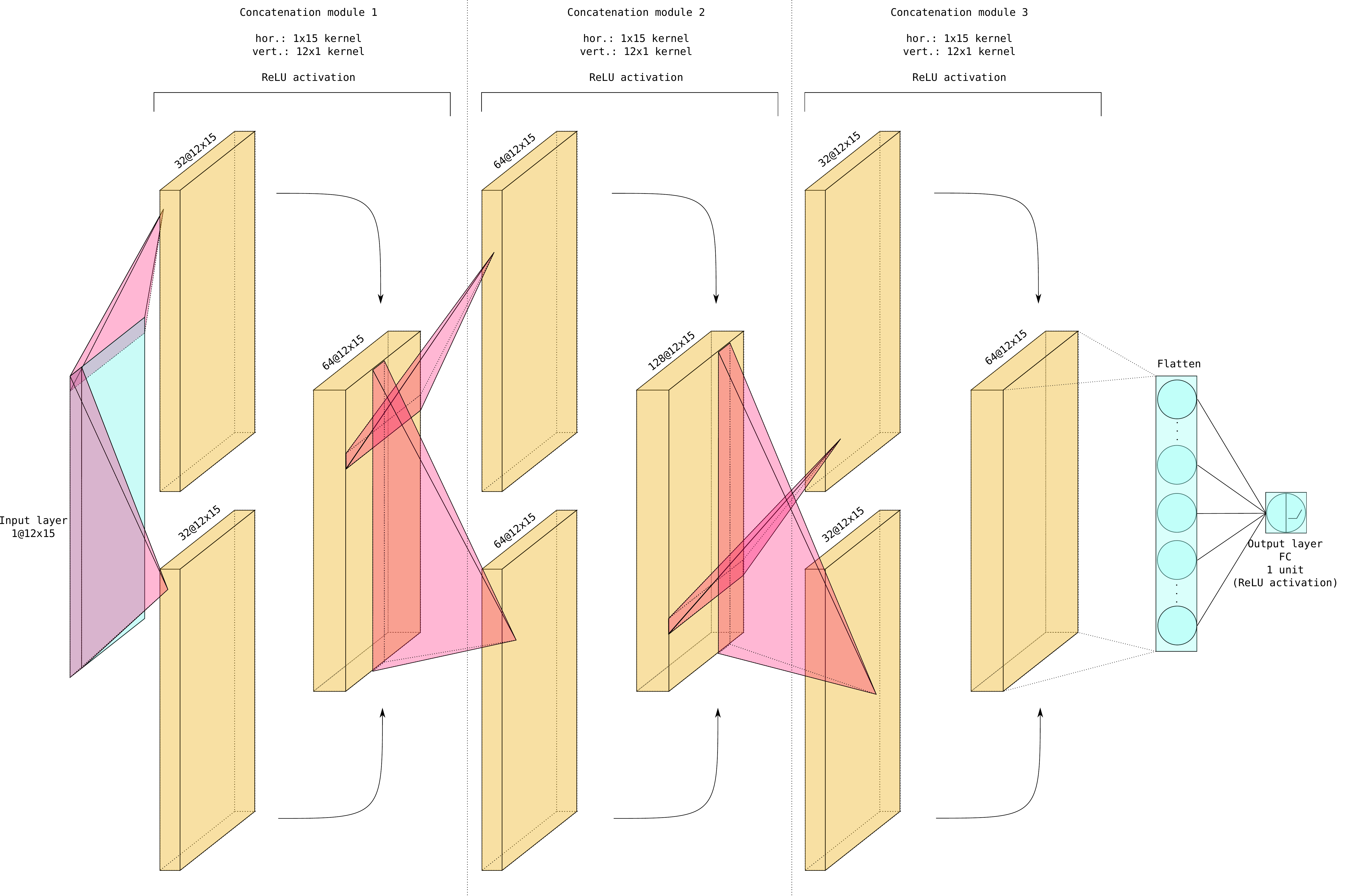}
    \caption{%
    	In each concatenation module (here shown for the ``old'' dataset) we operate with separate convolution operations over rows and columns, then concatenate the results. The overall architecture is composed of 3 ``inception'' modules made by two separate convolutions, a concatenation layer and a batch normalisation layer (strictly in this order), followed by a dropout layer, a flatten layer and the output layer with ReLU activation (in this order).}
    \label{fig:nn:inception}
\end{figure}

In \Cref{fig:nn:inception}, we show a schematic of our implementation.
Differently from the image classification task, we drop the pooling operation and implement two side-by-side convolution over rows ($12 \times 1$ kernel for the original dataset, $15 \times 1$ for the favourable) and one over columns ($1 \times 15$ and $1 \times 18$ respectively).\footnotemark{}
\footnotetext{%
	Pooling operations are used to shrink the size of the input.
	Similar to convolutions, they use a window of a given size to scan the input and select particular values inside.
	For instance, we could select the average value inside the small portion selected, performing an \emph{average pooling} operation, or the maximum value, a \emph{max pooling} operation.
	This usually improves image classification and object detection tasks as it can be used to sharpen edges and borders.
}%
We use \texttt{same} as padding option.
The output of the convolutions are then concatenated in the filter dimensions before repeating the ``inception'' module.
The results from the last module are directly connected to the output layer through a flatten layer.
In both datasets, we use batch normalisation layers (with momentum $0.99$) after each concatenation layer and a dropout layer (with rate $0.2$) before the FC network.\footnotemark{}
\footnotetext{%
	The position of the batch normalisation is extremely important as the parameters computed by such layer directly influence the following batch.
	We however opted to wait for the scan over rows and columns to finish before normalising the outcome to avoid biasing the resulting activation function.
}%

For both $h^{1,1}$ and $h^{2,1}$ (in both datasets), we used 3 modules made by 32, 64 and 32 filters for the first Hodge number, and 128, 128 and 64 filters for the second.
We also included $\ell_1$ and $\ell_2$ regularisation of magnitude $10^{-4}$ in all cases.
The number of parameters was thus restricted to $\num{2.3e5}$ parameters for $h^{1,1}$ in the original dataset and $\num{2.9e5}$ in the favourable set, and $\num{1.1e6}$ parameters for $h^{2,1}$ in the original dataset and $\num{1.4e6}$ in the favourable dataset.
In all cases, the number of parameters has decreased by a significant amount: in the case of $h^{1,1}$ they are roughly $\frac{1}{3}$ of the parameters used in the classical CNN and around $\frac{1}{6}$ of those used in the FC network.

For training we used the \emph{Adam} gradient descent with an initial learning rate of $10^{-3}$ and a batch size of $32$.
The callbacks helped to contain the training time (without optimisation) under 5 hours for each Hodge number in each dataset.

\paragraph{Results}

With these architectures, we were able to achieve more than \SI{99}{\percent} of accuracy for $h^{1,1}$ in the test set (same for the development set) and \SI{50}{\percent} of accuracy for $h^{2,1}$ (a slightly smaller value for the development set).
We report the results in \Cref{tab:res:ann}.

We therefore increased the accuracy for both Hodge numbers (especially $h^{2,1}$) compared to what can achieve a simple sequential network, while at the same time reducing significantly the number of parameters of the network.\footnotemark{}
This increases the robustness of the method and its generalisation properties.
\footnotetext{%
	In an attempt to improve the results for $h^{2,1}$ even further, we also considered to first predict $\ln( 1 + h^{2,1} )$ and then transform it back. However, the predictions dropped by almost $10\%$ in accuracy even using the ``inception'' network: the network seems to be able to approximate quite well the results (not better nor worse than simply $h^{2,1}$) but the subsequent exponentiation is taking apart predictions and true values.
	Choosing a correct rounding strategy then becomes almost impossible.
}

In \Cref{fig:nn:inception_errors}, we show the distribution of the residuals and their scatter plot, showing that the distribution of the errors does not present pathological behaviour and the variance of the residuals is well distributed over the predictions.

In fact, this neural network is much more powerful than the previous networks we considered, as can be seen by studying the learning curves (\Cref{fig:lc:inception}).
When predicting only $h^{1,1}$, it surpasses $97\%$ accuracy using only $30\%$ of the data for training.
While it seems that the predictions suffer when using a single network for both Hodge numbers, this remains much better than any other algorithm.
It may seem counter-intuitive that convolutions work well on this data since they are not translation or rotation invariant, but only permutation invariant.
However, convolution alone is not sufficient to ensure invariances under these transformations but it must be supplemented with pooling operations~\cite{Goodfellow:2016:DeepLearning}, which we do not use.
Moreover, convolution layers do more than just taking translation properties into account: they allow to make highly complicated combinations of the inputs and to share weights among components, which allow to find subtler patterns than standard fully connected layers.
This network is more studied in more details in~\cite{Erbin:2020:InceptionCICY}.

\begin{figure}[htp]
  \centering
  \begin{subfigure}[c]{0.45\textwidth}
    \centering
    \includegraphics[width=\textwidth, trim={0 0 6in 0}, clip]{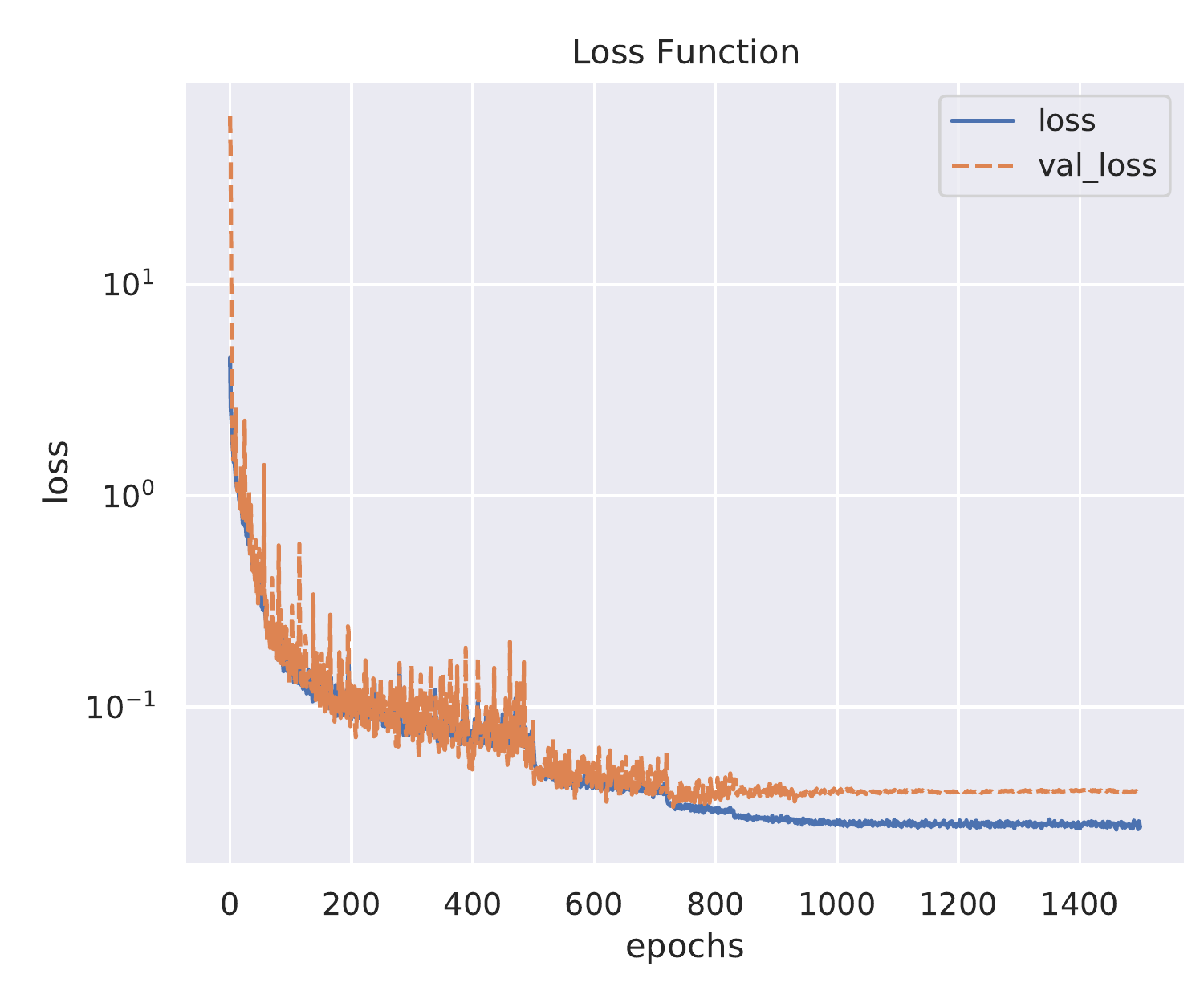}
    \caption{Loss of $h^{1,1}$.}
  \end{subfigure}
  \quad
  \begin{subfigure}[c]{0.45\textwidth}
    \centering
    \includegraphics[width=\textwidth, trim={0 0 6in 0}, clip]{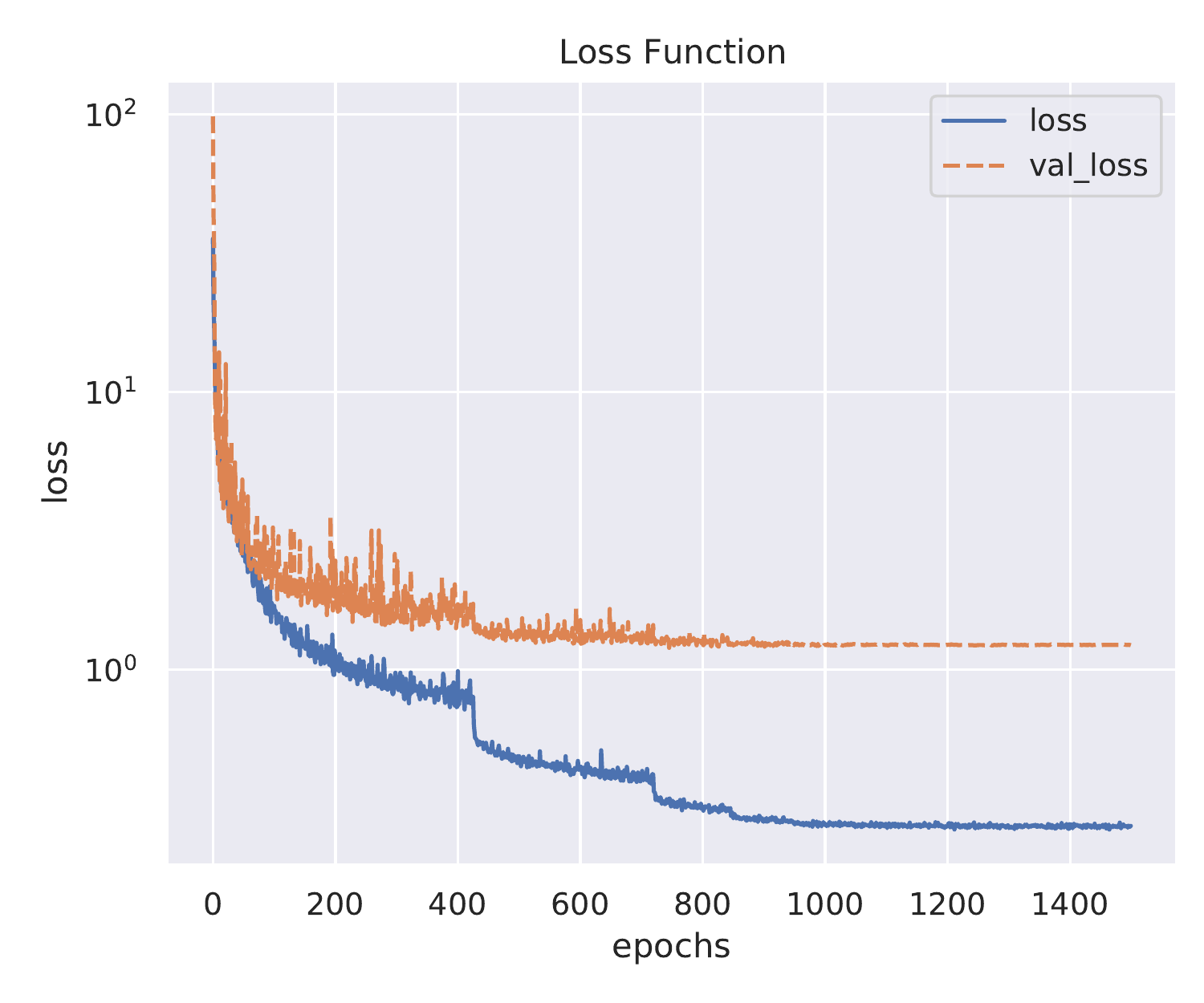}
    \caption{Loss of $h^{2,1}$.}
  \end{subfigure}
  \caption{The loss functions of ``inception'' network for $h^{1,1}$ and $h^{2,1}$ in the original dataset show that the number of epochs required for training is definitely larger than for simpler architectures, despite the reduced number of parameters.}
  \label{fig:cnn:inception-loss}
\end{figure}

\begin{figure}[htp]
  \centering
  \begin{subfigure}[c]{\textwidth}
    \centering
    \includegraphics[width=0.8\textwidth]{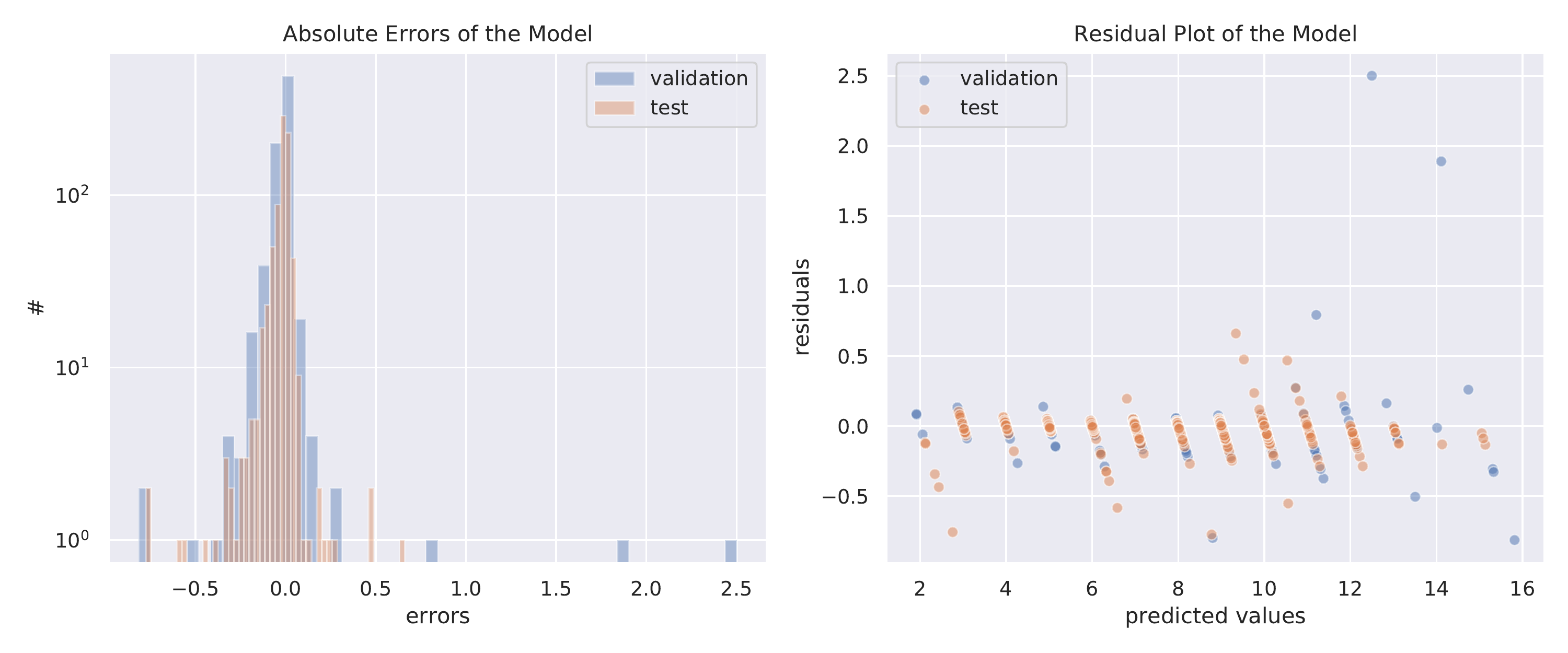}
    \caption{Residuals of $h^{1,1}$.}
  \end{subfigure}
  \quad
  \begin{subfigure}[c]{\textwidth}
    \centering
    \includegraphics[width=0.8\textwidth]{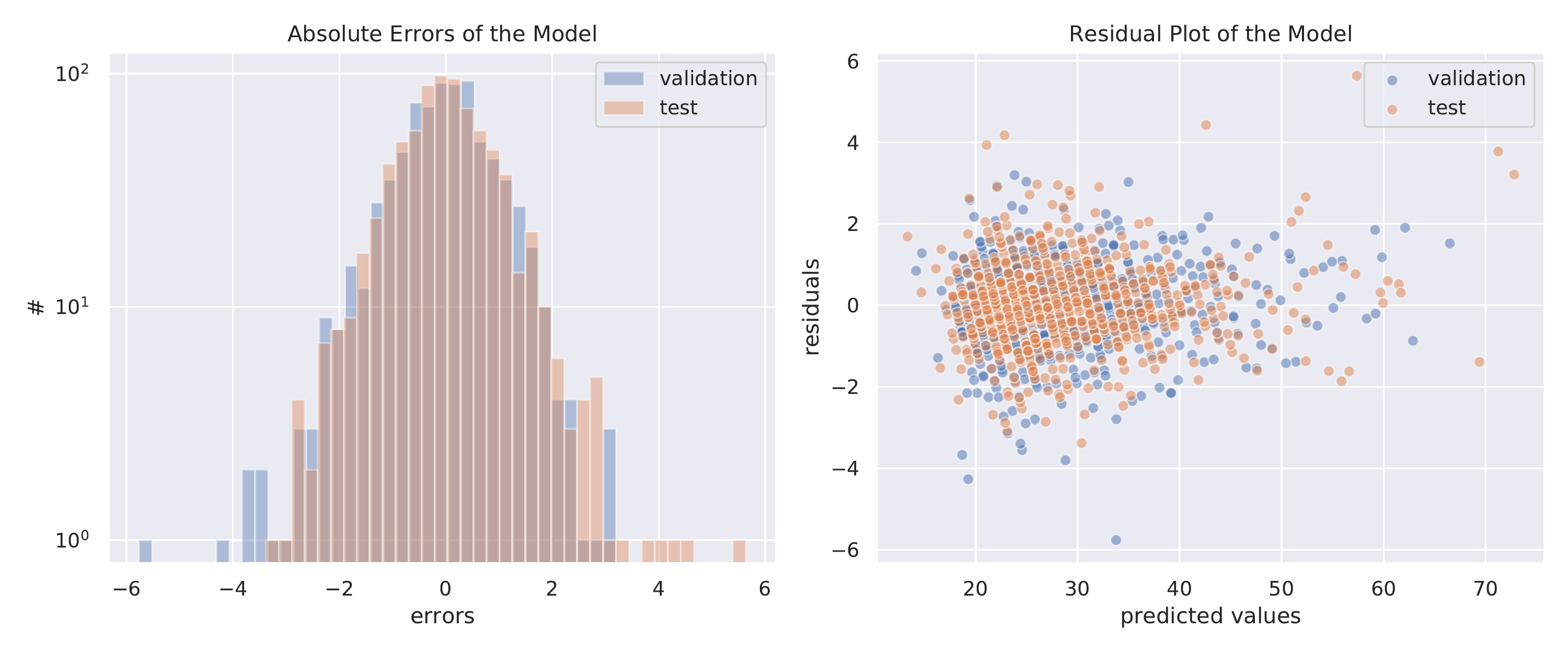}
    \caption{Residuals of $h^{2,1}$.}
  \end{subfigure}
  \caption{Histograms of the residual errors and residual plots of the Inception network.}
  \label{fig:nn:inception_errors}
\end{figure}

\begin{figure}[htp]
	\centering

	\begin{subfigure}[c]{0.45\linewidth}
		\centering
		\includegraphics[width=\textwidth]{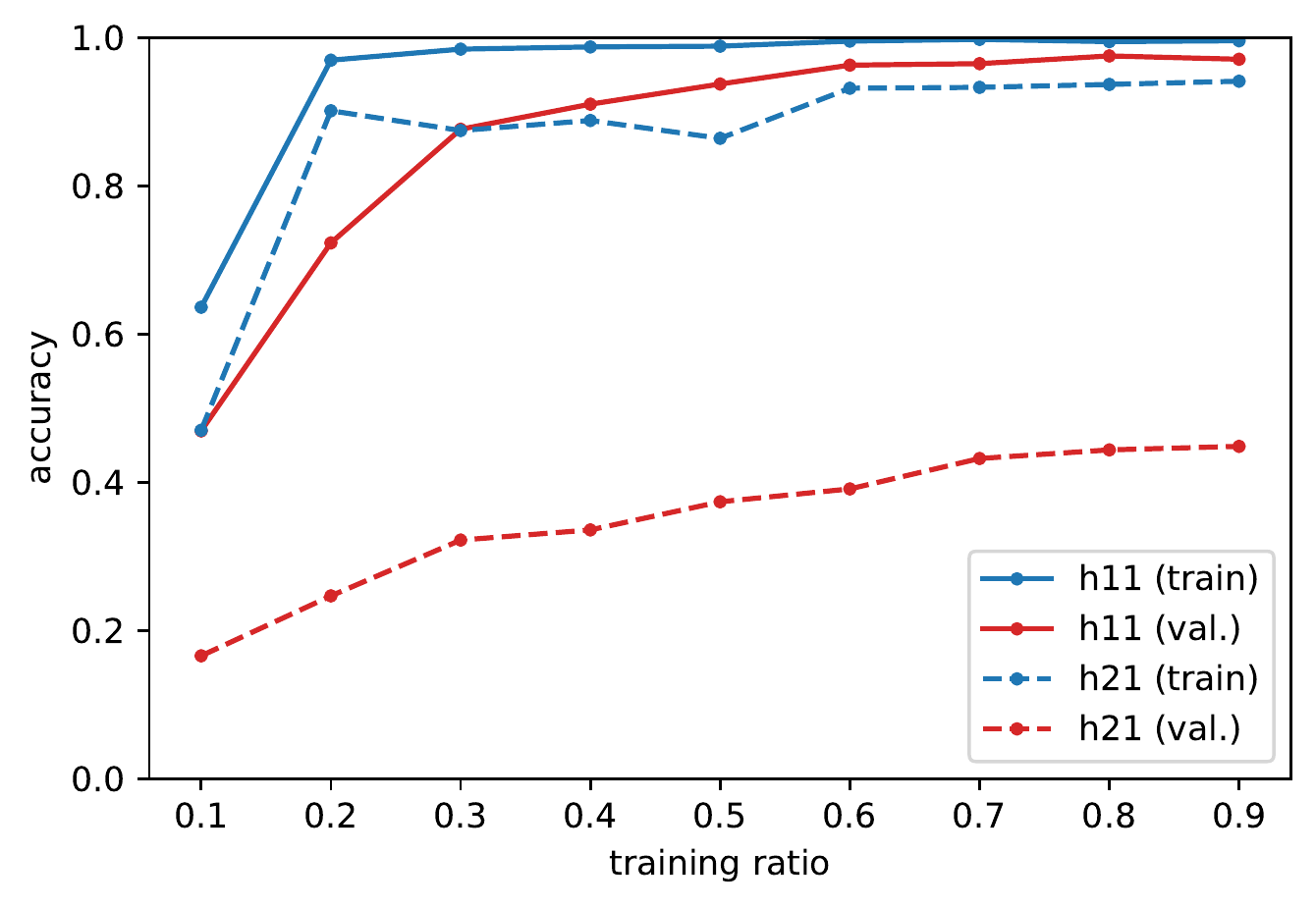}
		\caption{predicting both $h^{1,1}$ and $h^{2,1}$}
	\end{subfigure}
	\qquad
	\begin{subfigure}[c]{0.45\linewidth}
		\centering
		\includegraphics[width=\textwidth]{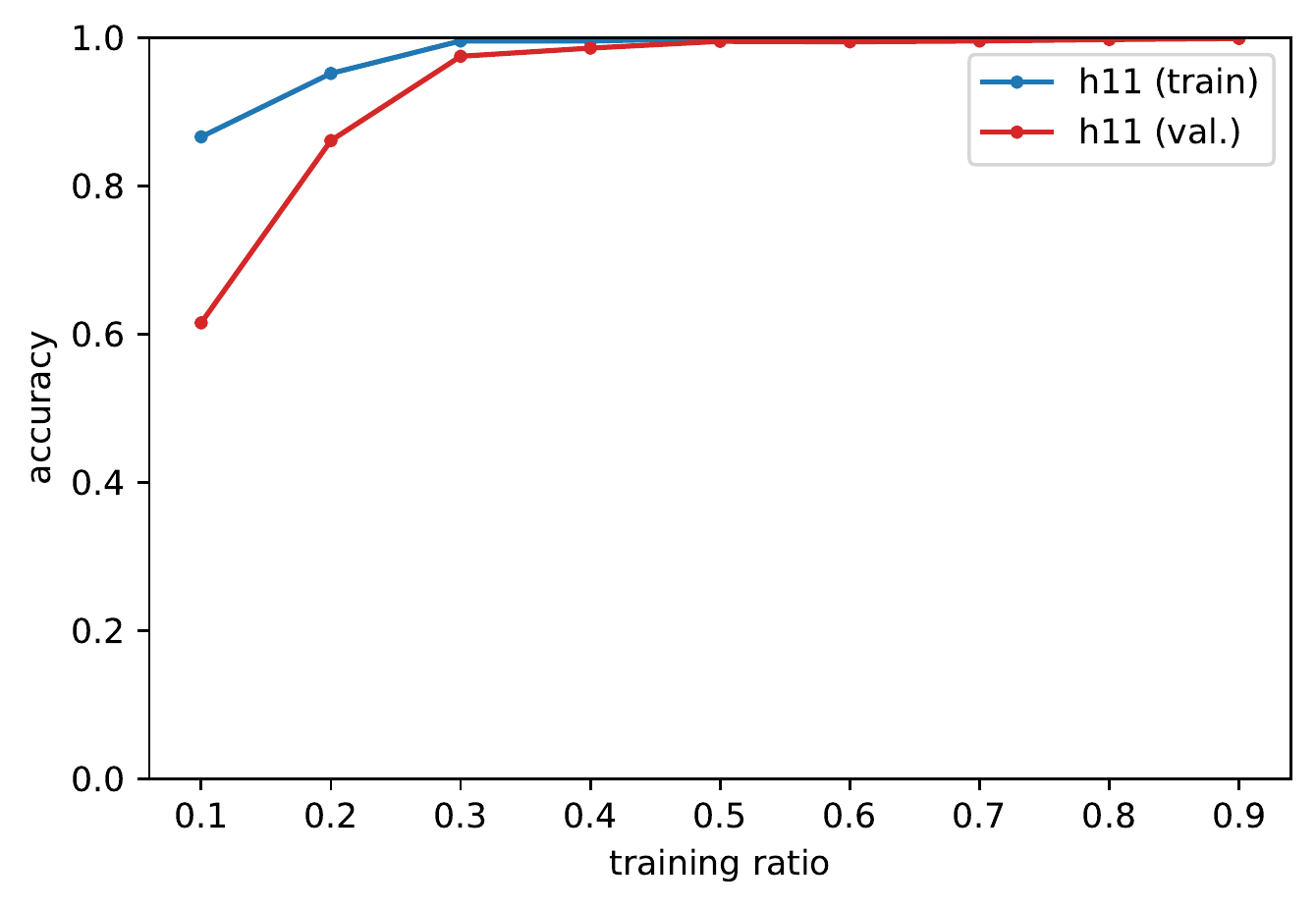}
		\caption{predicting $h^{1,1}$ only}
	\end{subfigure}

	\caption{Learning curves for the Inception neural network (original dataset).}
	\label{fig:lc:inception}
\end{figure}

\begin{table}[htb]
\centering
	\begin{tabular}{@{}ccccccc@{}}
		\toprule
		& \multicolumn{2}{c}{\textbf{DenseNet}}
		& \multicolumn{2}{c}{\textbf{classic ConvNet}}
		& \multicolumn{2}{c}{\textbf{inception ConvNet}}
		\\
		& \emph{old} & \emph{fav.}
		& \emph{old} & \emph{fav.}
		& \emph{old} & \emph{fav.}
		\\
		\midrule
		$h^{1,1}$
		& 77\%  & 97\%
		& 94\%  & 99\%
		& 99\%  & 99\%
		\\
		$h^{2,1}$
		& -     & -
		& 36\%  & 31\%
		& 50\%  & 48\%
		\\
		\bottomrule
\end{tabular}
\caption{Accuracy using \emph{rint} rounding on the predictions of the ANNs on $h^{1,1}$ and $h^{2,1}$ on the test set.}
\label{tab:res:ann}
\end{table}

\subsubsection{Boosting the Inception-like Model}

To improve further the accuracy of $h^{2,1}$, we have tried to modify the network by adding engineered features as auxiliary inputs.
This can be done by adding inputs to the inception neural network and merging the different branches at different stages.
There are two possibilities to train such a network: 1) train all the network directly, or 2) train the inception network alone, then freeze its weights and connect it to the additional inputs, training only the new layer.
We found that the architectures we tried did not improve the accuracy, but we briefly describe our attempts for completeness.

We focused in particular on the number of projective spaces, the vector of dimensions of the projective spaces and the vector of dimensions of the principal cohomology group) and predicting $h^{1,1}$ and $h^{2,1}$ at the same time.
The core of the neural network is the Inception network described in \Cref{sec:ml:nn:inception}.
Then, the engineered features are processed using fully connected layers and merged to the predictions from the Inception branch using a concatenation layer.
Obviously, output layers for $h^{1,1}$ and $h^{2,1}$ can be located on different branches, which allow for different processing of the features.

As mentioned earlier, a possible approach is to first train the Inception branch alone, before freezing its weights and connecting it to the rest of the network.
This can prevent spoiling the already good predictions and speed up the new learning process.
This is a common technique called \emph{transfer learning}: we can use a model previously trained on a slightly different task and use its weights as part of the new architecture.

Our trials involved shallow fully connected layers ($1$--$3$ layers with $10$ to $150$ units) between the engineered features and after the concatenation layer.
Since the EDA analysis (\Cref{sec:data:eda}) shows a correlation between both Hodge numbers, we tried architectures where the result for $h^{1,1}$ is used to predict $h^{2,1}$.

For the training phase, we also tried an alternative to the canonical choice of optimising the sum of the losses.
We first train the network and stop the process when the validation loss for $h^{1,1}$ does not longer improve, load back the best weights and save the results, keep training and stop when the loss for $h^{2,1}$ reaches a plateau.

With this setup we were able to slightly improve the predictions of $h^{1,1}$ in the original dataset, reaching almost \SI{100}{\percent} of accuracy in the predictions, while the favourable dataset stayed at around \SI{99}{\percent} of accuracy.
The only few missed predictions (4 manifolds out of 786 in the test set) are in very peculiar regions of the distribution of the Hodge number.
For $h^{2,1}$ no improvement has been noticed.

\subsection{Ensemble Learning: Stacking}

We conclude the ML analysis by describing a method very popular in ML competitions~\cite{Coursera:HowWinData}: ensembling.
This consists in taking several ML algorithms and combining together the predictions of each individual model to obtain a more precise predictions.
Using this technique it is possible to decrease the variance and improve generalization by compensating weaknesses of algorithms with strengths of others.
Indeed, the idea is to put together algorithms which perform best in different zones of the label distribution in order to combine them to build an algorithm better than any individual component.

\begin{figure}[htp]
	\centering
	\includegraphics[width=0.65\textwidth]{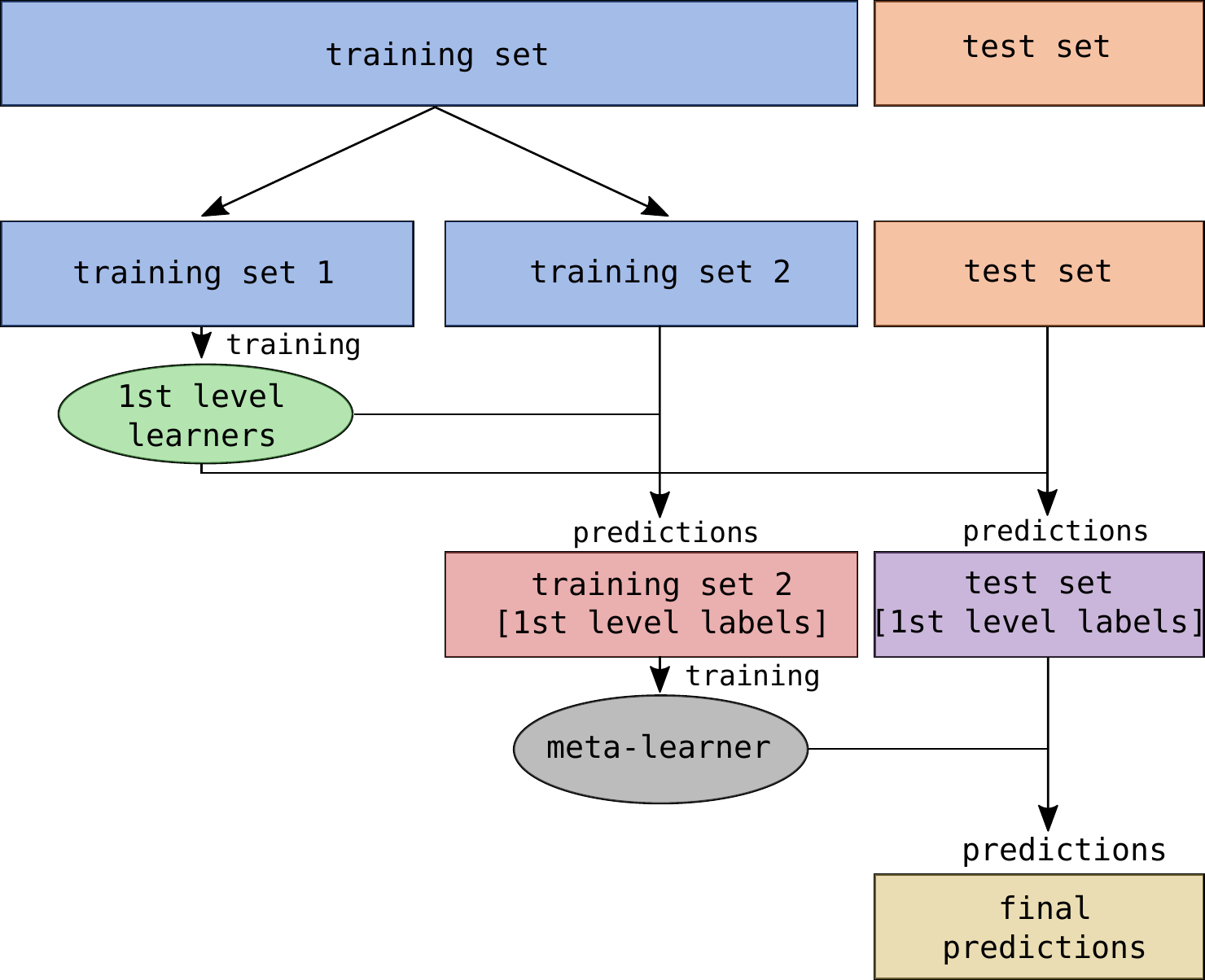}
	\caption{Stacking ensemble learning with two level learning.
	The original training set is split into two training folds and the first level learners are trained on the first.
	The trained models are then used to generate a new training set (here the ``1st level labels'') using the second split as input features.
	The same also applies to the test set.
	Finally a ``meta-learner'' uses the newly generated training set to produce the final predictions on the test set.}
	\label{fig:stack:def}
\end{figure}

The simplest such algorithm is \emph{stacking} whose principle is summarised in \Cref{fig:stack:def}.
First, the original training set is split in two parts (not necessarily even).
Second, a certain number of \emph{first-level learners} is trained over the first split and used to generate predictions over the second split.
Third, a ``meta learner'' is trained of the second split to combine the predictions from the first-level learners.
Predictions for the test set are obtained by applying both level of models one after the other.

We have selected the following models for the first level: linear regression, SVR with the Gaussian kernel, the random forest and the ``inception'' neural network.
The meta-learner is a simple linear regression with $\ell_1$ regularisation (Lasso).
The motivations for the first-level algorithms is that stacking works best with a group of algorithms which work in the most diverse way among them.

Also in this case, we use a cross-validation strategy with 5 splits for each level of the training: from \SI{90}{\percent} of total training set, we split into two halves containing each \SI{45}{\percent} of the total samples and then use 5 splits to grade the algorithm, thus using \SI{9}{\percent} of each split for cross correlation at each iteration) and the Bayes optimisation for all algorithms but the ANN (50 iterations for elastic net, SVR and lasso and 25 for the random forests).
The ANN was trained using a holdout validation set containing the same number of samples as each cross-validation fold, namely \SI{9}{\percent} of the total set.
The accuracy is then computed as usual using \texttt{numpy.rint} for SVR, neural networks, the meta learner and $h^{1,1}$ in the original dataset in general, and \texttt{numpy.floor} in the other cases.

In \Cref{tab:res:stack}, we show the accuracy of the ensemble learning.
We notice that accuracy improves slightly only for $h^{2,1}$ (original dataset) compared to the first-level learners.
However, this is much lower than what has been achieved in \Cref{sec:ml:nn:inception}.
The reason is that the learning suffers from the reduced size of the training set.
Another reason is that the different algorithms may perform similarly well in the same regions.

\begin{table}[htb]
\centering
\begin{tabular}{@{}cccccc@{}}
    \toprule
    &
    & \multicolumn{2}{c}{$h^{1,1}$}
    & \multicolumn{2}{c}{$h^{2,1}$}
    \\
    &
    & \emph{old}   & \emph{fav.}
    & \emph{old}   & \emph{fav.}
    \\
    \midrule
    \multirow{4}{*}{\emph{1st level}}
    & EN
        & 65\%  & 100\%
        & 19\%  & 19\%
    \\
    & SVR
        & 70\%  & 100\%
        & 30\%  & 34\%
    \\
    & RF
        & 61\%  & 98\%
        & 18\%  & 24\%
    \\
    & ANN
        & 98\%  & 98\%
        & 33\%  & 30\%
    \\
    \midrule
    \multirow{1}{*}{\emph{2nd level}}
    & Lasso
        & 98\%  & 98\%
        & 36\%  & 33\%
    \\
    \bottomrule
\end{tabular}
\caption{Accuracy of the first and second level predictions of the stacking ensemble for elastic net regression (EN), support vector with \texttt{rbf} kernel (SVR), random forest (RF) and the artificial neural network (ANN) as first level learners and lasso regression as meta learner.}
\label{tab:res:stack}
\end{table}

\section{Discussion}
\label{sec:conclusion}

In this paper, we have proved how a proper data analysis can lead to improvements in predictions of Hodge numbers $h^{1,1}$ and $h^{2,1}$ for CICY $3$-folds.
Moreover, considering more complex neural networks -- in particular, architectures inspired by the Inception model~\cite{Szegedy:2014:GoingDeeperConvolutions, Szegedy:2015:RethinkingInceptionArchitecture, Szegedy:2016:Inceptionv4InceptionResNetImpact} -- allowed us to reach close to $100\%$ accuracy for $h^{1,1}$ with much less data and less parameters than in previous works.

While our analysis improved the accuracy for $h^{2,1}$ over what can be expected from a simple sequential neural network, we barely reached $50\%$.
Hence, it would be interesting to push further our study to improve the accuracy.
Possible solutions would be to use a deeper Inception network, find a better architecture including engineered features, and refine the ensembling (for example using StackNet~\cite{Package:StackNet}).

Another interesting question to probe is related to representation learning, i.e.\ finding a better description of the Calabi--Yau.
Indeed, one of the main difficulty in making predictions is the redundancy of the possible descriptions of a single manifold.
For example, one could try to set up a map from any matrix to its favourable representation (if it exists).
Or, on the contrary, one could generate more matrices for the same manifold in order to increase the size of the training set.
Another possibility is to use the graph representation of the configuration matrix to which is automatically invariant under permutations~\cite{Hubsch:1992:CalabiYauManifoldsBestiary} (another graph representation has been decisive in~\cite{Krippendorf:2020:DetectingSymmetriesNeural} to get a good accuracy).
Techniques such as (variational) autoencoder~\cite{Kingma:2014:AutoEncodingVariationalBayes, Rezende:2014:StochasticBackpropagationApproximate, Salimans:2015:MarkovChainMonte}, cycle GAN~\cite{Zhu:2017:UnpairedImagetoImageTranslation}, invertible neural networks~\cite{Ardizzone:2019:AnalyzingInverseProblems}, graph neural networks~\cite{Gori:2005:NewModelLearning, Scarselli:2004:GraphicalBasedLearningEnvironments} or more generally techniques from geometric deep learning~\cite{Bronstein:2017:GeometricDeepLearning} could be helpful.

Finally, our techniques apply directly to CICY $4$-folds~\cite{Gray:2013:AllCompleteIntersection, Gray:2014:TopologicalInvariantsFibration}.
However, there are much more manifolds in this case, such that one can expect to reach a better accuracy for the different Hodge numbers (the different learning curves for the $3$-folds indicate that the model training would benefit from more data).
We hope to report soon on these issues.
Another interesting class of manifolds to explore with our techniques are generalized CICY $3$-folds~\cite{Anderson:2016:NewConstructionCalabiYau}.

We leave these questions to future explorations.

\section*{Acknowledgements}

We are grateful to Sven Krippendorf and Fabian Rühle for useful discussions on the applications of machine learning to string theory and for comments on the draft.
The work of H.E.\ has been conducted under a Carl Friedrich von Siemens Research Fellowship of the Alexander von Humboldt Foundation for postdoctoral researchers during part of this project.
H.E.\ and R.F.\ are partially supported by the \textsc{Miur Prin} Contract \textsc{2015Mp2cx4} “Non-perturbative Aspects of Gauge Theories and Strings”.

\appendix

\section{Machine Learning Algorithms}
\label{app:ml-algo}

\subsection{Linear regression}
\label{sec:app:linreg}

Considering a set of $F$ features $\{ x_n \}$ where $n = 1, \ldots, F$, a linear model learns a function
\begin{equation}
	f(x_n)
		= \sum_{n=1}^F w_n x_n + b,
\end{equation}
where $w$ and $b$ are the \emph{weights} and \emph{intercept} of the fit.

One of the key assumptions behind a linear fit is the independence of the residual error between the predicted point and the value of the model, which can therefore be assumed to be sampled from a normal distribution peaked at the average value~\cite{Lista:2017:StatisticalMethods,Coursera:DataScience}.
The parameters of the fit are then chosen to maximise their \emph{likelihood} function, or conversely to minimise its logarithm with a reversed sign (the $\chi^2$ function).
A related task is to minimise the mean squared error, without assuming a statistical distribution of the residual error: ML for regression usually implements this as loss function of the estimators.
In this sense, loss functions for regression are more general than a likelihood approach, but are nonetheless related.
For plain linear regression, the associated loss is
\begin{equation}
\mathcal{L}(w, b) =
\frac{1}{2N} \sum_{i=1}^N \sum_{n=1}^F
\left( y^{(i)} - (w_n x_n^{(i)} + b) \right)^2,
\end{equation}
where $N$ is the number of samples and $x_n^{(i)}$ the $n$th feature of the $i$th sample.
The values of the parameters will therefore be:
\begin{equation}
(w, b) = \underset{w,\,b}{\mathrm{argmin}}~ \mathcal{L}(w, b).
\end{equation}
This usually requires looping over all samples and all features, thus the \emph{least squares} method has a time complexity of $\mathrm{O}( F \times N )$: while the increase of the number of samples might be an issue, the number of engineered features and matrix components usually does not change and does not represent a huge effort in terms of rescaling the algorithm.

There are however different versions of possible regularisation which we might add to constrain the parameters of the fit and avoid adapting too well to the training set.
In particular we may be interested in adding a $\ell_1$ regularisation:
\begin{equation}
\mathcal{L}_1(w) = \sqrt{\sum\limits_{n=1}^F w_n^2},
\end{equation}
or the $\ell_2$ version:
\begin{equation}
\mathcal{L}_2(w) = \sum\limits_{n=1}^F w_n^2.
\end{equation}
Notice that in general we do not regularise the intercept.
These terms can be added to the plain loss function to try and avoid large parameters to influence the predictions and to keep better generalisation properties:
\begin{itemize}
	\item add both $\ell_1$ and $\ell_2$ regularisation (this is called \emph{elastic net}):
	\begin{equation}
	\mathcal{L}_{en}(w, b;~\alpha_{en}, L) = \mathcal{L}(w,b) + \alpha_{en} \cdot L \cdot \mathcal{L}_1(w) + \frac{\alpha_{en}}{2} \cdot (1 - L) \cdot \mathcal{L}_2(w),
	\end{equation}
	\item keep only $\ell_1$ regularisation (i.e.\ the \emph{lasso} regression):
	\begin{equation}
	\mathcal{L}_{lss}(w, b;~\alpha_{lss}) = \mathcal{L}(w,b) + \alpha_{lss} \cdot \mathcal{L}_1(w),
	\end{equation}
	\item keep only $\ell_2$ regularisation (\emph{ridge} regression):
	\begin{equation}
	\mathcal{L}_{rdg}(w, b;~\alpha_{rdg}) = \mathcal{L}(w,b) + \alpha_{rdg} \cdot \mathcal{L}_2(w).
	\label{eq:ridge:loss}
	\end{equation}
\end{itemize}
The role of the hyperparameter $L$ is to balance the contribution of the additional terms. For larger values of the hyperparameter $\alpha$, $w$ (and $b$) assume smaller values and adapt less to the particular training set.

\subsection{Support Vector Machines for Regression}
\label{sec:app:svr}

This family of supervised ML algorithms were created with classification tasks in mind~\cite{Cortes:1995:SVMClassification} but have proven to be effective also for regression problems~\cite{Drucker:1997:SVMRegression}.
Differently from the linear regression, instead of minimising the squared distance of each sample, the algorithm assigns a penalty to predictions of samples $x^{(i)} \in \R^F$ (for $i = 1, 2, \dots, N$) which are further away than a certain hyperparameter $\varepsilon$ from their true value $y$, allowing however a \textit{soft margin} of tolerance represented by the penalties $\zeta$ above and $\xi$ below.
This is achieved by minimising $w,\, b,\, \zeta$ and $\xi$ in the function\footnotemark
\footnotetext{In a classification task the training objective would be the minimisation of the opposite of the log-likelihood function of predicting a positive class, that is $y^{(i)} ( w_n \phi_n( x^{(i)} ) + b )$, which should equal the unity for good predictions (we can consider $\varepsilon = 1$), instead of the regression objective $y^{(i)} - w_n \phi_n( x^{(i)} ) - b$.
	The differences between SVR for classification purposes and regression follow as shown.}
\begin{equation}
\begin{split}
\mathcal{L}(w, b, \zeta, \xi)
& =
\frac{1}{2} \sum\limits_{n = 1}^{F'} w_n^2
+
C \sum\limits_{i = 1}^N \left( \zeta^{(i)} + \xi^{(i)} \right)
\\
& +
\sum\limits_{i = 1}^N \sum\limits_{n = 1}^{F'} \alpha^{(i)}
\left( y^{(i)} - w_n \phi_n(x^{(i)}) - b - \varepsilon - \zeta^{(i)} \right)
\\
& +
\sum\limits_{i = 1}^N \sum\limits_{n = 1}^{F'} \beta^{(i)}
\left( w_n \phi_n(x^{(i)}) + b - y^{(i)} - \varepsilon - \xi^{(i)} \right)
\\
& -
\sum\limits_{i = 1}^N \left( \rho^{(i)} \zeta^{(i)} + \sigma^{(i)} \xi^{(i)} \right)
\end{split}
\label{eq:svr:loss}
\end{equation}
where $\alpha^{(i)},\, \beta^{(i)},\, \rho^{(i)},\, \sigma^{(i)} \ge 0$ such that the previous expression encodes the constraints
\begin{equation}
\begin{cases}
y^{(i)} - \sum\limits_{n = 1}^{F'} w_n \phi_n( x^{(i)} ) - b & \le \varepsilon + \zeta^{(i)},
\qquad
\varepsilon \ge 0,
\quad
\zeta^{(i)} \ge 0,
\quad
i = 1, 2, \dots, N
\\
\sum\limits_{n = 1}^{F'} w_n \phi_n( x^{(i)} ) + b  - y^{(i)} & \le \varepsilon + \xi^{(i)},
\qquad
\varepsilon \ge 0,
\quad
\xi^{(i)} \ge 0,
\quad
i = 1, 2, \dots, N
\end{cases}
\label{eq:svr:constraints}
\end{equation}
and where $\phi( x^{(i)} ) \in \R^{F'}$ is a function mapping the feature vector $x^{(i)} \in \R^F$ in a higher dimensional space ($F' > F$), whose interpretation will become clear in an instant.
The minimisation problem leads to
\begin{equation}
\begin{cases}
w_n - \sum\limits_{i = 1}^N \left( \alpha^{(i)} - \beta^{(i)} \right) \phi_n( x^{(i)} ) = 0
\\
\sum\limits_{i = 1}^N \left( \alpha^{(i)} - \beta^{(i)} \right) = 0
\\
\sum\limits_{i = 1}^N \left( \alpha^{(i)} + \rho^{(i)} \right)
=
\sum\limits_{i = 1}^N \left( \beta^{(i)} + \sigma^{(i)} \right)
=
C
\end{cases}
\end{equation}
such that $0 \le \alpha^{(i)},\, \beta^{(i)} \le C,~\forall\, i = 1, 2, \dots, N$. This can be reformulated as a \textit{dual} problem in finding the extrema of $\alpha^{(i)}$ and $\beta^{(i)}$ in
\begin{equation}
W(\alpha, \beta)
=
\frac{1}{2} \sum\limits_{i, j = 1}^N \theta^{(i)} \theta^{(j)} \mathrm{K}( x^{(i)}, x^{(j)} )
-
\varepsilon \sum\limits_{i = 1}^N \left( \alpha^{(i)} + \beta^{(i)} \right)
+
\sum\limits_{i = 1}^N y^{(i)} \theta^{(i)},
\label{eq:svr:loss-v2}
\end{equation}
where $\theta = \alpha - \beta$ are called \textit{dual coefficients} (accessible through the attribute \texttt{dual\_coef\_} of \texttt{svm.SVR} in \texttt{scikit-learn}) and $\mathrm{K}( x^{(i)}, x^{(j)} ) = \sum\limits_{n = 1}^{F'} \phi_n( x^{(i)} ) \phi_n( x^{(j)} )$ is the \textit{kernel} function.
Notice that the Lagrange multipliers $\alpha^{(i)}$ and $\beta^{(i)}$ are non vanishing only for particular sets of vectors $l^{(i)}$ which lie outside the $\varepsilon$ dependent bounds of \eqref{eq:svr:constraints} and operate as landmarks for the others.
They are called \textit{support vectors} (accessible using the attribute \texttt{support\_vectors\_} in \texttt{svm.SVR}), hence the name of the algorithm. There can be at most $N$ when $\varepsilon \to 0^+$.
As a consequence any sum involving $\alpha^{(i)}$ or $\beta^{(i)}$ can be restricted to the subset of support vectors.
Using the kernel notation, the predictions will therefore be
\begin{equation}
y_{pred}^{(i)}
=
y_{pred}( x^{(i)} )
=
\sum\limits_{n = 1}^{F'} w_n \phi_n( x^{(i)} ) + b
=
\sum\limits_{a \in A} \theta^{(a)} \mathrm{K}( x^{(i)}, l^{(a)} ) + b,
\end{equation}
where $A \subset \lbrace 1, 2, \dots, N \rbrace$ is the subset of labels of the support vectors.

In \Cref{sec:res:svr} we consider two different implementations of the SVM algorithm:
\begin{itemize}
	\item the \textit{linear kernel}, namely the case when $K \equiv \mathrm{id}$ and the loss, in the \texttt{scikit-learn} implementation of \texttt{svm.LinearSVR}, can be simplified to
	\begin{equation}
	\mathcal{L}(w, b)
	=
	\mathrm{C} \sum\limits_{i = 1}^N \sum\limits_{n = 1}^{F'} \max\left( 0, \abs{ y^{(i)} - w_n \phi_n( x^{(i)} - b) } - \varepsilon \right) + \frac{1}{2} \sum\limits_{n = 1}^{F'} w_j^2,
	\end{equation}
	without resolving to the dual formulation of the problem.

	\item the Gaussian kernel (called \texttt{rbf}, from \textit{radial basis function}) in which
	\begin{equation}
	\mathrm{K}(x^{(i)}, l^{(a)}) = \exp\left( - \gamma \sum\limits_{n = 1}^F \left( x^{(i)}_n - l^{(a)}_n \right)^2 \right).
	\end{equation}
\end{itemize}
From the definition of the loss function \eqref{eq:svr:loss} and the kernels, we can appreciate the role of the main hyperparameters of the algorithm.
While the interpretation of $\varepsilon$ is straightforward as the margin allowed without penalty for the prediction, $\gamma$ represents the width of the normal distribution used to map the features in the higher dimensional space.
Furthermore, $C$ plays a similar role to the $l_2$ additional term in \eqref{eq:ridge:loss} by controlling the entity of the penalty for samples outside the $\varepsilon$-dependent bound, however its relation to the linear regularisation is $\alpha_{ridge} = C^{-1}$, thus $C > 0$ by definition.

Given the nature of the algorithm, SVMs are powerful tools which usually grant better results in both classification and regression tasks with respect to logistic and linear regression, but they scale poorly with the number of samples used during training.
In particular the time complexity is at worst\footnotemark $\mathrm{O}(F \times N^3)$ due to the quadratic nature of \eqref{eq:svr:loss-v2} and the computation of the kernel function for all samples: for large datasets ($N \gtrsim 10^4$) they are usually outperformed by ANNs.
\footnotetext{In general it is plausible that the time complexity is $\mathrm{O}(F \times N^2)$ based on good implementations of caching in the algorithm.}

\subsection{Decision Trees, Random Forests and Gradient Boosting}
\label{sec:app:trees}

Decision trees are supervised ML algorithms which model simple decision rules based on the input data~\cite{Quinlan:1986:DecisionTrees,Wittkowski:1986:CART}.
They are informally referred to with the acronym CART (from \textit{Classification And Regression Trees}) and their name descends from the binary tree structure coming from such decision functions separating the input data at each iteration (\textit{node}), thus creating a bifurcating structure with \textit{branches} (the different paths, or decisions made) and \textit{leaves} (the samples in each branch): the basic idea behind them is an \textit{if\dots then\dots else} structure.
In \texttt{scikit-learn} this is implemented in the classes \texttt{tree.DecisionTreeClassifier} and \texttt{tree.DecisionTreeRegressor}.

The idea behind it is to take input samples $x^{(i)} \in \R^F$ (for $i = 1, 2, \dots, N$) and partition the space in such a way that data with the same label $y^{(i)} \in \R$ is on the same subset of samples (while for classification this may be natural to visualise, for regression this amounts to approximate the input data with a step function whose value is constant inside the partition).
Let in fact $j = 1, 2, \dots, F$ be a feature and $x^{(i)}_j$ the corresponding value for the sample $i$, at each node $n$ of the tree we partition the set of input data $\mathcal{M}_n$ into two subsets:
\begin{equation}
\begin{split}
\mathcal{M}^{[1]}_n( t_{j,\, n} )
& =
\left\lbrace (x^{(i)}, y^{(i)}) \in \R^F \times \R \quad \vert \quad x^{(i)}_j < t_{j,\, n} \quad \forall i \in A_n \right\rbrace,
\\
\mathcal{M}^{[2]}_n( t_{j,\, n} )
& =
\mathcal{M}_n \setminus \mathcal{M}^{[1]}_n( t_{j,\, n} ),
\end{split}
\end{equation}
where $A_n$ is the full set of labels of the data samples in the node $n$ and $t_{j,\, n} \in \R$ is a threshold value for the feature $j$ at node $n$.

The measure of the ability of the split to reach the objective (classifying or creating a regression model to predict the labels) is modelled through an \textit{impurity} function (i.e. the measure of how often a random data point would be badly classified or how much it would be badly predicted).
Common choices in classification tasks are the Gini impurity, a special quadratic case of the Tsallis entropy (which in turn is a generalisation of the Boltzmann-Gibbs entropy, recovered as the first power of the Tsallis entropy) and the information theoretic definition of the entropy.
In regression tasks it is usually given by the $l_1$ and $l_2$ norms of the deviation from different estimators (mean and median) for each node $n$:
\begin{itemize}
	\item \textit{mean absolute error}
	\begin{equation}
	H^{[l]}_n(x;\, t_{j,\, n}) = \frac{1}{\Abs{\mathcal{M}^{[l]}_n( t_{j,\, n} )}} \sum\limits_{i \in A^{[l]}_n} \Abs{y^{(i)} - \tilde{y}^{[l]}_{pred,\, n}( x )},
	\quad
	( x^{(i)}, y^{(i)} ) \in \mathcal{M}_n( t_{j,\, n} ),
	\end{equation}
	\item \textit{mean squared error}:
	\begin{equation}
	H^{[l]}_n(x;\, t_{j,\, n}) = \frac{1}{\Abs{\mathcal{M}^{[l]}_n( t_{j,\, n} )}} \sum\limits_{i \in A^{[l]}_n} \left( y^{(i)} - \bar{y}^{[l]}_{pred,\, n}( x ) \right)^2,
	\quad
	( x^{(i)}, y^{(i)} ) \in \mathcal{M}_n( t_{j,\, n} ),
	\end{equation}
\end{itemize}
where $\Abs{\mathcal{M}^{[l]}_n( t_{j,\, n} )}$ is the cardinality of the set $\mathcal{M}^{[l]}_n( t_{j,\, n} )$ for $l = 1, 2$ and
\begin{equation}
\tilde{y}^{[l]}_{pred,\, n}( x ) = \underset{i \in A^{[l]}_n}{\mathrm{median}}~ y_{pred}( x^{(i)} ),
\qquad
\bar{y}^{[l]}_{pred,\, n}( x ) = \frac{1}{\Abs{A^{[l]}_n}} \sum\limits_{i \in A^{[l]}_n} y_{pred}( x^{(i)} ),
\end{equation}
where $A_n^{[l]} \subset A_n$ are the subset of labels in the left and right splits ($l = 1$ and $l = 2$, that is) of the node $n$.

The full measure of the impurity of the node $n$ and for a feature $j$ is then:
\begin{equation}
G_{j,\, n}(\mathcal{M};\, t_{j,\, n})
=
\frac{\Abs{\mathcal{M}_n^{[1]}( t_{j,\, n} )}}{\Abs{\mathcal{M}_n}} H^{[1]}_n( x;\, t_{j,\, n} )
+
\frac{\Abs{\mathcal{M}_n^{[2]}( t_{j,\, n} )}}{\Abs{\mathcal{M}_n}} H^{[2]}_n( x;\, t_{j,\, n} ),
\end{equation}
from which we select the parameters
\begin{equation}
\hat{t}_{j,\, n}
=
\underset{t_{j,\, n}}{\mathrm{argmin}}~ G_n( \mathcal{M}_n;\, t_{j,\, n} ).
\label{eq:trees:lossmin}
\end{equation}
We then recurse over all $\mathcal{M}_n^{[l]}( \hat{t}_{j,\, n} )$ (for $l = 1, 2$) until we reach the maximum allowed depth of the tree (at most $\Abs{\mathcal{M}_n} = 1$).

Other than just predicting a class or a numeric value, decision trees provide a criterion to assign the importance of each feature appearing in the nodes.
The implementation of the procedure can however vary between different libraries: in \texttt{scikit-learn} the importance of a feature is computed by the total reduction in the objective function due to the presence of the feature, normalised over all nodes.
Namely it is defined as the difference between the total impurity normalised by the total amount of samples in the node and the sum of the separate impurities of the left and right split normalised over the number of samples in the respective splits, summed over all the nodes.
Thus features with a high \textit{variable ranking} (or \textit{variable importance}) are those with a higher impact in reducing the loss of the algorithm and can be expected to be seen in the initial branches of the tree.
A measure of the variable importance is in general extremely useful for feature engineering and feature selection since it gives a natural way to pick features with a higher chance to provide a good prediction of the labels.

By nature decision trees have a query time complexity of $\mathrm{O}( \log(N) )$ as most binary search algorithms.
However their definition requires running over all $F$ features to find the best split for each sample thus increasing the time complexity to $\mathrm{O}( F \times N \log( N ) )$.
Summing over all samples in the whole node structure leads to the worst case scenario of a time complexity $\mathrm{O}( F \times N^2 \log( N ) )$.
Well balanced trees (that is, nodes are approximately symmetric with the same amount of data samples inside) can usually reduce that time by a factor $N$, but it may not always be the case.

Decision trees have the advantage to be very good at classifying or creating regression relations in the presence of ``well separable'' data samples and they usually provide very good predictions in a reasonable amount of time (especially when balanced).
However if $F$ is very large, a small variation of the data will almost always lead to a huge change in the decision thresholds and they are usually prone to overfit.
There are however smart ways to compensate this behaviour based on \textit{ensemble} learning such as \textit{bagging}\footnotemark and \textit{boosting} as well as \textit{pruning} methods such as limiting the depth of the tree or the number of splits and introducing a dropout parameter to remove certain nodes of the tree.
\footnotetext{The term \textit{bagging} comes from the contraction of \textit{bootstrap} and \textit{aggregating}: predictions are in fact made over randomly sampled partitions of the training set with substitution (i.e. samples can appear in different partitions, known as \textit{bootstrap} approach) and then averaged together (\textit{aggregating}).
	Random forests are an improvement to this simple idea and work best for decision trees: while it is possible to bag simple trees and take their predictions, using the random subsampling as described usually leads to better performance and results.}
Also random forests of trees provide a variable ranking system by averaging the importance of each feature across all base estimators in the bagging aggregator.

As a reference, \textit{random forests} of decision trees (\texttt{ensemble.RandomForestRegressor} in \texttt{scikit-learn}) are ensemble learning algorithms based on fully grown (deep) decision trees.
They were created to overcome the issues related to overfitting and variability of the input data and are based on random sampling of the training data~\cite{Ho:1995:RandomForests}.
The idea is to take $K$ random partitions of the training data and train a different decision tree for each of them and combine the results: for a classification task this would resort to averaging the \textit{a posteriori} (or conditional) probability of predicting the class $c$ given an input $x$ (i.e. the Bayesan probability $P(c \vert x)$) over the $K$ trees, while for regression this amount to averaging the predictions of the trees $y_{pred,\, \hat{n}}^{(i)\, \lbrace k \rbrace}$ where $k = 1, 2, \dots, K$ and $\hat{n}$ is the final node (i.e. the node containing the final predictions).
This defines what has been called a \textit{random forest} of trees which can usually help in improving the predictions by reducing the variance due to trees adapting too much to training sets.

\textit{Boosting} methods are another implementation of ensemble learning algorithms in which more \textit{weak learners}, in this case shallow decision trees, are trained over the training dataset~\cite{Friedman:2001:Boosting,Friedman:2002:Boosting}. In general parameters $\hat{t}_{j,\, n}$ in \eqref{eq:trees:lossmin} can be approximated by an expansion
\begin{equation}
t_{j,\, n}( x )
=
\sum\limits_{m = 0}^M t^{\{m\}}_{j,\, n}( x )
=
\sum\limits_{m = 0}^M \beta^{\{m\}}_{j,\, n} g( x;\, a^{\{m\}}_{j,\, n} ),
\label{eq:trees:par}
\end{equation}
where $g( x;\, a^{\{m\}}_{j,\, n})$ are called \textit{base learners} and $M$ is the number of iterations\footnotemark.
\footnotetext{Different implementations of the algorithm refer to the number of iterations in different way.
	For instance \texttt{scikit-learn} calls them \texttt{n\_estimators} in the class \texttt{ensemble.GradientBoostingRegressor} in analogy to the random forest where the same name is given to the number of trained decision trees, while \texttt{XGBoost} prefers \texttt{num\_boost\_rounds} and \texttt{num\_parallel\_tree} to name the number of boosting rounds (the iterations) and the number of trees trained in parallel in a forest.}
The values of $a^{\{m\}}_{j,\, n}$ and $\beta^{\{m\}}_{j,\, n}$ are enough to specify the value of $t_{j,\, n}( x )$ and can be compute by iterating \eqref{eq:trees:lossmin}:
\begin{equation}
( a^{\{m\}}_{j,\, n},\, \beta^{\{m\}}_{j,\, n} )
=
\underset{\{a_{j,\, n};\, \beta_{j,\, n}\}}{\mathrm{argmin}}~
G_{j,\, n}\left( \mathcal{M}_n;\, t^{\{m-1\}}_{j,\, n}( x ) + \beta_{j,\, n} g( x;\, a_{j,\, n} ) \right).
\label{eq:trees:iter}
\end{equation}
The specific case of boosted trees is simpler since the base learner predicts a constant value $g( x;\, a^{\{m\}}_{j,\, n} )$, thus \eqref{eq:trees:iter} simplifies to
\begin{equation}
\gamma^{\{m\}}_{j,\, n}
=
\underset{\gamma_{j,\, n}}{\mathrm{argmin}}~
G_{j,\, n}\left( \mathcal{M}_n;\, t^{\{m-1\}}_{j,\, n}( x ) + \gamma_{j,\, n} \right).
\end{equation}
Ultimately, the value of the parameters in \eqref{eq:trees:par} are updated using gradient descent as
\begin{equation}
t^{\{m\}}_{j,\, n}( x ) = t^{\{m-1\}}_{j,\, n}( x ) + \nu\, \gamma_{j,\, n}^{\{m\}},
\end{equation}
where $0 \le \nu \le 1$ is the \textit{learning rate} which controls the magnitude of the update.
Through this procedure, boosted trees can usually vastly improve the predictions of very small decision trees by increasing variance over bias.
Another way to prevent overfitting the training set is to randomly \textit{subsample} the features vector by taking a subset of them (in \texttt{scikit-learn} it is represented as a percentage of the total number of features).
Moreover \texttt{scikit-learn} introduces various ways to control the loss of gradient boosting: apart from the aforementioned \textit{least squares} and \textit{least absolute deviation}, we can have hybrid versions of these such as the \textit{huber} loss which combines the two previous losses with an additional hyperparameter $\alpha$ \cite{Fawcett:2001:Huber}. While more implementations are present, also the boosted trees provide a way to measure the importance of the variables as any decision tree algorithm.

\subsection{Artificial Neural Networks}
\label{sec:app:nn}

ANNs are a state of the art algorithm in ML.
They usually outperform any other algorithm in very large datasets (the size of our dataset is roughly at the threshold) and can learn very complicated decision boundaries and functions\footnotemark.
\footnotetext{Despite their fame in the face of the general public, even small networks can prove to be extremely good at learning complicated functions in a small amount of time.}
In the main text we used two types of neural networks: \textit{fully connected} (FC) networks and \textit{convolutional neural networks} (CNN).
They both rely on being built in a layered structure, starting from the input layers (e.g. the configuration matrix of CY manifolds or an RGB image or several engineered features) going towards the output layers (e.g. the Hodge numbers or the classification class of the image).

In FC networks the input of layer $l$ is a feature vector $a^{(i)\, \{l\}} \in \R^{n_l}$ (for $i = 1, 2, \dots, N$) and, as shown in \Cref{fig:nn:dense}, each layer is densely connected to the following\footnotemark.
\footnotetext{Clearly the input vector $x \in \R^F$ is equivalent to the vector $a^{\{0\}}$ and $n_0 = F$. Inputs to each layer are here represented as a matrix $a^{\{l\}}$ whose columns are made by samples and whose rows are filled with the values of the features.}
In other words, each entry of the vectors $a^{(i)\, \{l\}}_j$ (for $j = 1, 2, \dots, n_l$) is mapped through a function $\psi$ to all the components of the following layer $a^{\{l+1\}} \in \R^{n_{l+1}}$:
\begin{equation}
\begin{split}
\psi:~ & \R^{n_l}~~~ \longrightarrow \R^{n_{l+1}}
\\
& a^{(i)\, \{l\}} \longmapsto a^{(i)\, \{l+1\}} = \psi_j( a^{(i)\, \{l\}} ),
\end{split}
\end{equation}
such that
\begin{equation}
a^{(i)\, \{l+1\}}_j
=
\psi_j( a^{(i)\, \{l\}} )
=
\phi\left( \sum\limits_{k = 1}^{n_l} a^{(i)\, \{l\}}_k W^{\{l\}}_{kj} + b^{\{l\}}\, \mathbb{I}_{j} \right),
\end{equation}
where $\mathbb{I} \in \R^{n_{l+1}}$ is an identity vector.
The matrix $W^{\{l\}}$ is \textit{weight matrix} and $b^{\{l\}}$ is the \textit{bias} term.
The function $\phi$ is a non linear function and plays a fundamental role: without it the successive application of the linear map $a^{\{l\}} \cdot W^{\{l\}} + b\, \mathbb{I}$ would prevent the network from learning more complicated decision boundaries or functions as the ANN would only be capable of reproducing linear relations.
$\phi$ is known as \textit{activation function} and can assume different forms, as long as its non linearity is preserved (e.g. a \textit{sigmoid} function in the output layer of a network squeezes the results in the interval $[0, 1]$ thus reproducing the probabilities of of a classification).
A common choice is the \textit{rectified linear unit} ($\mathrm{ReLU}$) function
\begin{equation}
\phi( z ) = \mathrm{ReLU}( z ) = \max( 0, z ),
\end{equation}
which has been proven to be better at training deep learning architectures~\cite{Glorot:2011:ReLU}, or its modified version $\mathrm{LeakyReLU}( z ) = \max( \alpha z, z )$ which introduces a slope $\alpha > 0$ to improve the computational performance near the non differentiable point in the origin.

CNN architectures were born in the context of computer vision and object localisation~\cite{Tompson:2015:CNN}.
As one can suspect looking at \Cref{fig:nn:lenet} for instance, the fundamental difference with FC networks is that they use a convolution operation $K^{\{l\}} * a^{(i)\, \{l\}}$ instead of a linear map to transform the output of the layers, before applying the activation function\footnotemark.
\footnotetext{In general the input of each layer can be a generic tensor with an arbitrary number of axis.
	For instance, an RGB image can be represented by a three dimensional tensor with indices representing the width of the image, its height and the number of filters (in this case $3$, one for each colour channel).}
This way the network is no longer densely connected, as the results of the convolution (\textit{feature map}) depends only on a restricted neighbourhood of the original feature, depending on the size of the \textit{kernel} window $K^{\{l\}}$ used and the shape of the input $a^{(i) \{l\}}$, which is no longer limited to flattened vectors.
In turn, its size influences the convolution operator which we can compute: one way to see this is to visualise an image being scanned by a smaller window function over all pixels or by skipping some a certain number of them (the length of the \textit{stride} of the kernel).
In general the output will therefore be different than the input, unless the latter is \textit{padded} (with zeros usually) before the convolution. The size of the output is therefore:
\begin{equation}
O_n = \frac{I_n - k_n + 2 p_n}{S_n} + 1, \qquad n = 1, 2, \dots,
\end{equation}
where $O$ is the output size, $I$ the input size, $k$ the size of the kernel used, $p$ the amount of padding (symmetric at the start and end of the axis considered) and $S$ the stride.
In the formula, $n$ runs over the number of components of the input tensor.
While any padding is possible, we are usually interested in two kinds of possible convolutions:
\begin{itemize}
	\item ``same'' convolutions for which $O_n = I_n$, thus $p_n = \frac{I_n ( S_n - 1 ) - S_n + k_n}{2}$,
	\item ``valid'' convolutions for which $O_n < I_n$ and $p_n = 0$.
\end{itemize}

In both cases the learning process aims to minimise the loss function defined for the task: in our regression implementation of the architecture we used the mean squared error of the predictions.
The objective is to find best possible values of weight and bias terms $W^{\{l\}}$ and $b^{\{l\}}$) or to build the best filter kernel $K^{\{l\}}$ through \textit{backpropagation}~\cite{Rumelhart:1986:Backprop}, that is by reconstructing the gradient of the loss function climbing back the network from the output layer to the input and then using the usual gradient descent procedure to select the optimal parameters.
For instance, in the case of FC networks we need to find
\begin{equation}
( \widehat{W}^{\{l\}}, \hat{b}^{\{l\}} )
=
\underset{W^{\{l\}},\, b^{\{l\}}}{\mathrm{argmin}} \frac{1}{2 N} \sum\limits_{i = 1}^N \left( y^{(i)} - a^{(i)\, \{L\}} \right)^2
\quad
\forall l = 1, 2, \dots, L,
\end{equation}
where $L$ is the total number of layers in the network.
A similar relation holds in the case of CNN architectures.
In the main text we use the \textit{Adam}~\cite{Diederik:2014:Adam} implementation of gradient descent and add batch normalisation layers to improve the convergence of the algorithm.

As we can see from their definition, ANNs are capable of learning very complex structures at the cost of having a large number of parameters to tune.
The risk of overfitting the training set is therefore quite evident.
There are in general several techniques to counteract the tendency to adapt the training set, one of them being the introduction of regularisation ($l_2$ and $l_1$) in the same fashion of a linear model (we show it in \Cref{sec:app:linreg}).
Another successful way is to introduce \textit{dropout} layers~\cite{Srivastava:2014:Dropout} where connections are randomly switched off according to a certain retention probability (or its complementary, the dropout \textit{rate}): this regularisation technique allows to keep good generalisation properties since the prediction can rely in a less incisive way on the particular architecture since which is randomly modified during training (dropout layers however act as the identity during predictions to avoid producing random results).

\printbibliography[heading=bibintoc]

\end{document}